\documentclass[article,nojss]{jss}

\usepackage[utf8]{inputenc}
\usepackage{epsfig}
\usepackage{lmodern}
\pdfoutput=1

\newcommand{\rev}[1]{#1}

\usepackage{amsmath}
\usepackage{amssymb,amsfonts}
\usepackage{bbm}

\author{C\'ecile Proust-Lima\\ Inserm U1219 \\ Universit\'e de Bordeaux \And Viviane Philipps \\ Inserm U1219 \\ Universit\'e de Bordeaux \And Benoit Liquet \\ LMAP, UMR CNRS 5142 \\Universit\'e de Pau \\et des Pays de L'Adour}
\title{Estimation of Extended Mixed Models Using Latent Classes and Latent Processes: The \proglang{R} Package \pkg{lcmm}}

\Plainauthor{C\'ecile Proust-Lima, Viviane Philipps, Benoit Liquet} 
\Plaintitle{Estimation of extended mixed models using latent classes and latent processes: the R package lcmm} 
\Shorttitle{\pkg{lcmm}: Estimation of extended mixed models using latent classes and latent processes} 

\Abstract{
The \proglang{R} package \pkg{lcmm} provides a series of functions to estimate statistical models based on linear mixed model theory. It includes the estimation of mixed models and latent class mixed models for Gaussian longitudinal outcomes (\code{hlme}), curvilinear and ordinal univariate longitudinal outcomes (\code{lcmm}) and curvilinear multivariate outcomes (\code{multlcmm}), as well as joint latent class mixed models (\code{Jointlcmm}) for a (Gaussian or curvilinear) longitudinal outcome and a time-to-event that can be possibly left-truncated right-censored and defined in a competing setting. Maximum likelihood esimators are obtained using a modified Marquardt algorithm with strict convergence criteria based on the parameters and likelihood stability, and on the negativity of the second derivatives. The package also provides various post-fit functions including goodness-of-fit analyses, classification, plots, predicted trajectories, individual dynamic prediction of the event and predictive accuracy assessment. This paper constitutes a companion paper to the package by introducing each family of models, the estimation technique, some implementation details and giving examples through a dataset on cognitive aging. 
}
\Plainkeywords{curvilinearity, dynamic prediction,  Fortran 90, growth mixture model, joint model, psychometric tests, R}

\Address{
  C\'ecile Proust-Lima\\
  Inserm U1219\\
  Universit\'e de Bordeaux\\
  146 rue L\'eo Saignat\\
  33076 Bordeaux Cedex \\
  E-mail: \email{Cecile.Proust-Lima@inserm.fr} \\

 \rev{ Viviane Philipps}\\
 \rev{ Inserm U1219}\\
 \rev{  Universit\'e de Bordeaux}\\
 \rev{  146 rue L\'eo Saignat}\\
 \rev{  33076 Bordeaux Cedex} \\
 \rev{  E-mail: \email{viviane.philipps@u-bordeaux.fr}}\\

\rev{Benoit Liquet}\\
\rev{Laboratoire de Math\'ematiques et de leurs Applications,} \\
\rev{Universit\'e de Pau et des Pays de l'Adour,}\\
\rev{UMR CNRS 5142, Pau, France}\\
\rev{ARC Centre of Excellence for Mathematical and Statistical Frontiers}\\
\rev{Queensland University of Technology (QUT)}\\
\rev{Brisbane, Australia}\\
\rev{E-mail: \email{benoit.liquet@pau-univ.fr}} \\

}

\begin{document}

\setkeys{Gin}{width=\textwidth}

\section[Introduction]{Introduction}

The linear mixed model \citep{laird1982,verbeke2000,hedeker2006,fitzmaurice2009} has become a standard statistical method to analyze change over time of a longitudinal Gaussian outcome and assess the effect of covariates on it. Yet longitudinal data collected in cohort studies may be too complex to enter the framework of the linear mixed model:
\begin{itemize}
\item[-] the longitudinal outcomes are not necessarily Gaussian but possibly binary, ordinal or continuous but asymmetric (e.g., absence/presence of symptoms, psychological scale);
\item[-] not only one but several longitudinal outcomes may be collected, especially when the interest is in a biological or psychological process that cannot be measured directly (e.g., quality of life, cognition, immune response);
\item[-] the longitudinal process may be altered by the occurrence of one or multiple times-to-event (e.g., death, onset of disease, disease progression);
\item[-] non-observed heterogeneity may exist in the population (e.g., responders/non-responders, patterns of subjects linked to an unknown behaviour/disease risk/set of genes).
 \end{itemize}
The study of cognitive decline in the elderly combines all these complexities. Indeed, cognition which is the central longitudinal process in aging is not directly observed. It is measured by one or multiple psychometric tests collected repeatedly at cohort visits. These tests are not necessarily Gaussian variables; they can be sumscores, items or ratios. They are usually bounded with asymmetric distributions, or even ordinal \citep{proust-lima_misuse_2011}. In addition, cognitive decline is strongly associated with onset of dementia and death so it may be necessary to jointly model these processes \citep{proust-lima2009}. Finally, a population of elderly subjects usually combines groups of subjects with different types of cognitive trajectory (e.g., normal and pathological aging toward dementia) \citep{proust2005}.   

The aim of the \pkg{lcmm} package developed for \proglang{R} \citep{R} is to provide estimation functions that address these different extensions of the linear mixed model. The package is named \pkg{lcmm} in reference to the Latent Class Mixed Models as all the estimation functions can handle heterogeneity through latent classes of trajectory. However, beyond latent class mixed models, the package provides estimation functions for mixed models involving latent processes, and joint models. 

The estimation relies on the Maximum Likelihood Theory with a powerful modified Marquardt iterative algorithm \citep{marquardt1963}, a Newton-Raphson like algorithm \citep{fletcher1987}, and involves strict convergence criteria. Post-fit functions are also provided in order to assess the goodness-of-fit of the models, their predictive accuracy, as well as to compute outputs and average or individual predictions from the models. 
In its current version \rev{(V1.7.4)}, \pkg{lcmm} includes four main estimation functions: \code{hlme}, \code{lcmm}, \code{multlcmm} and \code{Jointlcmm}. 

The paper is organized as follows.\rev{Section 2 defines the statistical models implemented in the package and Section 3 details the common estimation process. Section 4 describes the implementation of the four estimation functions and gives details on the management of initial values. Section 5 details post-fit analyses and computations. Sections 2, 3 and 5 can be skipped if the reader is already familiar with the statistical methodology.} Finally, Section 6 provides a series of examples based on the \code{paquid} dataset available in the package and Section 7 concludes.

\section[Extended mixed models]{Extended mixed models}

\rev{This section describes each family of statistical models implemented in the package.}The first three subsections detail the standard linear mixed model, its extension to other types of longitudinal outcomes and its extension to the multivariate setting. The fourth subsection is dedicated to their extension to heterogeneous populations with the latent class mixed model theory that applies to all the models estimated within \pkg{lcmm}. Finally, the joint latent class mixed model for jointly analyzing a longitudinal marker and a time-to-event is introduced. 

\subsection[The linear mixed model]{The linear mixed model}\label{sectionLMM}

For each subject $i$ in a sample of $N$ subjects, let consider a vector of $n_i$ repeated measures $Y_i=(Y_{i1},...,Y_{ij},...,Y_{in_i})^\top$ where $Y_{ij}$ is the outcome value at occasion $j$ that is measured at time $t_{ij}$. We distinguish the time of measurement $t_{ij}$ from occasion $j$ because an asset of the linear mixed model is that the times and the number of measurements can vary from one subject to another. This makes it possible for example to include subjects with intermittent missing data and/or dropout, or to consider the actual individual time of measurement rather than the planned visit, which in some applications can greatly differ. 

Following \citet{laird1982}, we define the linear mixed model as follows:

\begin{equation}\label{LMM}
Y_{ij} = X_{Li}(t_{ij})^\top\beta + Z_{i}(t_{ij})^\top u_{i}+w_i(t_{ij})+ \epsilon_{ij}
\end{equation}

where $X_{Li}(t_{ij})$ and $Z_{i}(t_{ij})$ are two vectors of covariates at time $t_{ij}$ of respective length $p$ and $q$. The vector $X_{Li}(t_{ij})$ is associated with the vector of fixed effects $\beta$. The vector $Z_{i}(t_{ij})$, which typically includes functions of time $t_{ij}$, is associated with the vector of random effects $u_{i}$. Shapes of trajectories considered in $X_{Li}$ and $Z_{i}$ can be of any type (polynomial \citep{proust2005}, specifically designed to fit the trajectory \citep{proust-lima2009biostat}, or approximated using a basis of splines).

The vector $u_{i}$ of $q$ random effects has a zero-mean multivariate normal distribution with variance-covariance matrix $B$, where $B$ is an unspecified matrix. The measurement errors $\epsilon_{ij}$ are independent Gaussian errors with variance $\sigma_\epsilon^2$. Finally, the process $(w_i(t))_{t \in \mathbb{R}}$ is a zero-mean Gaussian stochastic process (e.g., Brownian motion with covariance $\text{cov}(w_i(t),w_i(s))=\sigma_w^2 \min(t,s)$ or a stationary process with covariance $\text{cov}(w_i(t),w_i(s))=\sigma_w^2 \exp(-\rho |t-s|$).
 
The vector of parameters to be estimated is $(\beta^\top,\text{vec}(B)^\top,\sigma_w,\rho,\sigma_\epsilon)^\top$ where $\text{vec}(B)$ is the vector of parameters involved for modelling the symmetric positive definite matrix $B$. In the package, $\text{vec}(B)$ is the $q$ standard errors of the random effects in the event of a diagonal matrix $B$ or the $\frac{q(q+1)}{2}$-vector of parameters in the Cholesky transformation of $B$.

\subsection[The latent process mixed model]{The latent process mixed model}\label{sectionLatent}

The linear mixed model applies to longitudinal markers that are continuous and have Gaussian random deviations (random effects, correlated errors and measurement errors). It also assumes that the covariate effects are constant ($\beta$) across the entire range of marker values. In practice, these assumptions do not hold for many longitudinal outcomes, especially psychological scales. The generalized linear mixed model extends the theory to binary, ordinal or Poisson longitudinal outcomes \citep{hedeker2006,fitzmaurice2009}. In order to study non-Gaussian longitudinal markers, we chose another direction by defining a family of mixed models called the latent process mixed models \citep{proust2006,proust-lima2012}. Coming from the latent variable framework, this approach consists in separating the structural model that describes the quantity of interest (a latent process) according to time and covariates from the measurement model that links the quantity of interest to the observations. 

The latent process $\Lambda_i(t)$ is defined in continuous time according to a standard linear mixed model without error of measurement:
\begin{equation}\label{LatentLMM}
\Lambda_i(t) = X_{Li}(t)^\top\beta + Z_{i}(t)^\top u_{i}+w_i(t)~~,~~~\forall t \in \mathbb{R}
\end{equation}

where $X_{Li}(t)$, $Z_{i}(t)$, $\beta$, $u_{i}$ and $w_i(t)$ are defined in Section \ref{sectionLMM}. 

In order to take into account different types of longitudinal markers, a flexible nonlinear measurement model is defined between the latent process $\Lambda_i(t_{ij})$ and the observed value $Y_{ij}$ at the measurement time $t_{ij}$:
\begin{equation}\label{measurement}
Y_{ij}=H(\tilde{Y}_{ij}; \eta)= H(\Lambda_i(t_{ij})+\epsilon_{ij} ; \eta)
\end{equation}
where $\epsilon_{ij}$ are independent Gaussian measurement errors with variance $\sigma_\epsilon^2$, $H$ is a parameterized link function and $\tilde{Y}_{ij}$ denotes the noisy latent process at time $t_{ij}$. 

For a quantitative marker, $H^{-1}$ is a monotonic increasing continuous function. The following are currently implemented:
\begin{itemize}
\item[-] the linear transformation that reduces to the Gaussian framework of the linear mixed model: $H^{-1}(Y_{ij})=\frac{Y_{ij}-\eta_1}{\eta_2}$;
\item[-] the rescaled cumulative distribution function (CDF) of a Beta distribution: $H^{-1}(Y_{ij};\eta)=\frac{h(Y^*_{ij};\eta_{1};\eta_{2})-\eta_{3}}{\eta_{4}}$ with $h(Y^*_{ij};\eta_{1};\eta_{2})=\displaystyle \int_0^{Y^*_{ij}} \dfrac{x^{\eta_{1}^*-1}(1-x)^{\eta_{2}^*-1}} {B(\eta_{1}^*,\eta_{2}^*)} dx$, $B(\eta_{1}^*,\eta_{2}^*)$ is the complete Beta function. For positiveness properties of canonical parameters $\eta_{1}^*$ and $\eta_{2}^*$ and computation reasons, the Beta distribution is parameterized as follows: $\eta_{1}^*=\dfrac{e^{\eta_1}}{e^{\eta_2}(1+e^{\eta_1})}$ 
and $\eta_{2}^*=\dfrac{1}{e^{\eta_1}(1+e^{\eta_2})}$. 
 In addition, $Y_{ij}$ is rescaled in $(0,1)$ using $Y^*_{ij} = \frac{Y_{ij}-\text{min}(Y)+\epsilon_Y}{\text{max}(Y)-\text{min}(Y) + 2\epsilon_Y}$ with the constant $\epsilon_Y>0$ and \text{min}(Y) and \text{max}(Y) the (theoretical or observed) minimum and maximum values of Y. 

\item[-] a basis of quadratic I-splines with $m$ knots: $H^{-1}(Y_{ij};\eta)= \eta_{0} + \sum_{l=1}^{m+1} \eta_{l}^2 B^I_l(Y_{ij})$ with $(B^I_1,...,B^I_{m+1})$ the basis of I-splines \citep{ramsay1988}. 
\end{itemize}

For an ordinal or binary marker (with $M$ levels), Equation~\ref{measurement} reduces to a probit (cumulative) model with $Y_{ij}=H(\Lambda_i(t_{ij})+\epsilon_{ij};\eta)=M_0+l$ if $\Lambda_i(t_{ij})+\epsilon_{ij} \in [\eta_l^*,\eta_{l+1}^*]$ for $l=0,...,M-1$, $M_0$ the minimum value of the marker, $\eta_0=\eta_0^*=-\infty$, $\eta_{M}=\eta_{M}^*=+\infty$ and $\eta_{1}^*=\eta_{1}$, $\eta_{l}^*=\eta_1 + \sum_{j=2}^l\eta_j^2$ for $l>1$ to ensure increasing thresholds $\eta_0^*\leq \eta_1^* \leq ... \leq \eta_{M-1}^* \leq \eta_{M}^*$ for the noisy latent process. 

Latent process mixed models need two constraints to be identified: one on the location of the latent process managed by the mean intercept $\beta_0=0$ and the other for the scale of the latent process managed by $\sigma_\epsilon^2=1$. So the vector of parameters to be estimated is $(\beta^\top,\text{vec}(B)^\top,\sigma_w,\rho,\eta^\top)^\top$ where $\text{vec}(B)$ is defined in Section \ref{sectionLMM}.

\subsection[The latent process mixed model for multivariate longitudinal markers]{The latent process mixed model for multivariate longitudinal markers}\label{sectionMultLatent}

The concept of separation between the structural model for the underlying quantity of interest and the measurement model for its observations applies naturally to the case where multiple longitudinal markers of the quantity of interest are observed. It assumes that the underlying latent process $(\Lambda_i(t))$ defined in Section \ref{sectionLatent} generated $K$ longitudinal markers instead of a unique one. In this case, the latent process can be seen as the common factor underlying the markers \citep{proust2006,proust-lima2012}. 

In this multivariate mixed model, the structural model for $(\Lambda_i(t))$ according to time and covariates is exactly the same as defined in \eqref{LatentLMM} but the measurement model defined in \eqref{measurement} is extended to the multivariate setting in order to take into account the specific relationship between the underlying latent process and each longitudinal marker. 

Let $Y_{kij}$ be the measure of marker $k$ ($k=1,...,K$) for subject $i$ ($i=1,...,N$) at occasion $j$ ($j=1,...,n_{ki}$). The corresponding time of measurement is denoted $t_{kij}$. Note that the number of repeated measurements $n_{ki}$ and the times of measurement can differ according to the subject and the marker for more flexibility. 

The measurement model now takes into account two aspects of the measure: 
\begin{itemize}
\item[-] the measurement error is accounted for through the definition of the intermediate variable $\tilde{Y}_{kij}$:
\begin{equation}\label{tildeY}
\tilde{Y}_{kij}= \Lambda_i(t_{kij})+X_{Yi}(t_{kij})^\top\gamma_k+b_{ki}+\epsilon_{kij}
\end{equation}
where $X_{Yi}(t_{kij})$ are covariates with a marker-specific effect $\gamma_k$ called contrast since $\sum_{k=1}^K \gamma_k = 0$. As in Item Response Theory with Differential Items Functioning \citep{clauser1998}, they capture a differential marker functioning that could have induced a measurement bias if not taken into account. The random intercept $b_{ki}$ also captures a systematic deviation for each subject that would not be captured by covariates; $b_{ki} \sim \mathcal{N}(0,\sigma_{\alpha_k}^2)$. The independent random measurement error $\epsilon_{kij} \sim \mathcal{N}(0,\sigma_{\epsilon_k}^2)$. 

\item[-] the marker-specific nonlinear relationship with the underlying quantity of interest is modelled through the marker-specific link function $H_k$:
\begin{equation}\label{Hk}
Y_{kij}=H_k(\tilde{Y}_{kij}; \eta_k)
\end{equation}
where each $H_k$ is defined using a rescaled Beta CDF, I-splines or a linear transformation as detailed in Section \ref{sectionLatent}. 
\end{itemize}

As in the univariate version of the latent process mixed model, two constraints are required to obtain an identified model. In the multivariate setting, the dimension of the latent process is constrained by the intercept $\beta_0=0$ (for the location) and the variance of the random intercept $\text{Var}(u_{i0})=1$ (for the scale) rather than the standard error of one marker-specific residual error. As a consequence, a random intercept is required and no mean intercept is allowed in the structural model defined in \eqref{LatentLMM}.

The vector of parameters to be estimated is now 
$(\beta^\top,\text{vec}(B)^\top,\sigma_w,\rho,(\gamma_k^\top)_{k=1,K-1},(\sigma_{\alpha_k})_{k=1,K},$ $(\eta_k^\top)_{k=1,K},(\sigma_{\epsilon_k})_{k=1,K})^\top$ where $\text{vec}(B)$ is defined in Section \ref{sectionLMM}.

\subsection[The latent class linear mixed model]{The latent class linear mixed model}\label{sectionLCMM}

The linear mixed model assumes that the population of $N$ subjects is homogeneous and described at the population level by a unique profile $X_{Li}(t)^\top\beta$. In contrast, the latent class mixed model consists in assuming that the population is heterogeneous and composed of $G$ latent classes of subjects characterized by $G$ mean profiles of trajectories.  

Each subject belongs to one and only one latent class so latent class membership is defined by a discrete random variable $c_i$ that equals $g$ if subject $i$ belongs to latent class $g$ ($g=1,...,G$). The variable $c_i$ is latent; its probability is described using a multinomial logistic model according to covariates $X_{ci}$:

\begin{equation} \label{proba}
\pi_{ig}=P(c_i=g | X_{ci})=\dfrac{e^{\xi_{0g}+X_{ci}^\top\xi_{1g}}}{\sum_{l=1}^{G}
e^{\xi_{0l}+X_{ci}^\top\xi_{1l}}}
\end{equation}

\noindent where $\xi_{0g}$ is the intercept for class $g$ and $\xi_{1g}$ is the $q_1$-vector of class-specific parameters associated with the $q_1$-vector of time-independent covariates $X_{ci}$. For identifiability, the scalar $\xi_{0G}=0$ and the vector $\xi_{1G}=0$. When no covariate predicts the latent class membership, this model reduces to a class-specific probability. 

The $G$ mean profiles are defined according to time and covariates through latent class-specific mixed models. The difference with a standard linear mixed model is that both fixed effects and the distribution of the random effects can be class-specific. For a Gaussian outcome, the linear mixed model defined in \eqref{LMM} becomes for class $g$: 

\begin{equation}\label{longitudinal}
Y_{ij}|_{c_i=g} = X_{L1i}(t_{ij})^\top\beta + X_{L2i}(t_{ij})^\top\upsilon_g+ Z_{i}(t_{ij})^\top u_{ig}+w_i(t_{ij})+\epsilon_{ij}
\end{equation}

where $X_{Li}(t_{ij})$ previously defined is split in $X_{L1i}(t_{ij})$ with common fixed effects $\beta$ over classes and $X_{L2i}(t_{ij})$ with class-specific fixed effects $\upsilon_g$. The vector $Z_{i}(t_{ij})$ is still associated with the individual random effects $u_{i}|_{c_i=g}$ called $u_{ig}$ in Equation~\ref{longitudinal} whose distributions are now class-specific. In class $g$, they have a zero-mean multivariate normal distribution with variance-covariance matrix $\omega_g^2B$, where $B$ is an unspecified variance covariance matrix and $\omega_g$ is a proportional coefficient ($\omega_G$=1 for identifiability) allowing for a class-specific intensity of individual variability. The autocorrelated process $w_i(t)$ and the errors of measurement $\epsilon_{ij}$ are the same as in Section \ref{sectionLMM}. 

This extension of the linear mixed model also applies to the latent process mixed model described in Section \ref{sectionLatent} and Section \ref{sectionMultLatent} by replacing the structural model in \eqref{LatentLMM} by:

\begin{equation}\label{LatentLCMM}
\Lambda_i(t)|_{c_i=g} = X_{L1i}(t)^\top\beta + X_{L2i}(t)^\top\upsilon_g+ Z_{i}(t)^\top u_{ig}+w_i(t_{ij}).
\end{equation}

The location constraint for this model becomes $\beta_{01}=0$, that is, the mean intercept in the first class is constrained to 0. The scale constraint remains unchanged. The measurement models remain the same by assuming that the heterogeneity in the population only affects the underlying latent process of interest. The vector of parameters to be estimated defined in Section \ref{sectionLMM}, Section \ref{sectionLatent} and Section \ref{sectionMultLatent} now also includes $((\xi_{0g},\xi_{1g}^\top)_{g=1,G-1},(\upsilon_g^\top)_{g=1,G},(\omega_g)_{g=1,G-1})$.

\subsection[The joint latent class mixed model]{The joint latent class mixed model}\label{sectionJLCMM}

The linear mixed model assumes that the missing data are missing at random \citep{little1995}, that is, the probability that a value of $Y$ is missing is explained by the observations (dependent markers $Y$ and covariates $X$). When this assumption does not hold, the longitudinal process and the missing data process can be simultaneously modelled in a so-called joint model. More generally, it is usual that the longitudinal process is associated with a survival process (e.g., disease onset, death, disease progression), and the joint model captures this correlation to provide valid inference. 

Among joint models, one can distinguish two families: the \emph{shared random effect models} \citep[see][]{Rizopoulos2012} in which functions of the random effects from the linear mixed model are included in the survival model, and the \emph{joint latent class model} \citep{lin2002,proust-lima2014}. The latter is a direct extension of the latent class mixed model described in Section \ref{sectionLCMM}.

It assumes that each of the $G$ latent classes of subjects is characterized by a class-specific linear mixed model for the longitudinal process and a class-specific survival model for the survival process. As such, it is composed of three submodels: the multinomial logistic model defined in \eqref{proba}, the class-specific linear mixed model defined in \eqref{longitudinal} or the class-specific latent process mixed model defined in \eqref{LatentLCMM} and \eqref{measurement}, and finally the class-specific survival model defined below. 

Let $T^*_i$ denote the time-to-event of interest, $\tilde{T}_{i}$ the
censoring time, $T_i=min(T^*_i,\tilde{T}_{i})$ and $E_i=\mathbbm{1}_{T_i^*\leq \tilde{T}_{i}}$. In latent class $g$, the risk of event is described using a proportional hazard model:

\begin{equation}\label{surv_JLCM}
\lambda_i(t)|_{c_i=g}=\lambda_{0g}(t)e^{X_{Si1}^\top \nu+ X_{Si2}^\top \delta_g}
\end{equation}

where $X_{Si1}$ and $X_{Si2}$ are vectors of covariates respectively associated with the vector of parameters common over classes $\nu$ and of class-specific parameters $\delta_g$. The class-specific baseline hazard is defined according to a vector of parameters $\zeta_g$. It can be stratified on the latent class structure ($\lambda_{0g}(t) = \lambda_{0}(t;\zeta_{g})$) or be proportional in each latent class ($\lambda_{0g}(t)= \lambda_{0}(t;\zeta^*)e^{\zeta_{g}}$ with $e^{\zeta_g}$ the proportional factor and  $\zeta_{G}=0$). A series of parametric baseline risk functions parameterized by a vector $\zeta$ are considered: 
\begin{itemize}
\item[-] Weibull specified either by $\lambda_{0}(t;\zeta)=\zeta_1\zeta_2t^{\zeta_2-1}$ or $\lambda_{0}(t;\zeta)=\zeta_1\zeta_2(\zeta_1 t)^{\zeta_2-1}$ depending on the transformation used to ensure positivity of parameters;
\item[-] piecewise constant specified by $\lambda_{0}(t;\zeta)=\sum_{l=1}^{n_z-1}\zeta_l\mathbbm{1}_{t\in[t_l,t_{l+1}]}$ with $n_z$ the number of knots;
\item[-] cubic M-splines specified by $\lambda_{0}(t;\zeta)=\sum_{l=1}^{n_z+2}\zeta_lM_l(t)$ with $n_z$ the number of knots and $(M_l(t))_{l={1,...n_z+2}}$ the basis of cubic M-splines \citep{proust-lima2009}.
\end{itemize}

In these three families of baseline risk functions, parameters are restricted to be positive. This was ensured in practice by a square transformation or an exponential transformation (see paragraph \ref{calljointlcmm}).\\

Instead of a unique cause of event, multiple causes of event can be considered in a competing setting \citep{proust-lima2015}. This is achieved by denoting $T^*_{ip}$ the time to the event of cause $p$ ($p=1,...,P$), and $\tilde{T}_{i}$ the time to censoring so that $T_i=\min(\tilde{T}_i,T^*_{i1},...,T^*_{iP})$ is observed with indicator $E_i=p$ if the event of nature $p$ occurred first or $E_i=0$ if the subject was censored before any occurrence. In this case, the cause-specific and class-specific proportional hazard model is:

\begin{equation}\label{surv_JLCMcompet}
\lambda_{ip}(t)|_{c_i=g}=\lambda_{0gp}(t)e^{{X_{Si1}^{(p)}}^\top \nu_p+ {X_{Si2}^{(p)}}^\top \delta_{gp}}
\end{equation}
where covariates $X_{Si1}^{(p)}$ and $X_{Si2}^{(p)}$ and effects $\nu_p$ and $\delta_{gp}$ can be cause-specific, as well as baseline risk functions $\lambda_{0gp}$. The cause-specific baseline risk functions can be stratified on the latent classes or proportional across latent classes like when one cause is modelled. The covariate effects can be cause-specific or the same over causes. \\


In the joint latent class model, the vector of parameters to be estimated is 
  $((\xi_{0g},\xi_{1g}^\top)_{g=1,G-1},$\\
$(\zeta_{p}^{totT})_{p=1,P},(\nu_{p}^\top)_{p=1,P},(\delta_{gp}^\top)_{p=1,P,g=1,G},\beta^\top,(\upsilon_g^\top)_{g=1,G},\text{vec}(B)^\top,(\omega_g)_{g=1,G-1}),\sigma_w,\rho,\sigma_\epsilon)^\top$  \\
where $\text{vec}(B)$ is defined in Section \ref{sectionLMM} and $\zeta_{p}^{tot}$ includes the vector of class-specific parameters involved in the $G$ $\lambda_{0gp}$.

\section[Estimation]{Estimation}

All these extended mixed models can be estimated within the maximum likelihood framework. For each model, we note $\theta_G$ the entire vector of parameters involved in the model as estimation is performed at a fixed number $G$ of latent classes ($G=1$ for the homogeneous case). The log-likelihood $l(\theta_G)=\sum_{i=1}^N \text{log}(L_i(\theta_G))$ with $L_i$ the individual contribution to the likelihood of the model considered.

\subsection[Individual contributions to the likelihoods]{Individual contributions to the likelihoods}

\subsubsection{Linear mixed model}

The individual contribution to the likelihood of a linear mixed model as defined in Section \ref{sectionLMM} is: 
\begin{equation}\label{contribLMM}
L_i(\theta_1)=\phi_i(Y_i;\theta_1)
\end{equation}
with $\phi_i$ the density function of a multivariate normal distribution with mean $\mu_i=X_{Li}\beta$ and variance $V_i=Z_{i}BZ_i^\top +R_i+\Sigma_i$ with $X_{Li}$ and $Z_i$ the design matrices with row $j$ vectors $X_{Li}(t_{ij})^\top$ and $Z_{i}(t_{ij})^\top$, $\Sigma_i=\sigma_\epsilon^2 \text{I}_{n_i}$ with $\text{I}_n$ the identity matrix of size $n$ and $R_i$ the variance-covariance matrix for the stochastic process $(w_i(t))$. For example, for element $j,j'$ and a Brownian motion, $R_i[j,j']=\sigma_w^2 \min(t_{ij},t_{ij'})$.

\subsubsection{Latent process mixed model}

For continuous link functions, the individual contribution to the likelihood of a latent process mixed model as defined in Section \ref{sectionLatent} is: 
\begin{equation}\label{contribLatentCont}
L_i(\theta_1)=\phi_i(\tilde{Y}_i;\theta_1) \prod_{j=1}^{n_{i}} J(H^{-1}(Y_{ij};\theta_1)) 
\end{equation}
where $\phi_i$ is the same density function of a multivariate normal variable as defined in Equation~\ref{contribLMM}, and $J$ is the Jacobian determinant of the inverse of the link function, that is, the derivative of the linear transformation, the rescaled Beta CDF or the quadratic I-splines. 

For discrete link functions (ordinal data with $M$ levels), the individual contribution to the likelihood of a latent process mixed model as defined in Section \ref{sectionLatent} is written conditionally to the random effects and as such, no stochastic process is considered for the moment ($\forall t,~ w_i(t)=0$):
\begin{equation}\label{contribLatentOrd}
L_i(\theta_1)= \displaystyle \int \prod_{j=1}^{n_{i}}  P(Y_{ij} |u_{i};\theta_1) \phi_u(u_{i})du_{i}
\end{equation}

where $P(Y_{ij} |u_{i};\theta_1)=\prod_{l=0}^{M-1} \left (\Phi(\eta_{l+1}-\Lambda_i(t_{ij});\theta_1)-\Phi(\eta_{l}-\Lambda_i(t_{ij});\theta_1) \right )^{\mathbbm{1}_{Y_{ij}=l}}$ with $\Phi$ the CDF of a standard Gaussian variable, and $\phi_u$ is the density function of a zero-mean multivariate normal variable with variance-covariance matrix $B$.

In the presence of random effects, the integral over the random effects distribution in \eqref{contribLatentOrd} needs to be evaluated numerically. This is done using either the univariate Gauss-Hermite quadrature with 30 points in the presence of a unique random effect or using the multivariate Gauss-Hermite quadrature implemented by \cite{genz1996} otherwise. 

\subsubsection{Latent process mixed model for multivariate longitudinal markers}

Currently, only continuous link functions are considered in the multivariate version of the latent process mixed model defined in Section \ref{sectionMultLatent}. In this case, the individual contribution to the likelihood is: 
\begin{equation}\label{contribLatentContMult}
L_i(\theta_1)=\phi_i(\tilde{Y}_i;\theta_1) \prod_{k=1}^K \prod_{j=1}^{n_{ik}} J_k(H_k^{-1}(Y_{kij})) 
\end{equation}
where $J_k$ is the Jacobian determinant of the link function $H_k^{-1}$ and $\phi_i$ is the density function of a multivariate normal variable with mean $\mu^*_i=(\mu_{1i}^\top,...,\mu_{Ki}^\top)^\top$, $\mu_{ki}=X_{Lik} \beta + X_{Yik} \gamma_k$  and variance-covariance matrix $V^*_i=Z^*_i B {Z^*_i}^\top + R^*_i + \Sigma^*_i$. In these definitions, the matrices $Z_{ik}$, $X_{Lik}$ and $X_{Yik}$ have row vectors $Z_i(t_{kij})$, $X_{Li}(t_{kij})$ and $X_{Yi}(t_{kij})$ for $j=1,...,n_{ik}$, $Z^*_{i}=(Z_{i1}^\top,...,Z_{iK}^\top)^\top$, $R^*_i$ defines the covariance matrix of the stochastic process $(w_i(t))_{t \in \mathbb{R}^+}$ at times $t^*_i=\{t_{kij},k=1,...,K,j=1,...n_{ik}\}$, and $\Sigma^*_{i}$ is the $K$-block diagonal matrix with $k^{\text{th}}$ block $\Sigma_{ik}=\sigma_{\alpha_k}^2J_{n_{ik}}+\sigma_{\epsilon_k}^2I_{n_{ik}}$, $J_n$ the $n \times n$- matrix of elements 1, and $I_{n}$ the $n \times n$- identity matrix.

\subsubsection{Latent class linear mixed model}

The individual contribution to the likelihood of a latent class linear mixed model as defined in Section \ref{sectionLCMM} is: 
\begin{equation}\label{contribLCMM}
L_i(\theta_G)=\sum_{g=1}^G \pi_{ig} \phi_{ig}(Y_i|c_i=g;\theta_G)
\end{equation}
where $\pi_{ig}$ is given in \eqref{proba} and $\phi_{ig}$ is the density function of a multivariate normal distribution with mean $\mu_{ig}=X_{L1i}\beta+X_{L2i}\upsilon_g$ and variance $V_{ig}=Z_{i}B_gZ_i^\top +R_i+\Sigma_i$ and $X_{L.i}$ is the matrix with row $j$ vector $X_{L.i}(t_{ij})$.

Individual contributions to latent process mixed models with latent classes for one or multiple longitudinal markers are obtained by replacing $\phi_{ig}$ in  \eqref{contribLCMM} by the individual contribution given in \eqref{contribLatentCont}, \eqref{contribLatentOrd} or \eqref{contribLatentContMult} with appropriate class-specific parameters.

\subsubsection{Joint latent class mixed model}\label{est_jlcm}

The individual contribution to the likelihood of a joint latent class mixed model as defined in Section \ref{sectionJLCMM} for a $P$-cause right-censored time to event is: 
\begin{equation}\label{contribJLCMM}
L_i(\theta_G)=\sum_{g=1}^G \pi_{ig} \phi_{ig}(Y_i|c_i=g;\theta_G) e^{-\sum_{p=1}^P A_p(T_i \mid c_i=g;\theta_G)}\prod_{p=1}^P \lambda_p(T_i \mid c_i=g;\theta_G)^{\mathbbm{1}_{E_i=p}}
\end{equation}
where $\pi_{ig}$ and $\phi_{ig}$ are defined in \eqref{contribLCMM}, $\lambda_p(t \mid c_i=g;\theta_G)$ is the cause-$p$-specific instantaneous hazard defined in \eqref{surv_JLCMcompet} and $A_p(t \mid c_i=g;\theta_G)$ is the corresponding cumulative hazard. 

With curvilinear longitudinal outcomes, $\phi_{ig}$ is replaced by the individual contribution given in \eqref{contribLatentCont} with appropriate class-specific parameters.

In the case of a left-truncated time-to-event with delayed entry at time $T_{0i}$, the contribution for the truncated data becomes $L_i^{T_0}(\theta_G)=\dfrac{L_i(\theta_G)}{S_i(T_{i0};\theta_G)}$ with the marginal survival function in $T_{i0}$: $S_i(T_{i0};\theta_G)=\sum_{g=1}^G \pi_{ig} e^{-\sum_{p=1}^P A_p(T_{i0} \mid c_i=g;\theta_G)}$.

\subsection[Iterative Marquardt algorithm]{Iterative Marquardt algorithm}

The log-likelihoods of models based on the mixed model theory can be maximized using algorithms in the EM family (e.g., \citet{verbeke1996,muthen1999,xu2001} for latent class mixed models) or the Newton-Raphson family (e.g., \citet{proust2005} for latent class mixed models). In our work,  whatever the type of model, we chose the latter using an extended Marquardt algorithm because of the better convergence rate and speed found in previous analyses. 

In the extended Marquardt algorithm, the vector of parameters $\theta_G$ is updated until convergence using the following equation for iteration $l+1$ :
\begin{equation}\label{eq:update}
\theta_G^{(l+1)}=\theta_G^{(l)}-\delta (\tilde{\mathcal{H}}^{(l)})^{-1}\nabla(L(\theta_G^{(l)}))
\end{equation}
Step $\delta$ equals 1 by default but is internally modified to ensure that the log-likelihood is improved at each iteration. The matrix $\tilde{\mathcal{H}}$ is a diagonal-inflated Hessian to ensure positive-definiteness: if necessary, diagonal terms $\tilde{\mathcal{H}}_{ii}$ are inflated so that $\tilde{\mathcal{H}}_{ii}=\mathcal{H}_{ii}+\lambda[(1-\eta)|\mathcal{H}_{ii}|+\eta \text{tr}(\mathcal{H})]$ where $\mathcal{H}$ is the Hessian matrix with diagonal terms $\mathcal{H}_{ii}$, and $\lambda$ and $\eta$ are initially fixed at $0.01$ and are reduced if $\tilde{\mathcal{H}}$ is positive-definite and increased if not. 
$\nabla(L(\theta_G^{(l)}))$ is the gradient of the log-likelihood at iteration $l$. First derivatives are computed by central finite differences with steps $2 \times \max(10^{-7},10^{-4}|\theta_{Gv}|)$ for parameter $v$. Second derivatives are computed by forward finite differences with steps $\max(10^{-7},10^{-4}|\theta_{Gu}|)$ and $\max(10^{-7},10^{-4}|\theta_{Gv}|)$ for parameters $u$ and $v$.

Three convergence criteria are used: 
\begin{itemize}
\item[-] one based on parameter stability $\sum_{j=1}^{n_\theta} (\theta_{G}(j)^{(l)}-\theta_{G}(j)^{(l-1)})^2 \leq \epsilon_a $;
\item[-] one based on log-likelihood stability $|L^{(l)}-L^{(l-1)}| \leq \epsilon_b $;
\item[-] one based on the size of the derivatives $\frac{\nabla(L(\theta_G^{(l)}))^\top\mathcal{H}^{(l)-1}\nabla(L(\theta_G^{(l)}))^\top}{n_\theta} \leq \epsilon_d $ with $n_\theta$ the length of $\theta_{G}$. 
\end{itemize}
The default values are $\epsilon_a=\epsilon_b=\epsilon_d=10^{-4}$. The thresholds might seem relatively large but the three convergence criteria must be simultaneously satisfied for convergence and the criterion based on derivatives is very stringent so that it ensures a good convergence even at $\epsilon_d=10^{-4}$. A drawback of other algorithms may be that they only converge according to likelihood or parameter stability, and that in complex settings such as latent class mixed models or joint latent class mixed models, the log-likelihood can be relatively flat in some areas of the parameters space so that likelihood or parameter stability does not systematically ensure convergence to an actual maximum.

An estimate of the variance-covariance matrix of the maximum likelihood estimates (MLE) $\widehat{V(\hat{\theta}_G)}$ is provided by the inverse of the Hessian matrix.

\section[Implementation]{Implementation}\label{Implementation}

\rev{The package currently includes 4 estimation functions: 
\begin{itemize}
\item[-] Linear mixed models and latent class linear mixed models are estimated with \code{hlme};
\item[-] Univariate latent process mixed models possibly including latent classes are estimated with \code{lcmm};
\item[-] Multivariate latent process mixed models possibly including latent classes are estimated with \code{multlcmm} or \code{mlcmm};
\item[-] Joint latent class mixed models are estimated with \code{Jointlcmm} or \code{jlcmm}.
\end{itemize}
}

The four estimation functions rely on estimation programs (log-likelihood computation, optimization algorithm) written in \proglang{Fortran 90}. \rev{This section describes the calls of these functions and details the initialization of the iterative algorithm.} 
 
\rev{The package also includes other functions (generic, post-fit, etc). Table \ref{Generics} gives the list of the main functions. The exhaustive list can be obtained with code \texttt{ls("package:lcmm")}. The post-fit functions are detailed in section \ref{sec-post-fit}.} 

\begin{table}
\begin{center}
\begin{tabular}{l p{11cm}}
\hline
Function & Description \\
\hline
\multicolumn{2}{l}{\emph{Estimation functions:}} \\
\code{hlme} & Estimation of latent class linear mixed models\\
\code{lcmm} & Estimation of univariate latent process (and latent class) mixed models \\
\code{multlcmm} or \code{mlcmm} & Estimation of multivariate latent process (and latent class) mixed models\\
\code{Jointlcmm} or \code{jlcmm} & Estimation of joint latent class models for longitudinal and time to event data\\

\hline
\multicolumn{2}{l}{\emph{Generic functions:}} \\
\code{print} & Brief summary of the estimation\\
\code{summary} & Summary of the estimation and tables of maximum likelihood estimates with standard errors and Wald tests.  \\
\code{plot} & Different types of plots (residuals, fit, link functions, hazard, etc) \\
\code{coef} or \code{estimates} & Vector of maximum likelihood estimates (MLE)\\
\code{vcov} or \code{VarCov} & Variance-covariance matrix of the MLE \\
\code{ranef} & Matrix of best linear unbiased predictors for the random effects \\
\code{residuals} or \code{resid} & Subject-specific residuals\\
\code{fitted} & Subject-specific longitudinal predictions\\
\code{fixef} & Vectors of fixed effects by submodel\\
\hline
\multicolumn{2}{l}{\emph{Post-fit functions:}} \\
\code{WaldMult} & Univariate and multivariate Wald tests for combinations of parameters\\
\code{VarCovRE} & Estimates, standard errors and Wald tests of the parameters constituting the variance-covariance matrix of the random effects.\\
\code{postprob}& Posterior classification stemming from latent class models\\
\code{predictY}& Marginal predictions (possibly class-specific) in the natural scale of the markers for a profile of covariates\\
\code{predictL}& Marginal predictions (possibly class-specific) in the latent process scale for a profile of covariates\\
\code{fitY} & Marginal predictions of the longitudinal observations in their natural scale\\
\code{predictlink} & Confidence intervals for estimated link functions\\
\code{cuminc}&  Predictive cumulative incidence of event according to a profile of covariates\\
\code{dynpred}& Individual dynamic prediction from a joint latent class model\\
\code{epoce}& Estimators of the expected prognostic cross-entropy \\
\code{Diffepoce}& Difference of expected prognostic cross-entropy estimators \\
\code{VarExpl}& Percentage of variance explained by the (latent class) linear mixed model regression\\
\hline
\multicolumn{2}{l}{\emph{Other functions:}} \\
\code{gridsearch}& Automatic grid search for latent class models\\
\hline
\end{tabular}
\end{center}
\caption{\rev{Brief description of main functions available in \pkg{lcmm} package.}}\label{Generics}
\end{table}
\subsection[hlme call]{\code{hlme} call}\label{callhlme}

The call of \code{hlme} is 
\begin{verbatim}
hlme(fixed, mixture, random, subject, classmb, ng = 1, idiag = FALSE, 
    nwg = FALSE, cor = NULL, data, B, convB = 0.0001, convL = 0.0001,
    convG = 0.0001, prior, maxiter = 500, subset = NULL, na.action = 1, 
    posfix = NULL)
 \end{verbatim}

Argument \code{fixed} defines the two-sided formula for the linear regression at the population level with the dependent variable ($Y$) on the left-hand side and the combination of covariates ($X_{L1}$ and $X_{L2}$) with fixed effects on the right-hand side. 
Argument \code{random} defines the one-sided formula with the covariates having a random effect ($Z$). 
Argument \code{subject} provides the name of the identification variable for the random effects.  
Argument \code{ng} indicates the number of latent classes $G$. When $G>1$, \code{mixture} indicates a one-sided formula with the subset of covariates having a class-specific effect ($X_{L2}$) and \code{classmb} provides the optional covariates explaining the latent class membership ($X_c$). An optional argument \code{prior} provides a vector of \emph{a priori} known class-memberships when relevant (very rare).

Argument \code{idiag} indicates whether the variance-covariance matrix for the random effects ($B$) is diagonal (\code{TRUE}) or unstructured (\code{FALSE} by default), the boolean \code{nwg} indicates whether the matrix $B$ is proportional over classes ($\omega_g \neq 1~,~g=1,G-1$), and \code{cor} indicates the nature of the optional zero-mean Gaussian stochastic process, either a Brownian motion with \code{cor=BM(time)} or a stationary process with \code{cor=AR(time)} with \code{time} the time variable (by default, none is included). Argument \code{data} provides the name of the dataframe containing the data in the longitudinal format, that is, with $n_i$ rows by subject (or $\max_k(n_{ik})$ for \code{multlcmm}). Optional \code{subset} provides the vector of rows to be selected in the dataframe and \code{na.action} is an indicator for the management of missing data which are omitted by default. 

Argument \code{B} specifies the vector of initial values. \rev{This argument is described in detail in section \ref{sec-initvalues}. When a vector is specified in \code{B}, argument \code{posfix} can be used to fix some parameters to the value indicated in \code{B}. These parameters are not estimated.} Arguments \code{convB}, \code{convL}, \code{convG} indicate the thresholds for convergence criteria on the parameters, the log-likelihood and the derivatives, respectively. Argument \code{maxiter} indicates the maximum number of iterations in the optimization algorithm.

\subsubsection{Example of call}

\rev{The functionalites of \code{hlme} are detailed in section \ref{ex_hlme} using the \code{paquid} dataset. We give here two simple examples of calls with \code{data$\_$hlme}, a simulated dataset available in the package:}

\begin{Schunk}
\begin{Sinput}
R> hlme1 <- hlme(Y ~ Time * X1, random =~ Time, subject = 'ID', ng = 1,
+    data = data_hlme)
R> hlme2 <- hlme(Y ~ Time * X1, random =~ Time, subject = 'ID', ng = 2,
+    data = data_hlme, mixture =~ Time, classmb =~ X2 + X3, B = hlme1)
\end{Sinput}
\end{Schunk}

\rev{
From dataset \code{data$\_$hlme}, the first call (\code{hlme1}) fits a standard linear mixed model in which the dependent variable \code{Y} is explained according to \code{Time}, \code{X1} and the interaction betwen \code{Time} and \code{X1}. Two correlated random effects are assumed for the intercept and \code{Time}. These random effects are grouped by \code{ID}, the identification variable for the subjects. }

\rev{The second call fits a 2-class linear mixed model  (\code{hlme2}) in which the dependent variable \code{Y} is explained again according to \code{Time}, \code{X1} and the interaction betwen \code{Time} and \code{X1} but the intercept and the effect of \code{Time} are different in class 1 and 2. The same two correlated random effects are assumed for the intercept and \code{Time} grouped by \code{ID}. The latent-class membership is explained according to two time-independent covariates \code{X2} and \code{X3}. Finally, the iterative algorithm starts from automatic initial values generated from the estimates of model \code{m1} (see section \ref{sec-initvalues}.)}

\subsection[lcmm call]{\code{lcmm} call}

The call of \code{lcmm} has the same structure as the one of \code{hlme}. It is

\begin{verbatim}
lcmm(fixed, mixture, random, subject, classmb, ng = 1, idiag = FALSE, 
    nwg = FALSE,  link = "linear", intnodes = NULL, epsY = 0.5, cor = NULL, 
    data, B, convB = 1e-04, convL = 1e-04,  convG = 1e-04, maxiter = 100, 
    nsim = 100, prior, range = NULL, subset = NULL, na.action = 1, 
    posfix = NULL, partialH = FALSE)
\end{verbatim}

Most of the arguments are detailed in Section \ref{callhlme}. Arguments are added to specify the link function ($H$ or $H^{-1}$) in Equation~\ref{measurement}. Argument \code{link} indicates the nature of the link function, either \code{link="linear"} for a linear transformation, \code{link="beta"} for a rescaled Beta CDF, \code{link="thresholds"} for the cumulative probit model or \code{link="splines"} for a I-splines transformation. In the case of a splines transformation, the number and location of the knots (5 knots placed at quantiles of Y by default) can be specified by \code{link="X-type-splines"} where \code{X} is the total number of knots (\code{X}$>2$) and \code{type} is \code{equi}, \code{quant} or \code{manual} for knots placed at regular intervals, at the percentiles of Y distribution or knots entered manually. In the latter case, argument \code{intnodes} provides the vector of internal knots. Optional argument \code{epsY} provides the constant used to rescale $Y$ with the rescaled Beta CDF. Optional \code{range} indicates the range of $Y$ that should be considered when different from the observed one in the data and when using Splines or Beta CDF transformations only. Finally, \code{nsim} indicates the number of equidistant values within the range of $Y$ at which the estimated link function should be computed in output. \rev{When link functions are Beta CDF or Splines, option \code{partialH=TRUE} indicates that the corresponding parameters should not be considered in the final Hessian matrix computation. This might solve problems of convergence due to parameters at the hedge of the parameter space. However, in such situations, fixing the problematic parameters using \code{posfix} usually works better.}

\subsubsection{Example of call}

\rev{The functionalites of \code{lcmm} are detailed in section \ref{ex_lcmm} using the \code{paquid} dataset. We give here two simple examples of calls with \code{data$\_$lcmm}, a simulated dataset available in the package:}

\begin{Schunk}
\begin{Sinput}
R> lcmm1 <- lcmm(Ydep2 ~ poly(Time, degree = 2, raw = TRUE), random =~ Time,
+    subject = 'ID', data = data_lcmm)
R> lcmm2 <- lcmm(Ydep2 ~ poly(Time, degree = 2, raw = TRUE), random =~ Time,
+    subject = 'ID', data = data_lcmm, link = "5-quant-splines")
\end{Sinput}
\end{Schunk}

\rev{
From dataset \code{data$\_$lcmm}, the first call (\code{lcmm1}) fits a standard linear mixed model in which the dependent variable \code{Ydep2} is explained according to a quadratic function of \code{Time} at the population level (fixed effects), and a linear function of Time at the individual level with 2 correlated random effects on the intercept and \code{Time} (the random effect on the quadratic function of \code{Time} is not relevant on these data). The random effects are grouped by \code{ID}, the identification variable for the subjects. This model could also have been fitted with \code{hlme} function.}

\rev{The second call (\code{lcmm2}) fits exactly the same model except that a nonlinear link function is considered to normalize \code{Ydep2}. The nonlinear function is a basis of quadratic I-splines with 5 knots placed at the quantiles of \code{Ydep2} distribution.}

\subsection[multlcmm call]{\code{multlcmm} call}

The call of \code{multlcmm} \rev{(or of its shortcut \code{mlcmm})}  uses the same structures as those of \code{hlme} and \code{lcmm}. It is

\begin{verbatim}
multlcmm(fixed, mixture, random, subject, classmb, ng = 1, idiag = FALSE,
    nwg = FALSE, randomY = FALSE, link = "linear", intnodes = NULL, 
    epsY = 0.5, cor = NULL, data, B, convB = 1e-04, convL = 1e-04, 
    convG = 1e-04, maxiter = 100, nsim = 100, prior, range = NULL, 
    subset = NULL, na.action = 1, posfix = NULL, partialH = FALSE)
\end{verbatim}

To account for the multivariate nature of the model estimated by \code{multlcmm}, the left-hand side of \code{fixed} formula includes the sum of all the dependent variables' names, and the right-hand side can now include covariates with a mean effect on the common factor or marker-specific effects in addition to the mean effect (with \code{contrast(X)} instead of \code{X}). 
When the family of link functions is not the same for all dependent variables, a vector of link function names is provided in \code{link}. Argument \code{intnodes} now provides the vector of internal knots for all the Splines link functions involving knots entered manually. Argument \code{range} possibly indicates the range of the dependent variables with transformations Splines or Beta CDF when it differs from the one observed in the data. 

The boolean argument \code{randomY} is the only specific argument of \code{multlcmm}. It indicates whether marker-specific random intercepts ($b_{ki}$) should be considered in the measurement model defined in \eqref{tildeY}.

\subsubsection{Example of call}

\rev{The functionalites of \code{multlcmm} are detailed in section \ref{ex_multlcmm} using the \code{paquid} dataset. We give here an example of call with \code{data$\_$lcmm}, a simulated dataset available in the package:}

\begin{Schunk}
\begin{Sinput}
R> mlcmm1 <- multlcmm(Ydep1 + Ydep2 + Ydep3 ~ X1 * poly(Time, degree = 2, 
+    raw = TRUE), random =~ Time, subject = 'ID', data = data_lcmm,
+    link=c("linear","3-quant-splines","3-quant-splines"))
\end{Sinput}
\end{Schunk}

\rev{From dataset \code{data$\_$lcmm}, the call fits a latent process mixed model for three dependent variables \code{Ydep1}, \code{Ydep2}, \code{Ydep3} in which the latent process underlying the three dependent variables is explained at the population level (fixed effects) according to a quadratic function of \code{Time} for each level of the binary covariate \code{X1}, and at the individual level according to a linear function of \code{Time} with 2 correlated random effects on the intercept and \code{Time}. The random effects are grouped by \code{ID}, the identification variable for the subjects. The three tests are transformed using a linear transformation for \code{Ydep1} and bases of quadratic I-splines with one internal knot placed at the median for \code{Ydep2} and \code{Ydep3}. No marker-specific random intercept or marker-specific effects of covariates are considered here. }

\subsection[Jointlcmm call]{\code{Jointlcmm} call}\label{calljointlcmm}

The call of \code{Jointlcmm} \rev{(or of its shortcut \code{jlcmm})} is

\begin{verbatim}
Jointlcmm(fixed, mixture, random, subject, classmb, ng = 1, idiag = FALSE, 
    nwg = FALSE, survival, hazard = "Weibull", hazardtype = "Specific", 
    hazardnodes = NULL, TimeDepVar = NULL, link = NULL, intnodes = NULL, 
    epsY = 0.5, range = NULL, cor = NULL, data, B, convB = 1e-4, convL = 
    1e-4, convG = 1e-4, maxiter = 100, nsim = 100, prior, logscale = FALSE, 
    subset = NULL, na.action = 1, posfix = NULL, partialH = FALSE)
\end{verbatim}

Most arguments of \code{Jointlcmm} call are the same as those of \code{hlme} or \code{lcmm} calls. Arguments defining the class-specific survival model are added. Argument \code{survival} is a two-sided formula that defines the structure of the survival model. The left-hand side includes a \code{Surv} object as defined in \pkg{survival} package \citep{survival-package}. The right-hand side indicates the covariates involved in the survival model with \code{mixture(X)} when covariate \code{X} has a class-specific effect, \code{cause(X)} when \code{X} has a different effect on all the causes of event or \code{cause1(X)} when \code{X} has an effect on type 1 cause (similar functions for causes 2 to $P$). 

Argument \code{hazard} indicates the family of the baseline risk functions or the vector of families of the cause-specific baseline risk functions in the presence of competing events. The program includes \code{"Weibull"} for 2-parameter Weibull hazards, \code{"piecewise"} for piecewise constant hazards or \code{"splines"} for hazard approximated using M-splines. By default, \code{"piecewise"} and \code{"splines"} consider 5 regular knots within the range of event times. The first knot is at the minimum time of entry and the last knot is at the maximum observed time. The number and locations of the knots can be specified by indicating \code{"X-type-piecewise"} or \code{"X-type-splines"} where \code{X} is the total number of knots (\code{X}$>2$) and \code{type} is \code{equi}, \code{quant} or \code{manual} for knots placed at regular intervals, at the percentiles of the event times distribution or knots entered manually. In the latter case, argument \code{hazardnodes} provides the corresponding vector of internal knots. Argument \code{hazardtype} indicates whether the baseline risk functions are stratified on the latent classes (\code{hazardtype="Specific"}) or proportional across latent classes (\code{hazardtype="PH"}) or common over latent classes (\code{hazardtype="Common"}). 

Two parameterizations are implemented to ensure the positivity of the parameters of the baseline risk functions: \code{logscale=TRUE} uses the exponential transformation and \code{logscale=FALSE} uses the square transformation. For Weibull, these parameterizations also imply two different specifications of the baseline hazard $\lambda_{0}(t;\zeta)=\zeta_1\zeta_2t^{\zeta_2-1}$ with \code{logscale=TRUE} (by noting $\zeta^*$ the vector of unconstrained parameters to be estimated, $\zeta=\exp(\zeta^*)$) and $\lambda_{0}(t;\zeta)=\zeta_1\zeta_2(\zeta_1 t)^{\zeta_2-1}$ with \code{logscale=FALSE}  (by noting $\zeta^*$ the vector of unconstrained parameters to be estimated, $\zeta=(\zeta^*)^2$). Indeed, depending on the range of times to events, one specification or the other may be better suited to ensure convergence of the program. The optional argument \code{nsim} indicates the number of points within the range of event times at which the estimated baseline risk functions and cumulative risk functions should be computed in output.

\subsubsection{Example of call}

\rev{The functionalites of \code{Jointlcmm} are detailed in section \ref{ex_Jointlcmm} using the \code{paquid} dataset. We give here an example of calls with \code{data$\_$lcmm}, a simulated dataset available in the package:}

\begin{Schunk}
\begin{Sinput}
R> jlcmm1 <- Jointlcmm(Ydep1 ~ X1 * Time, random =~ Time, subject = 'ID',
+    survival = Surv(Tevent, Event) ~ X1 + X2, hazard = "3-quant-splines",
+    data = data_lcmm)
R> jlcmm2 <- Jointlcmm(Ydep1 ~ Time * X1, random =~ Time, subject = 'ID', 
+    mixture =~ Time, survival = Surv(Tevent, Event) ~ X1 + mixture(X2),
+    hazard = "3-quant-splines", hazardtype = "PH", ng = 2, data = data_lcmm, 
+    B = jlcmm1)
\end{Sinput}
\end{Schunk}

\rev{From dataset \code{data$\_$lcmm}, the first call (\code{jlcmm1}) fits a linear mixed model and a survival model independently (since there is a unique latent class by default). The linear mixed model explains the dependent variable \code{Ydep1} according to a linear trajectory of \code{Time} specific to each level of \code{X1} at the population level and accounts for two random effects on the intercept and \code{Time} at the individual level. The grouping variable is \code{ID}. The survival model for the censored observed time \code{Tevent} (with \code{Event} the indicator of event) involves a baseline risk function approximated by a basis of cubic M-splines with one internal knot placed at the quantiles of the times of event and an effect of \code{X1} and \code{X2}. }

\rev{The second call (\code{jlcmm2}) fits the model in the case of two latent classes. The linear mixed model has the same definition as above except that the linear trajectory according to \code{Time} is now specific to each latent class while the effects of \code{X1} and \code{X1:Time} remain the same in the two classes. No covariate explains the latent class membership. In the survival model, the baseline risk function is still approximated by cubic M-splines but the risk of event is now proportional in each latent class and the effect of \code{X2} is class-specific. The effect of \code{X1} remains common over classes. Initial values for the iterative algorithm are automatically specified from \code{jlcmm1} estimates (see below). }

\subsection[Initial Values]{Initial values}\label{sec-initvalues}

Iterative estimation algorithms need to be initialized using a set of initial values for the vector of parameters $\theta$. \rev{In each estimation function, argument \code{B} specifies the initialization of the algorithm. Depending on the number of latent classes, several techniques are available.}

\subsubsection[One class]{In the presence of a unique latent class}

\rev{The default initial values are defined in Table \ref{initial_values}. Alternatively, the user can enter any vector of specific initial values in argument \code{B}.}

\begin{table}
\begin{center}
\begin{tabular}{lc}

Parameters & Initial value \\
\hline

\multicolumn{2}{l}{\emph{\bf{Linear mixed model in \code{hlme}, \code{lcmm}, \code{multlcmm}, \code{Jointlcmm}}}}\\
$\beta$ \& $\upsilon$ (intercepts in \code{hlme}) & $\bar{Y}$ \\
 $\beta$ \& $\upsilon$ (coefficients) & 0 \\
 $B$ & $I_q$ \\
  $\sigma_w$ & 1 \\
  $\rho$ & 0  \\
  $\sigma_{\epsilon}$ (except for \code{multcmm})  & 1 \\
 & \\
\multicolumn{2}{l}{\emph{\bf{Outcome-specific regression in \code{multlcmm}}}}\\
  $\sigma_{\alpha_k}$, $k=1,...,K$ & 1  \\
  $\gamma_k$, $k=1,...,K$ & 0  \\
  $\sigma_{\epsilon_k}$, $k=1,...,K$  & 1 \\
 & \\
\multicolumn{2}{l}{\emph{\bf{Link functions in \code{lcmm}, \code{multlcmm}, \code{Jointlcmm} (for $k=1,...,K$)}}}\\
$\eta_k$ for \code{link="linear"}   & ($\bar{Y}$,1) \\
$\eta_k$ for \code{link="Splines"}  & (-2,0.1,...,0.1) \\
$\eta_k$ for \code{link="Beta"}  & $(0,-\log(2),0.7,0.1)$ \\
$\eta_k$ for \code{link="thresholds"} \& $M=2$  & 0 \\
$\eta_k$ for \code{link="thresholds"} \& $M>2$ & $\left ( \frac{2\mathcal{U}_{0.98}(-\text{med}(Y)+\min{(Y)}+1)}{M-2},\sqrt{\frac{2\mathcal{U}_{0.98}}{M-2}},...,\sqrt{\frac{2\mathcal{U}_{0.98}}{M-2}}\right )$ \\
 & \\
\multicolumn{2}{l}{\emph{\bf{Survival or cause-specific model in \code{Jointlcmm} (for $p=1,...,P$)}}}\\
 $\nu_p$  \& $\delta_{p}$ & 0  \\
 $\zeta_p$ for \code{hazard="Weibull"} \& \code{logscale=T}  & $\left ( \log \left (\frac{\sum \mathbbm{1}_{E_i=p}}{\sum T_i\mathbbm{1}_{E_i=p}} \right ),0 \right )$\\ 
 $\zeta_p$ for \code{hazard="Weibull"} \& \code{logscale=F}  & $\left ( \sqrt{ \frac{\sum \mathbbm{1}_{E_i=p}}{\sum T_i\mathbbm{1}_{E_i=p}}},1 \right )$ \\ 
 $\zeta_p$ for \code{hazard="piecewise"} \& \code{logscale=T}  & $\left (- \log (n_z-1)\right )_{\{1,n_z-1\}}$ \\
 $\zeta_p$ for \code{hazard="piecewise"} \& \code{logscale=F}  & $\left (\sqrt {\frac{1}{n_z-1}} \right )_{\{1,n_z-1\}}$\\ 
 $\zeta_p$ for \code{hazard="splines"} \& \code{logscale=T}  & $\left (- \log (n_z+2)\right )_{\{1,n_z+2\}}$ \\
 $\zeta_p$ for \code{hazard="splines"} \& \code{logscale=F}  & $\left ( \sqrt{\frac{1}{n_z+2}} \right )_{\{1,n_z+2\}}$\\ 
\hline
\end{tabular}
\end{center}
\caption{Automatic choice of initial values for the iterative estimation process when $G=1$. $\bar{Y}$, med($Y$) and $\min(Y)$ respectively indicates the mean, median and minimum of the dependent variable, $(T_i,E_i)$ the couple of survival data, and $n_z$ the number of knots for the baseline risk functions.}\label{initial_values}
\end{table}

\subsubsection[At least two class]{In the presence of at least two latent classes}

In the presence of mixture, initial values are crucial for the correct convergence of the program so specific attention should be paid to this section. Indeed, in mixture modelling, the log-likelihood may have multiple maxima and algorithms based on maximisation of the likelihood might converge to local maxima \citep{redner1984}. This means that convergence towards the global maximum of the log-likelihood is not ensured when running the algorithm once. To ensure the convergence to the global maximum, we thus strongly recommend running each model several times from different sets of initial values (typically from a grid of initial values).

 \rev{There are currently four different ways of initializing the algorithm when $G>1$:} 

\begin{enumerate}

\item \rev{{\bf automatic specification from $G=1$ model estimates (\code{B=m1} with \code{m1} the model with 1 class):} all common parameters over classes are those obtained with $G=1$. For each class-specific parameter generically called $\theta_g$, initial value $\theta_g^{(0)}$ is automatically set at $\theta_g^{(0)}=\hat{\theta}_{G=1}+\left (g-\dfrac{G+1}{2}\right )\times \widehat{SE(\hat{\theta}_{G=1})}$ where $\hat{\theta}_{G=1}$ and $\widehat{SE(\hat{\theta}_{G=1})}$ are the corresponding estimated parameter and standard error under $G=1$ assumption. There are exceptions for $\xi_{0g}$ and $\xi_{1g}$ (for $g=1,...,G-1$) set to 0, $\omega_g$  (for $g=1,...,G-1$) set to 1 and the proportional coefficients over classes for \code{hazardtype="PH"} in \code{Jointlcmm} set to $\frac{g}{2}$ for $g=1,...,G-1$. Note that this automatic specification does not ensure convergence towards the global maximum, may be inapropriate for a specific analysis, and thus should only be used for first attempts.}

\item \rev{{\bf no specification  (\code{B=NULL}, by default):} for an easier discovery of the program, the program can run without specifying any initial values. The same strategy as above is used except that the model with $G=1$ is first estimated internally. This should be avoided whenever possible as it substantially increases the estimation time. }

\item \rev{{\bf random draws from $G=1$ model estimates (\code{B=random(m1)} with \code{m1} the model with 1 class):} instead of using an automatic specification of the initial values from $G=1$ model, the initial values can be drawn from the asymptotic distribution of the MLE of $G=1$ model ($\mathcal{N}(\hat{\theta}_{G=1}, \widehat{V(\hat{\theta}_{G=1})})$). This permits the use of an automatic grid search as implemented in the \code{gridsearch} function (see below).}

\item \rev{{\bf specification of the initial values  (\code{B=Binit} with \code{Binit} a vector of initial values)}: the user can provide any set of initial values. This is particularly useful to manually change initial values or constraining parameters to zero. The most difficult part is to enter the correct number of parameters. }
\end{enumerate}

\rev{An automatic grid search is also implemented in the generic function \code{gridsearch}. It consists in running the estimation function for a maximum of $m$ iterations from $B$ random sets of initial values. The parameters corresponding to the best log-likelihood after $m$ iterations are used as initial values for the final estimation of the parameters. This procedure is derived from the emEM technique \citep{biernacki2003}.}

\rev{The different specifications of initial values described in this section are illustrated in Sections \ref{ex_hlme} and \ref{ex_Jointlcmm}, including the grid search.}

\section[Post-fit Computations]{Post-fit computations}\label{sec-post-fit}

\rev{
A series of post-fit analyses and computations is available in the package, most of which are common to the four estimation functions (\code{hlme}, \code{lcmm}, \code{multlcmm}, \code{Jointlcmm}). The next subsections describe the post-fit computations. In the following, the hat symbol ($\hat{~}$) denotes the value of a parameter/vector/matrix/function computed at the maximum likelihood estimates $\hat{\theta}_G$. 
}

\subsection[Maximum likelihood estimates]{Maximum likelihood estimates}\label{MLE}

This subsection applies to the four estimation functions. 
The table of the maximum likelihood estimates along with their estimated standard errors are given in function \code{summary}. The vector is directly given by function \code{estimates} or in output value \code{best}. 

The estimated variance-covariance matrix of the maximum likelihood estimates is given in function \code{VarCov} and in output value \code{V}. In the latter, the upper triangular matrix is given as a vector. 

The parameters of the variance-covariance matrix of the random effects are not directly estimated although they are provided in the summaries. The Cholesky parameters used for the estimation are available in output vector \code{cholesky} or in function \code{estimates}. Estimated standard errors of the parameters of the variance-covariance matrix are computed in function \code{VarCovRE}.

Function \code{WaldMult} provides univariate and multivariate Wald tests for combinations of parameters from \code{hlme}, \code{lcmm}, \code{multlcmm} or \code{Jointlcmm} objets.

\subsection[Posterior classification]{Posterior classification} \label{sec-postprob}

In models involving latent classes, a posterior classification of the subjects in each latent class can be made. It is based on the posterior calculation of the class-membership probabilities and is used to characterize the classification of the subjects as well as to evaluate the goodness-of-fit of the model \citep{proust-lima2014}. 

\subsubsection{Class-membership posterior probabilities and classification}

The posterior class-membership probabilities are computed using the Bayes theorem as the probability of belonging to a latent class given the information collected. In a longitudinal model, they are defined for subject $i$ and latent class $g$ as 

\begin{equation}\label{postp}
\hat{\pi}_{ig}^{(Y)}=P(c_i=g | X_{Li},X_{ci},Y_i,\hat{\theta}_G)
=\dfrac{\pi_{ig} \phi_{ig}(Y_i|c_i=g;\theta_G)}{\sum_{l=1}^G \pi_{il} \phi_{il}(Y_i|c_i=l;\theta_G)}.
\end{equation}

In a joint latent class model, the complete information also includes the time-to-event so that for subject $i$ and latent class $g$, the posterior class-membership probability can also be defined for subject $i$ and latent class $g$ as 

\begin{equation}\label{postpJoint}
\begin{split}
\hat{\pi}_{ig}^{(Y,T)}&=P(c_i=g | X_{Li},X_{ci},X_{Si},Y_i,T_i,\E_i,\hat{\theta}_G)\\
&=\dfrac{\hat{\pi}_{ig} \phi_{ig}(Y_i|c_i=g;\hat{\theta}_G) e^{-\sum_{p=1}^P A_p(T_i \mid c_i=g;\hat{\theta}_G)}\prod_{p=1}^P \lambda_p(T_i \mid c_i=g;\hat{\theta}_G)^{\mathbbm{1}_{E_i=p}}}{\sum_{l=1}^G \pi_{il} \phi_{il}(Y_i|c_i=l;\hat{\theta}_G) e^{-\sum_{p=1}^P A_p(T_i \mid c_i=l;\hat{\theta}_G)}\prod_{p=1}^P \lambda_p(T_i \mid c_i=l;\hat{\theta}_G)^{\mathbbm{1}_{E_i=p}}}.
\end{split}
\end{equation}
 
A posterior classification can be obtained from these posterior probabilities by assigning for each subject the latent class in which he has the highest posterior class-membership probability ($\hat{c}_i=\text{argmax}_{g}(\hat{\pi}_{ig}^{(Y)})$ or $\hat{c}_i=\text{argmax}_{g}(\hat{\pi}_{ig}^{(Y,T)})$).

In \code{hlme}, \code{lcmm} and \code{multlcmm} objects, the output table \code{pprob} provides the posterior probabilities $\hat{\pi}_{ig}^{(Y)}$
and the corresponding posterior classification. 
In \code{Jointlcmm}, the output table \code{pprob} provides the posterior probabilities $\hat{\pi}_{ig}^{(Y,T)}$ and the corresponding posterior classification while \code{pprobY} provides the posterior probabilities based only on the longitudinal model $\hat{\pi}_{ig}^{(Y)}$. 

\subsubsection{Posterior classification}

The posterior classification can be used to assess the goodness-of-fit of the model (for the selection of the number of latent classes for instance) and the discrimination of the latent classes. Many indicators can be derived from it \citep{proust-lima2014}. The package \pkg{lcmm} provides two indicators in the function \code{postprob}:
\begin{itemize}
\item[-] the proportion of subjects classified in each latent class with a posterior probability above 0.7, 0.8 and 0.9. This indicates the proportion of subjects not ambiguously classified in each latent class.

\item[-] the posterior classification table as defined in Table \ref{tableclassif} which computes the mean of the posterior probabilities of belonging to the latent class among the subjects classified \emph{a posteriori} in each latent class. A perfect classification would provide ones in the diagonal and zeros elsewhere. In practice, high diagonal terms indicate a good discrimination of the population. 

\begin{table}
\begin{center}
\begin{tabular}{ccccccc}
\hline
Final  & $\sharp$ &\multicolumn{5}{c}{Mean of the probabilities of belonging to each class}\\
class $\hat{c}_i$&  & 1      &  $\hdots$ & $g$  & $\hdots$   & $G$ \\
\hline
1 & $N_1$ & $\dfrac{1}{N_1} \sum_{i=1}^{N_1} \hat{\pi}_{i1}^{(.)}$ & $\hdots$ & $\dfrac{1}{N_1} \sum_{i=1}^{N_1} \hat{\pi}_{ig}^{(.)}$ & $\hdots$ & $\dfrac{1}{N_1} \sum_{i=1}^{N_1} \hat{\pi}_{iG}^{(.)}$  \\
$\vdots$ & & $\vdots$ & $\ddots$ & & & $\vdots$ \\
$g$ & $N_g$ &  $\dfrac{1}{N_g} \sum_{i=1}^{N_g} \hat{\pi}_{i1}^{(.)}$ & $\hdots$ &$\dfrac{1}{N_g} \sum_{i=1}^{N_g} \hat{\pi}_{ig}^{(.)}$ & $\hdots$ & $\dfrac{1}{N_g} \sum_{i=1}^{N_g} \hat{\pi}_{iG}^{(.)}$  \\
$\vdots$ & & $\vdots$ & & & $\ddots$ & $\vdots$ \\
$G$ & $N_G$ &  $\dfrac{1}{N_G} \sum_{i=1}^{N_G} \hat{\pi}_{i1}^{(.)}$ & $\hdots$ & $\dfrac{1}{N_G} \sum_{i=1}^{N_G} \hat{\pi_{ig}}^{(.)}$ & $\hdots$ & $\dfrac{1}{N_G} \sum_{i=1}^{N_G} \hat{\pi}_{iG}^{(.)}$ \\
\hline
\end{tabular}
\end{center}
\caption{Posterior classification table provided in function \code{postprob}. $\hat{\pi}_{ig}^{(.)}$ refers to $\hat{\pi}_{ig}^{(Y)}$ except for a \code{Jointlcmm} object in which case it refers to $\hat{\pi}_{ig}^{(Y,T)}$. }\label{tableclassif}
\end{table}

\end{itemize}

\subsection[Longitudinal predictions and residuals]{Longitudinal predictions and residuals} \label{sec-pred}

The four estimation functions rely on the linear mixed model theory, and as such empirical bayes estimates and longitudinal predictions are naturally derived. 

\subsubsection{Empirical Bayes Estimates of the random effects}

Empirical bayes estimates of the random effects $u_i$ are provided in output of the four estimation functions with the output table \code{predRE} \rev{and generic function \code{ranef}}. 

For a standard linear mixed model defined in Equation~\ref{LMM}, these empirical Bayes estimates are $\hat{u}_i=\hat{B} Z_{i}^\top \hat{V}_{i}^{-1}(Y_i - X_{Li}\hat{\beta})$. 

In the latent process mixed models defined by Equation~\ref{LatentLMM} and Equation~\ref{measurement} (for the univariate case) or \eqref{tildeY} (for the multivariate case), the empirical bayes estimates are computed when only continuous link functions are assumed. In these cases, the random effects are predicted in the latent process scale by $\hat{u}_i=\hat{B} Z_{i}^\top \hat{V}_{i}^{-1}(\hat{\tilde{Y}}_i - X_{Li}\hat{\beta})$ where $\hat{\tilde{Y}}_i$ is the vector of transformed marker values $\hat{\tilde{Y}}_{ij}=H^{-1}(Y_{ij};\hat{\eta})$ for $j=1,...,n_i$ in the univariate case, or of transformed markers values $\hat{\tilde{Y}}_{kij}=H_k^{-1}(Y_{kij};\hat{\eta}_k)$ with $k=1,...,K$ and $j=1,...,n_{ki}$ in the multivariate case. 

In models involving latent classes, class-specific empirical bayes estimates are defined as $\hat{u}_{ig}=\hat{\omega}_g^2 \hat{B} Z_{i}^\top \hat{V}_{ig}^{-1}(Y_i - X_{L1i}\hat{\beta}-X_{L2i}\hat{\upsilon}_g)$ for linear mixed models (defined in \eqref{longitudinal}) or $\hat{u}_{ig}=\hat{\omega}_g^2 \hat{B} Z_{i}^\top \hat{V}_{ig}^{-1}(\hat{\tilde{Y}}_i - X_{L1i}\hat{\beta}-X_{L2i}\hat{\upsilon}_g)$ for latent process mixed models with continuous link functions (defined in \eqref{LatentLCMM}). Marginal empirical bayes estimates are obtained as $\hat{u}_{i}=\sum_{g=1}^G \hat{\pi}_{ig}^{(Y)} \hat{u}_{ig}$.

\subsubsection{Longitudinal predictions and residuals}

Predictions and residuals of the linear mixed model are computed in the four estimation functions and provided in output table \code{pred}. Both subject-specific and marginal predictions/residuals are computed. 

For \code{hlme} and \code{Jointlcmm} \rev{(with \code{link=NULL})}, marginal and subject-specific predictions are respectively $\hat{Y}_{ij}^{(M)}=X_{Li}(t_{ij})^\top\hat{\beta}$ and $\hat{Y}_{ij}^{(SS)}=X_{Li}(t_{ij})^\top\hat{\beta}+Z_{i}(t_{ij})^\top \hat{u}_{i}$ when $G=1$. Marginal and subject-specific residuals are $R_{ij}^{(M)} = Y_{ij}-\hat{Y}_{ij}^{(M)}$ and $R_{ij}^{(SS)} = Y_{ij}-\hat{Y}_{ij}^{(SS)}$. 

For $G>1$, class-specific marginal and subject-specific predictions are respectively $\hat{Y}_{ijg}^{(M)}=X_{L1i}(t_{ij})^\top\hat{\beta}+X_{L2i}(t_{ij})^\top\hat{\upsilon}_g$ and $\hat{Y}_{ijg}^{(SS)}=X_{L1i}(t_{ij})^\top\hat{\beta}+X_{L2i}(t_{ij})^\top\hat{\upsilon}_g+Z_{ij}^\top\hat{u}_{ig}$. To compute residuals, class-specific marginal and subject-specific predictions are averaged over latent classes as $\hat{Y}_{ij}^{(M)}= \sum_{g=1}^G \hat{\pi}_{ig} \hat{Y}_{ijg}^{(M)}$ and $\hat{Y}_{ij}^{(SS)}= \sum_{g=1}^G \hat{\pi}_{ig}^{(Y)} \hat{Y}_{ijg}^{(SS)}$, and corresponding residuals are $R_{ij}^{(M)} = Y_{ij}-\hat{Y}_{ij}^{(M)}$ and $R_{ij}^{(SS)} = Y_{ij}-\hat{Y}_{ij}^{(SS)}$.

For \code{lcmm}, \code{multlcmm} and \code{Jointlcmm} \rev{(with \code{link!=NULL})}, the exact same predictions are computed and provide marginal and subject-specific predictions ($\hat{\tilde{Y}}_{ijg}^{(M)}$ and $\hat{\tilde{Y}}_{ijg}^{(SS)}$ for \code{lcmm} or \code{Jointlcmm}, $\hat{\tilde{Y}}_{kijg}^{(M)}$ and $\hat{\tilde{Y}}_{kijg}^{(SS)}$ for \code{multlcmm}) in the latent process scale. Residuals in the latent process scale are: $R_{ij}^{(M)} = \hat{\tilde{Y}}_{ij}-\hat{\tilde{Y}}_{ij}^{(M)}$ and $R_{ij}^{(SS)} = \hat{\tilde{Y}}_{ij}-\hat{\tilde{Y}}_{ij}^{(SS)}$ for \code{lcmm} or \code{Jointlcmm};  $R_{kij}^{(M)} = \hat{\tilde{Y}}_{kij}-\hat{\tilde{Y}}_{kij}^{(M)}$ and $R_{kij}^{(SS)} = \hat{\tilde{Y}}_{kij}-\hat{\tilde{Y}}_{kij}^{(SS)}$ for \code{multlcmm}. Note that in these two functions, variable \code{obs} in table \code{pred} contains the transformed data $\hat{\tilde{Y}}$.

By default (option \code{which="residuals"}), function \code{plot} provides graphs of the marginal and subject-specific residuals. With option \code{which="fit"}, function \code{plot} provides graphs of the class-specific marginal and subject-specific mean evolutions with time and the observed class-specific mean evolution and its 95\% confidence bounds. In these graphs, time is split in time intervals provided in input. When $G>1$, the class-specific mean evolutions are weighted by the class-membership probabilities.

\subsubsection{Special case of longitudinal predictions in the marker scale for latent process mixed models}\label{predfitY}

For \code{Jointlcmm}, \code{multlcmm} objects and \code{lcmm} objects with continuous link functions, \code{fitY} computes the marginal longitudinal predictions in the marker scale ($\hat{Y}_{ij}^{(M)}$ or $\hat{Y}_{kij}^{(M)}$). When G=1, they are computed using a numerical integration of $H(\tilde{y}_{ij};\hat{\eta})$ for a \code{lcmm} or \code{Jointlcmm} object or of $H_k(\tilde{y}_{kij};\hat{\eta_k})$ for a \code{multlcmm} object over the multivariate Gaussian distribution of $\tilde{y}_{i}$ at the maximum likelihood. In these formula, a Newton algorithm is used to compute $H$ or $H_k$ values. When $G>1$, the same method is used but conditional to each latent class $g$.
Numerical integrations are managed by a MonteCarlo method.

For \code{lcmm} and thresholds link functions, \code{fitY} computes the marginal longitudinal predictions in the marker scale ($\hat{Y}_{ij}$) as: $\hat{Y}_{ij} = M_0+M-1 - \sum_{l=1}^{M-1} \Phi \left (\frac{\hat{\eta}_l^*-\hat{\tilde{Y}}_{ij}}{\sqrt{\hat{V}(\tilde{Y}_{ij})}} \right )$ where $\Phi$ is the standard Gaussian cumulative distribution function.

\subsubsection{Predicted mean trajectory according to a profile of covariates}

The predicted mean trajectory of the markers $Y$ according to an hypothetical profile of covariates can be computed (and represented). This is provided by the functions \code{predictY} and the \code{plot} function applied on \code{predictY} objects.

The computations are exactly the same as described previously, except that the longitudinal predictions are computed for an hypothetical (new) subject from a table containing the hypothetical covariate information required in $X_{Li}$, $Z_{i}$, $X_{ci}$, $X_{Si}$ and referred to as $X$. The predicted mean vector of the marker $E(Y|X=x,\hat{\theta}_G)$ is computed as the (class-specific) marginal vector of predictions for \code{hlme} and \code{Jointlcmm}. For \code{lcmm}, \code{Jointlcmm} and \code{multlcmm} objects, the predicted mean vector of values for the latent process are computed in function \code{predictL} and the predicted mean vector of values for the markers are computed in function \code{predictY}. 
In the latter case, two numerical integrations can be specified: either a MonteCarlo method or a Gauss-Hermite technique. The latter neglects the correlation between the repeated measurements. 

Instead of the mean trajectory at the point estimate $\hat{\theta}_G$, the posterior prediction distribution can be approximated by a Monte Carlo method by computing the quantity for a large number of draws from the asymptotic distribution of the parameters $\mathcal{N}(\hat{\theta}_G,\widehat{V(\hat{\theta}_G)}))$. Then the 2.5\%, 50\% and 97.5\% provide the mean prediction and its 95\% confidence interval. 

\subsection[Link functions]{Link functions} \label{sec-link}

This section is specific to \code{lcmm}, \code{multlcmm} \rev{and \code{Jointlcmm} (with \code{link!=NULL})}. 

\subsubsection{Predicted link functions} 

Table \code{estimlink} provides the (inverse of the) link functions computed for a vector of marker values at the maximum likelihood estimates $\hat{\eta}$ in outputs of \code{lcmm} and \code{multlcmm}. These estimated link functions can be plotted using function \code{plot} with option \code{which="link"} or  \code{which="linkfunction"}. 
Function \code{predictlink} further computes the 50\%, 2.5\% and 97.5\% percentiles of the posterior distribution of the estimated (inverse) link functions using a Monte Carlo method (large set of draws from the asymptotic distribution of the parameters). This function can also be used to compute the estimated link function at specific marker values. 

\subsubsection{Discrete log-likelihood and derived criteria} 

In the case of ordinal outcomes with a large number of levels, the latent process mixed model estimated in function \code{lcmm} with continuous nonlinear link functions constitutes an approximation of the cumulative probit mixed model (much easier to estimate as it does not involve any numerical integration over the random effects). However, it is important to assess whether the approximation is acceptable. This can be done by using the discrete log-likelihood and the derived information criteria: discrete AIC \citep{proust-lima2012} and UACV \citep{commenges2012} that are computed with respect to the counting measure instead of the Lebesgue measure. These measures are computed in the \code{summary} and in output of \code{lcmm}. 

\subsection[Prediction of the event]{Prediction of the event} \label{sec-predsurv}

This section is specific to \code{Jointlcmm} function. 

\subsubsection{Profile of survival functions according to covariates} 

Class-specific baseline risks are plotted in \code{plot} function with option \code{which="baselinerisk"} or \code{which="hazard"}. Class-specific survival functions in the category of reference can also be plotted with \code{plot} and option \code{which="survival"} when there is a unique cause of event. Otherwise, predicted cumulative incidences of a specific cause of event can be computed with \code{cuminc} and plotted with the associated \code{plot.cuminc} function for any profile of covariates given in input. 
Again, a Monte Carlo method is implemented to provide the 2.5\%, 50\% and 97.5\% of the posterior distribution of the predicted cumulative incidences.

\subsubsection{Individual dynamic predictions} 

Individual dynamic predictions as developed in \cite{proust-lima2009,proust-lima2014} can be computed with \code{dynpred} and plotted with \code{plot.dynpred}. They consist of the predicted probability of event (of a cause $p$ if multiple causes) in a window of time $(s,s+t)$ computed for any subject according to his/her own longitudinal information collected up to time $s$ that is $Y_i ^{(s)}=\{Y_{ij},\, j=1,..., n_{i}, \text{such as}\; t_{ij}\leq s\}$, $X_{i}^{(s)}=\{X_{L1i}(t_{ij}),X_{L2i}(t_{ij}),Z_{i}(t_{ij}),\,j=1,..., n_{i}, \text{such as}\; t_{ij}\leq s\}$, $X_{Si}=\{X_{Si1},X_{Si2}\}$ and $X_{ci}$. For cause $p$ ($p=1,...,P$), it is: 

\begin{equation}
\begin{split}
P(T_i &\leq s+t, \delta_i=p | T_i \ge s,Y_i^{(s)},X_{i}^{(s)},X_{Si},X_{ci};\theta_G)=\\
 &= \frac{\sum_{g=1}^G P(c_i=g|X_{ci};\theta_G)P(T_i \in (s,s+t],\delta_i=p | X_{Si},c_i=g;\theta_G)f(Y_i^{(s)}|X_{i}^{(s)},c_i=g ;\theta_G)}
{\sum_{g=1}^G P(c_i=g|X_{ci};\theta_G)S_i(s| X_{Si},c_i=g;\theta_G)f(Y_i^{(s)}|X_{i}^{(s)},c_i=g ;\theta_G)}
\end{split}
\end{equation}

where the density of the longitudinal outcomes $f(Y_i^{(s)}|X_{i}^{(s)},c_i=g ;\theta_G)$ in class $g$, the class-specific membership probability $P(c_i=g|X_{ci};\theta_G)$ and the class-specific survival function $S_i(s| X_{Si},c_i=g;\theta_G)$  are defined similarly as in Section \ref{est_jlcm}. Finally, with a unique cause of event, the class-specific cumulative incidence is: 
\begin{equation}
\begin{split}
P(T_i \in (s,s+t],\delta_i=1 | X_{Si},c_i=g;\theta_G) &=P(T_i \in (s,s+t]| X_{Si},c_i=g;\theta_G)\\
&= S_i(s| X_{Si},c_i=g;\theta_G) - S_i(s+t| X_{Si},c_i=g;\theta_G). 
\end{split}
\end{equation} 
With multiple causes of event ($P>1$), the class-specific cause-specific cumulative incidence is:
\begin{equation}
\begin{split}
P(T_i \in (s,s+t]&,\delta_i=p | X_{Si},c_i=g;\theta_G)=\\
& \int_s^{s+t} \lambda_p(u \mid c_i=g;\theta_G) \exp \left ( - \sum_{l=1}^P A_l(u \mid c_i=g;\theta_G) \right) du
\end{split}
\end{equation} 
with $\lambda_p(t \mid c_i=g;\theta_G)$ the cause-$p$-specific instantaneous hazard defined in \eqref{surv_JLCMcompet} and $A_p(t \mid c_i=g;\theta_G)$ the corresponding cumulative hazard. When $P>1$, the cause-specific cumulative incidence requires the numerical computation of the integral. This is achieved by a 50-point Gauss-Legendre quadrature.

Individual dynamic predictions can be computed for any subject (included or not in the dataset used for estimating the model). Times $s$ and $t$ are respectively called the landmark time and the horizon of prediction. Computation in \code{dynpred} is performed for any vectors of landmark times and of horizons.  

Individual predictions are computed either in $\hat{\theta}$ or the posterior distribution is approximated using a Monte Carlo method with a large number of draws, in which case the 2.5\%, 50\% and 97.5\% provide the median prediction and its 95\% confidence interval.

\subsubsection{Assessment of predictive accuracy} \label{sec-predacc}

Predictive accuracy of dynamic predictions based on \code{Jointlcmm} objects can be assessed using the prognostic information criterion EPOCE \citep{commenges_choice_2012} implemented in the function \code{epoce}. Predictive accuracy is computed from a vector of landmark times on the subjects still at risk at the landmark time and the biomarker history up to the landmark time. 

On external data, \code{epoce} provides the Mean Prognostic Observed Log-likelihood (MPOL) for each landmark time. When applied to the same dataset as used for the estimation, the function provides for each landmark time both the MPOL and the Cross-Validated Observed Log-Likelihood (CVPOL). The latter corrects the MPOL estimate for possible over-optimism by approximated cross-validation. Further details on these estimators can be found in \cite{commenges_choice_2012} and  \cite{proust-lima2014}. 

Predictive accuracy of two models can also be compared using \code{Diffepoce} which computes the difference in EPOCE estimators along with a 95\% tracking interval. Functions \code{epoce} and \code{Diffepoce} include a plot functionality.

Further predictive accuracy measures can be computed by using \pkg{timeROC} \citep{blanche2014} on individual dynamic predictions computed by \code{dynpred} from a \code{Jointlcmm} object.

\section[Examples]{Examples}

\rev{
This section details a series of examples of models estimated with \code{hlme}, \code{lcmm}, \code{multlcmm} and \code{Jointlcmm} functions described in section \ref{Implementation}. All the examples are based on the \code{paquid} dataset provided with \pkg{lcmm} \proglang{R} package. Examples also illustrate the initial values specification described in section \ref{sec-initvalues} and the post-fit computations and generic functions described in sections \ref{MLE} to \ref{sec-predacc} and listed in table \ref{Generics}. }

The first step consists in loading \pkg{lcmm}. \rev{This automatically loads the datasets.}

\begin{Schunk}
\begin{Sinput}
R> library("lcmm")
\end{Sinput}
\end{Schunk}

\subsection[Paquid data]{Paquid data}

\code{paquid} dataset consists of a random subsample of 500 subjects (identified by \code{ID}) from the Paquid prospective cohort study \citep{letenneur1994} that aimed at investigating cerebral and functional aging in southwestern France. Repeated measures of three cognitive tests (\code{MMSE}, \code{IST}, \code{BVRT}), physical dependency (\code{HIER}, a 4-level factor) and depression symptomatology (\code{CESD}) were collected over a maximum period of 20 years along with age at the visit (\code{age}), age at dementia diagnosis or last visit (\code{agedem}) and dementia diagnosis (\code{dem}). Three time-independent socio-demographic variables are provided: education (\code{CEP}), gender (\code{male}), age at entry in the cohort (\code{age_init}). In some cases, owing to the very asymmetric distribution of MMSE, a normalized version of MMSE (\code{normMMSE}) obtained with package \pkg{NormPsy} \citep{philipps2014} will be analyzed instead of crude MMSE scores. 
For computation and interpretation purposes, \code{age} will usually be replaced by \code{age65}, which is the age minus 65 and divided by 10. Centering around 65 makes the interpretation of the intercepts easier and division by 10 reduces numerical problems due to too large ages in quadratic models (and so too small effects or variances of random effects). 

The following lines create \code{normMMSE} and \code{age65} variables, and display the first lines of \code{paquid} dataset:

\begin{Schunk}
\begin{Sinput}
R> library("NormPsy")
R> paquid$normMMSE <- normMMSE(paquid$MMSE)
R> paquid$age65 <- (paquid$age-65)/10
R> head(paquid)
\end{Sinput}
\begin{Soutput}
  ID MMSE BVRT IST HIER CESD      age  agedem dem age_init CEP male normMMSE
1  1   26   10  37    2   11 68.50630 68.5063   0  67.4167   1    1    61.18
2  2   26   13  25    1   10 66.99540 85.6167   1  65.9167   1    0    61.18
3  2   28   13  28    1   15 69.09530 85.6167   1  65.9167   1    0    74.61
4  2   25   12  23    1   18 73.80720 85.6167   1  65.9167   1    0    55.98
5  2   24   13  16    3   22 84.14237 85.6167   1  65.9167   1    0    51.44
6  2   22    9  15    3   NA 87.09103 85.6167   1  65.9167   1    0    43.74
     age65
1 0.350630
2 0.199540
3 0.409530
4 0.880720
5 1.914237
6 2.209103
\end{Soutput}
\end{Schunk}

\subsection[hlme examples]{hlme}\label{ex_hlme}

The latent class linear mixed models implemented in \code{hlme} are illustrated by the study of the quadratic trajectories of \code{normMMSE} with \code{age65} adjusted for \code{CEP} and assuming correlated random effects for the functions of \code{age65}. The next line estimates the corresponding standard linear mixed model (1 latent class) in which \code{CEP} is in interaction with age functions:

\begin{Schunk}
\begin{Sinput}
R> m1a <- hlme(normMMSE ~ poly(age65, degree = 2, raw = TRUE)*CEP, 
+    random =~ poly(age65, degree = 2, raw = TRUE), subject = 'ID', 
+    data = paquid, ng = 1)
R> summary(m1a)
\end{Sinput}
\begin{Soutput}
Heterogenous linear mixed model 
     fitted by maximum likelihood method 
 
hlme(fixed = normMMSE ~ poly(age65, degree = 2, raw = TRUE) * 
    CEP, random = ~poly(age65, degree = 2, raw = TRUE), subject = "ID", 
    ng = 1, data = paquid)
 
Statistical Model: 
     Dataset: paquid 
     Number of subjects: 500 
     Number of observations: 2214 
     Number of observations deleted: 36 
     Number of latent classes: 1 
     Number of parameters: 13  
 
Iteration process: 
     Convergence criteria satisfied 
     Number of iterations:  27 
     Convergence criteria: parameters= 2.2e-06 
                         : likelihood= 5.7e-08 
                         : second derivatives= 3.9e-14 
 
Goodness-of-fit statistics: 
     maximum log-likelihood: -8919.93  
     AIC: 17865.87  
     BIC: 17920.66  
 
 
Maximum Likelihood Estimates: 
 
Fixed effects in the longitudinal model:

                                               coef      Se     Wald p-value
intercept                                  66.42132 3.53110   18.810 0.00000
poly(age65, degree = 2, raw = TRUE)1        2.13956 4.69993    0.455 0.64894
poly(age65, degree = 2, raw = TRUE)2       -4.72910 1.51977   -3.112 0.00186
CEP                                        11.28973 3.94099    2.865 0.00417
poly(age65, degree = 2, raw = TRUE)1:CEP    4.75517 5.33204    0.892 0.37249
poly(age65, degree = 2, raw = TRUE)2:CEP   -1.90682 1.75031   -1.089 0.27597


Variance-covariance matrix of the random-effects:
                                     intercept poly(age65, degree = 2, raw = TRUE)1
intercept                             211.7965                                     
poly(age65, degree = 2, raw = TRUE)1 -214.0782                             451.5095
poly(age65, degree = 2, raw = TRUE)2   54.7010                            -143.7151
                                     poly(age65, degree = 2, raw = TRUE)2
intercept                                                                
poly(age65, degree = 2, raw = TRUE)1                                     
poly(age65, degree = 2, raw = TRUE)2                             58.48017

                               coef      Se
Residual standard error:   10.07493 0.20284
\end{Soutput}
\end{Schunk}

The first part of the summary provides information about the dataset, the number of subjects, observations, observations deleted (since by default, missing observations are deleted), number of latent classes and number of parameters. Then, it details the convergence process with the number of iterations, the convergence criteria and the most important information which is whether the model converged correctly: "convergence criteria satisfied". The next block provides the maximum log-likelihood, Akaike criterion and Bayesian Information criterion. Finally, tables of estimates are given with the estimated parameter, the estimated standard error, the Wald Test statistics (with Normal approximation) and the corresponding p-value. 
For the random-effect distribution, the estimated matrix of covariance of the random effects is displayed (see Section \ref{MLE} for details). Finally, the standard error of the residuals is given along with its estimated standard error. 

The effect of CEP does not seem to be associated with change over age of \texttt{normMMSE}. This is formally assessed using a multivariate Wald test: 

\begin{Schunk}
\begin{Sinput}
R> WaldMult(m1a, pos = c(5, 6),name = "CEP interaction with age65 & age65^2")
\end{Sinput}
\begin{Soutput}
                                     Wald Test p_value
CEP interaction with age65 & age65^2   1.38243 0.50097
\end{Soutput}
\end{Schunk}

Based on this, we now consider the model with an adjustment for \code{CEP} only on the intercept:

\begin{Schunk}
\begin{Sinput}
R> m1 <- hlme(normMMSE ~ poly(age65, degree = 2, raw = TRUE) + CEP, 
+    random =~ poly(age65, degree = 2, raw = TRUE), subject = 'ID', ng = 1, 
+    data = paquid)
\end{Sinput}
\end{Schunk}

The next lines provide the estimation of corresponding models for 2 and 3 latent classes using the automatic specification for the initial values when $G>1$:

\begin{Schunk}
\begin{Sinput}
R> m2 <- hlme(normMMSE ~ poly(age65, degree = 2, raw = TRUE) + CEP, 
+    random =~ poly(age65, degree = 2, raw = TRUE), mixture =~ poly(age65, 
+    degree = 2, raw = TRUE), subject = 'ID', ng = 2, data = paquid, B = m1)
R> m3 <- hlme(normMMSE ~ poly(age65, degree = 2, raw = TRUE) + CEP, 
+    random =~ poly(age65, degree = 2, raw = TRUE), mixture =~ poly(age65, 
+    degree = 2, raw = TRUE), subject = 'ID', ng = 3, data = paquid, B = m1)
\end{Sinput}
\end{Schunk}

\rev{
Option \code{B=m1} automatically generates initial values from the maximum likelihood estimates of a 1-class model (here, \code{m1}). An alternative option is not to specify option \code{B} or specify \code{B=NULL} but this is not recommended since it induces the internal pre-estimation of the model with $G=1$ (i.e., \code{m1}). As the model with $G=1$ is generally estimated first to define the structure of the model, this option uselessly slows the estimation procedure. 
}

\rev{
With mixture models, convergence toward global maximum is never guaranteed because of the existence of local maxima. It is thus recommended to run the model several times from different sets of initial values. This can be done by pre-specifying different vectors of initial values or by randomly and repeatedly generating initial values. In the following example, the initial values are pre-specified by the user: parameters of the variance covariance were taken at the estimated values of the linear mixed model and arbitrary initial values were tried for the class-specific trajectories:}

\begin{Schunk}
\begin{Sinput}
R> m2b <- hlme(normMMSE ~ poly(age65, degree = 2, raw = TRUE) + CEP, 
+    random =~ poly(age65, degree = 2, raw = TRUE), mixture =~ poly(age65, 
+    degree = 2, raw = TRUE), subject = 'ID', ng = 2, data = paquid, 
+    B = c(0, 60, 40, 0, -4, 0, -10, 10, 212.869397, -216.421323, 
+    456.229910, 55.713775, -145.715516, 59.351000, 10.072221))
\end{Sinput}
\end{Schunk}

\rev{An alternative is to randomly generate the initial values from the asymptotic distribution of the estimates of the 1-class model (here, \code{m1}). Note that the seed was defined here for replication purposes only. }

\begin{Schunk}
\begin{Sinput}
R> set.seed(1)
R> m2c <- hlme(normMMSE ~ poly(age65, degree = 2, raw = TRUE) + CEP, 
+    random =~ poly(age65, degree = 2, raw = TRUE), mixture =~ poly(age65, 
+    degree = 2, raw = TRUE), subject = 'ID', data = paquid, ng = 2, 
+    B = random(m1))
\end{Sinput}
\end{Schunk}

\rev{Finally, \code{gridsearch} function can be used to run an automatic grid search. In the next examples with $G=2$ and $G=3$ classes (\code{m2d} and \code{m3b}, respectively), \code{hlme} is run for a maximum of 15 iterations from 30 random vectors of initial values. The estimation procedure is then finalized only for the departure that provided the best log-likelihood after 15 iterations.}

\begin{Schunk}
\begin{Sinput}
R> m2d <- gridsearch(normMMSE ~ poly(age65, degree = 2, raw = TRUE) + CEP, 
+    random =~ poly(age65, degree = 2, raw = TRUE), mixture =~ poly(age65, 
+    degree = 2, raw = TRUE), subject = 'ID', data = paquid, ng = 2),
+    rep = 30, maxiter = 15, minit = m1)
R> m3b <- gridsearch(normMMSE ~ poly(age65, degree = 2, raw = TRUE) + CEP, 
+    random =~ poly(age65, degree = 2, raw = TRUE), mixture =~ poly(age65, 
+    degree = 2, raw = TRUE), subject = 'ID', data = paquid, ng = 3),
+    rep = 30, maxiter = 15, minit = m1)
\end{Sinput}
\end{Schunk}

The estimation process of a set of models (usually with a varying number of latent classes) can be summarized with \code{summarytable}. The function gives the log-likelihood, the number of parameters, the Bayesian Information Criterion, and the posterior proportion of each class: 
\begin{Schunk}
\begin{Sinput}
R> summarytable(m1, m2, m2b, m2c, m2d, m3, m3b)
\end{Sinput}
\begin{Soutput}
    G    loglik npm      BIC %class1 %class2 %class3
m1  1 -8920.623  11 17909.61   100.0                
m2  2 -8899.228  15 17891.67    12.4    87.6        
m2b 2 -8899.228  15 17891.67    87.6    12.4        
m2c 2 -8899.228  15 17891.67    12.4    87.6        
m2d 2 -8899.228  15 17891.67    87.6    12.4        
m3  3 -8891.351  19 17900.78     4.0    85.8    10.2
m3b 3 -8891.351  19 17900.78    85.8     4.0    10.2
\end{Soutput}
\end{Schunk}

In this example, the optimal number of latent classes is two according to the BIC. The posterior classification, \rev{defined in Section \ref{sec-postprob}, is obtained with:}

\begin{Schunk}
\begin{Sinput}
R> postprob(m2)
\end{Sinput}
\begin{Soutput}
Posterior classification: 
  class1 class2
N   62.0  438.0
%   12.4   87.6
 
Posterior classification table: 
     --> mean of posterior probabilities in each class 
        prob1  prob2
class1 0.8054 0.1946
class2 0.1270 0.8730
 
Posterior probabilities above a threshold (%): 
         class1 class2
prob>0.7  61.29  90.18
prob>0.8  58.06  69.18
prob>0.9  43.55  47.95
\end{Soutput}
\end{Schunk}

The first class includes a posteriori 62 subjects (12.4\%) while class 2 includes 438 (87.6\%) subjects. Subjects were classified in class 1 with a mean posterior probability of 80.5\%, and in class 2 with a mean posterior probability of 87.3\%. In class 1, 61.3\% were classified with a posterior probability above 0.7 while 90.2\% of the subjects were classified in class 2 with a posterior probability above 0.7. 

\begin{figure}[!ht]
\begin{center}
\includegraphics{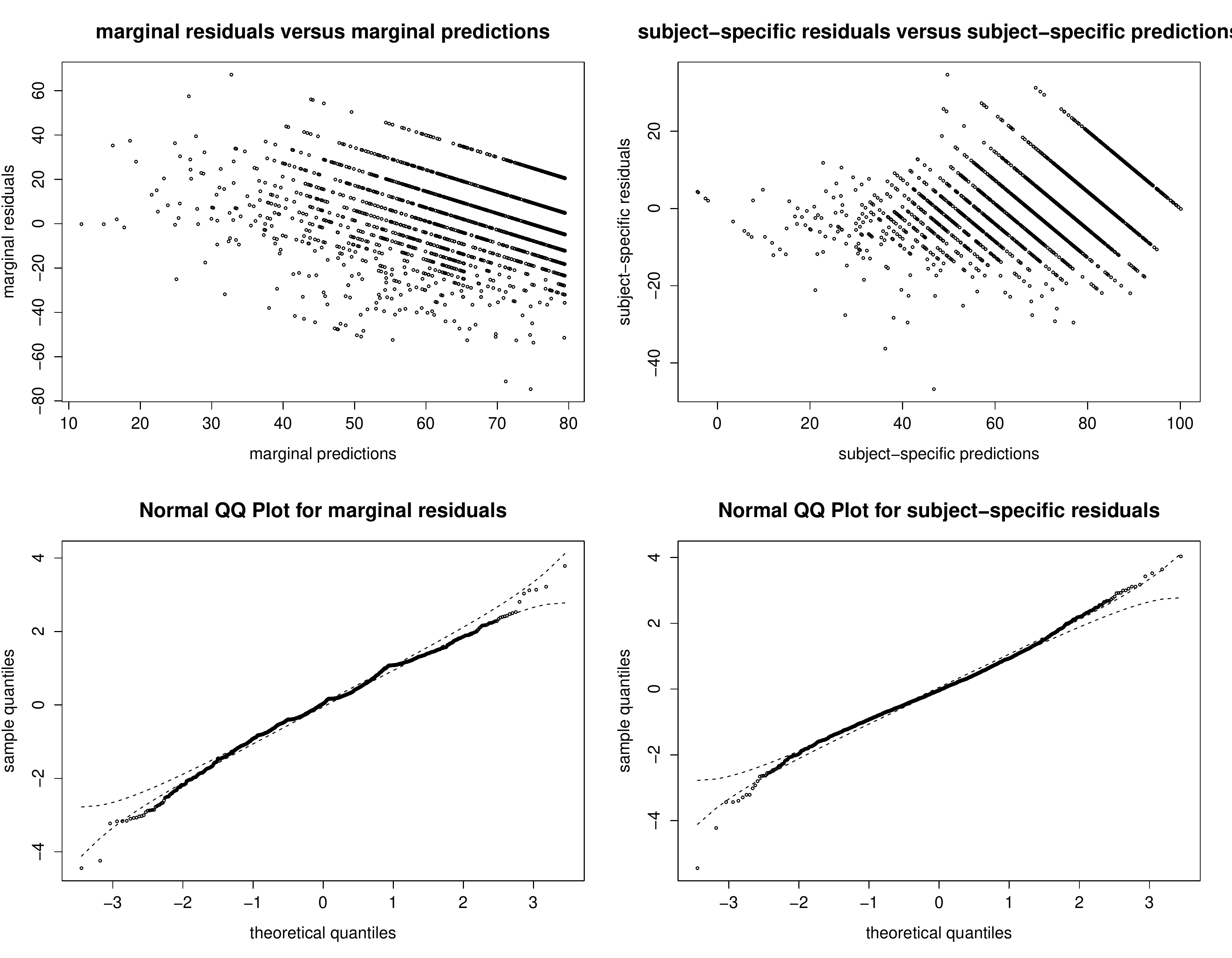}
\caption{\label{fig_hlme_res} Subject-specific and marginal residuals in the two-class linear mixed model using \code{plot}.}
\end{center}
\end{figure}

\begin{figure}[!ht]
\begin{center}
\includegraphics{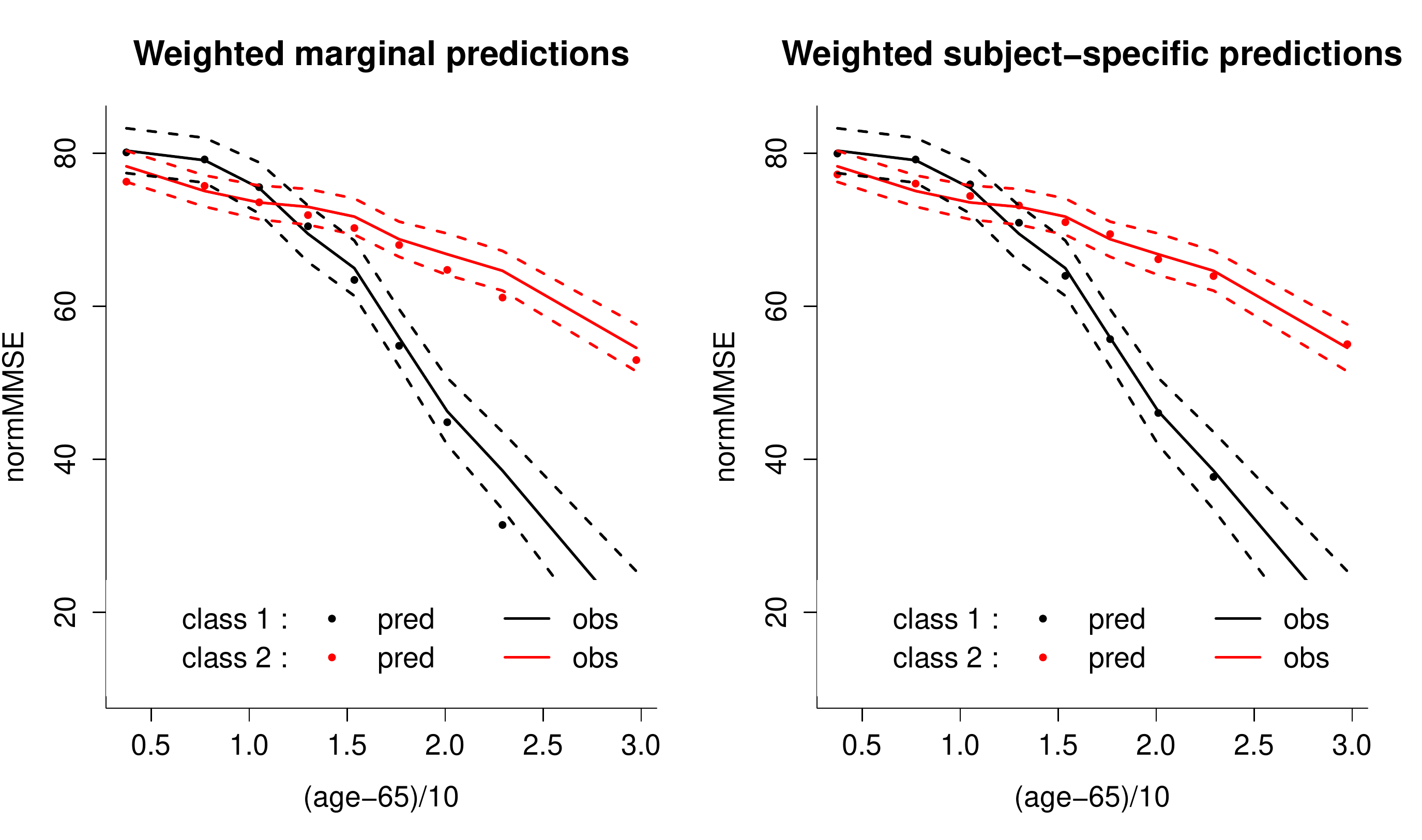}
\caption{\label{fig_hlme_meanP} Weighted mean (marginal on the left, subject-specific on the right) predictions and weighted mean observations according to intervals of age in the two-class linear mixed model using \code{plot} and option \code{which="fit"}.}
\end{center}
\end{figure}

The goodness-of-fit of the model can be assessed by displaying the residuals as in Figure \ref{fig_hlme_res} and the mean predictions of the model as in Figure \ref{fig_hlme_meanP} according to the time variable given in \code{var.time} \rev{(see Section \ref{sec-pred} for computation details)}:

\begin{Schunk}
\begin{Sinput}
R> plot(m2)  
R> plot(m2, which = "fit", var.time = "age65", bty = "l", ylab = "normMMSE",
+    xlab = "(age-65)/10", lwd = 2)
R> plot(m2, which = "fit", var.time = "age65", bty = "l", ylab= "normMMSE",
+    xlab = "(age-65)/10", lwd = 2, marg = FALSE)
\end{Sinput}
\end{Schunk}

\begin{figure}[!ht]
\begin{center}
\includegraphics[width=0.9\textwidth]{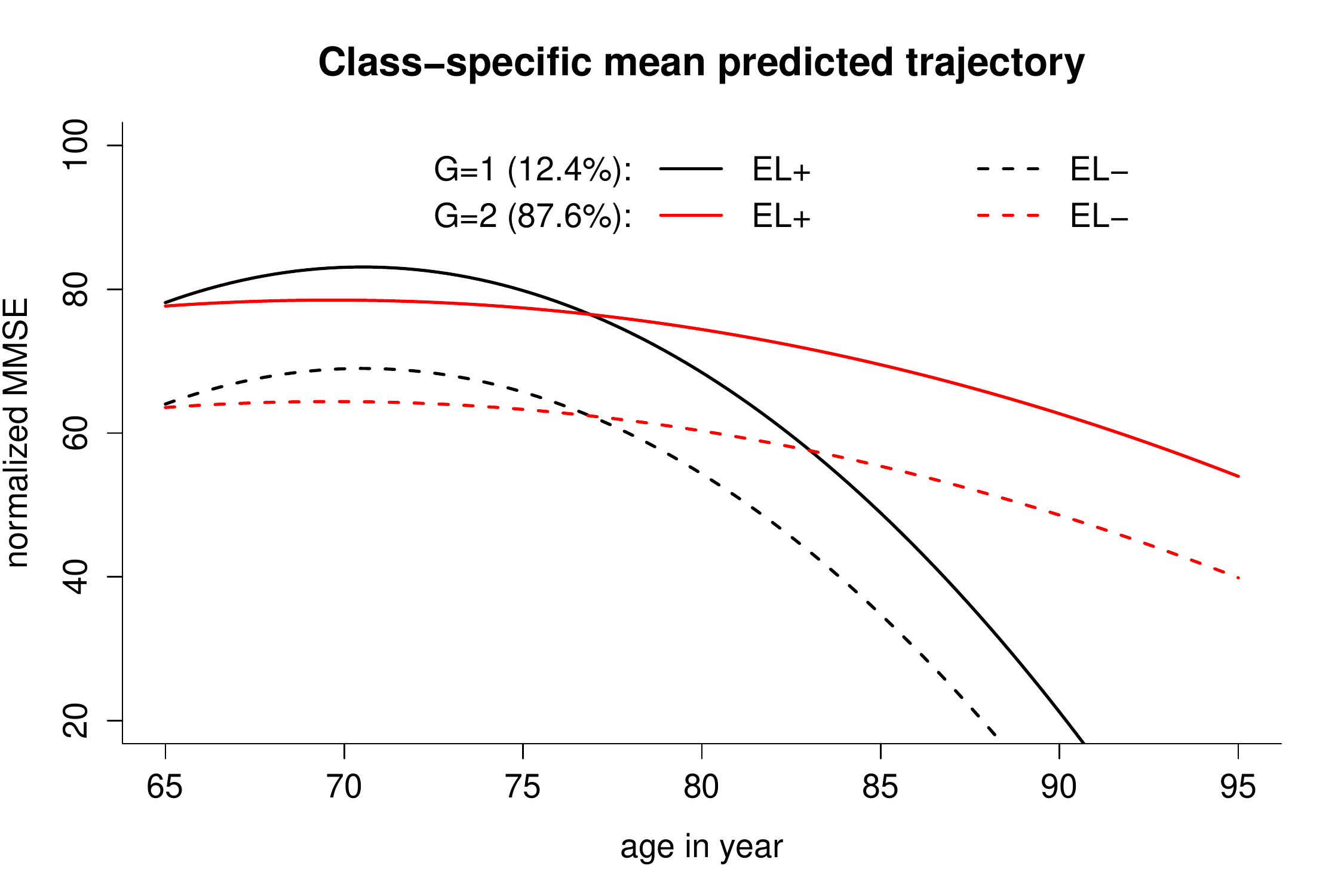}
\caption{\label{fig_hlme_pred} Predicted trajectories with age according to CEP in the two-class linear mixed model.}
\end{center}
\end{figure}

Class-specific predictions, \rev{defined in Section \ref{sec-pred},} can be computed for any data contained in a dataframe as soon as all the covariates specified in the model are included in the dataframe. In the next lines, such a dataframe is created by generating a vector of \code{age} values between 65 and 95 and defining \code{CEP} at 1 or 0. The predictions are computed with \code{predictY} and plotted with the associated \code{plot} functionality or by using standard \proglang{R} tools as illustrated below and in Figure \ref{fig_hlme_pred}.

\begin{Schunk}
\begin{Sinput}
R> datnew <- data.frame(age = seq(65, 95, length = 100))
R> datnew$age65 <- (datnew$age - 65) / 10
R> datnew$CEP <- 0
R> CEP0 <- predictY(m2, datnew, var.time = "age")
R> datnew$CEP <- 1
R> CEP1 <- predictY(m2, datnew, var.time = "age")
R> plot(CEP1, lty = 1,lwd = 2, type = "l", col = 1:2 , ylim = c(20, 100), 
+    bty = "l", xlab = "age in year", ylab = "normalized MMSE", 
+    legend = NULL)
R> plot(CEP0, lty = 2, lwd = 2, type = "l", col = 1 : 2, ylim = c(20, 100), 
+    add = TRUE)
R> legend(x = "topright", bty = "n", ncol = 3, lty = c(NA, NA, 1, 1, 2, 2), 
+    col = c(NA, NA, 1, 2, 1, 2), legend = c("G=1 (12.4%):", "G=2 (87.6%):", 
+    "EL+", "EL+", "EL-", "EL-"), lwd = 2)
\end{Sinput}
\end{Schunk}

\subsection[lcmm examples]{lcmm}\label{ex_lcmm}

The latent process mixed models implemented in \code{lcmm} are illustrated by the study of the linear trajectory of depressive symptoms (as measured by \code{CES-D} scale) with \code{age65} adjusted for \code{male} and assuming correlated random effects for the intercept and \code{age65}. The next lines estimate the corresponding latent process mixed model with different link functions:

\begin{Schunk}
\begin{Sinput}
R> mlin <- lcmm(CESD ~ age65 * male, random =~ age65, subject = 'ID', 
+    data = paquid)
R> mbeta <- lcmm(CESD ~ age65 * male, random =~ age65, subject = 'ID',
+    data = paquid, link = 'beta')
R> mspl <- lcmm(CESD ~ age65 * male, random =~ age65, subject = 'ID', 
+    data = paquid, link = 'splines')
R> mspl5q <- lcmm(CESD ~ age65 * male, random =~ age65, subject = 'ID', 
+    data=paquid, link = '5-quant-splines')
\end{Sinput}
\end{Schunk}

Objects \code{mlin}, \code{mbeta}, \code{mspl} and \code{mspl5q} are latent process mixed models that assume the exact same trajectory for the underlying latent process but different link functions: linear, BetaCDF, I-splines with 5 equidistant knots (default with \code{link='splines'}) and I-splines with 5 knots at percentiles, respectively. 
Note that \code{mlin} reduces to a standard linear mixed model (\code{link='linear'} by default). The only difference with a \code{hlme} object is the parameterization for the intercept and the residual standard error that are considered as rescaling parameters.  

CES-D is an ordinal scale with more than 50 levels so it might be estimated with a cumulative probit mixed model, even if it is rarely done in practice because of the very high number of parameters induced as well as the substantial additional numerical complexity. 

Owing to the numerical integration at each evaluation of the log-likelihood when assuming a threshold link function, estimation of the cumulative probit mixed model can be very long. We thus recommend estimating the model first without random effects to obtain satisfactory inital values for the thresholds before any inclusion of random effects, as shown in the next lines:

\begin{Schunk}
\begin{Sinput}
R> mord0 <- lcmm(CESD ~ age65 * male, random =~ -1, subject = 'ID', 
+    data = paquid, link = 'thresholds')
R> binit <- NULL
R> binit[1:6] <- mspl$best[1:6]
R> binit[7:56] <- mord0$best[4:53]
R> mord <- lcmm(CESD ~ age65 * male, random =~ age65, subject = 'ID',
+    data = paquid, link = 'thresholds', B = binit)
\end{Sinput}
\end{Schunk}

Note here than \code{mord} takes a lot of time to be estimated (can be more than 1 hour depending on the computer).

The output of a \code{lcmm} object is very similar to that of a \code{hlme} object, as shown for \code{mspl5q} for example:

\begin{Schunk}
\begin{Sinput}
R> summary(mspl5q)
\end{Sinput}
\begin{Soutput}
General latent class mixed model 
     fitted by maximum likelihood method 
 
lcmm(fixed = CESD ~ age65 * male, random = ~age65, subject = "ID", 
    link = "5-quant-splines", data = paquid)
 
Statistical Model: 
     Dataset: paquid 
     Number of subjects: 500 
     Number of observations: 2104 
     Number of observations deleted: 146 
     Number of latent classes: 1 
     Number of parameters: 13  
     Link function: Quadratic I-splines with nodes  
0 2 6 12 52  
 
Iteration process: 
     Convergence criteria satisfied 
     Number of iterations:  19 
     Convergence criteria: parameters= 7.6e-09 
                         : likelihood= 9.2e-08 
                         : second derivatives= 1.8e-14 
 
Goodness-of-fit statistics: 
     maximum log-likelihood: -6320.08  
     AIC: 12666.17  
     BIC: 12720.96  
 
     Discrete posterior log-likelihood: -6309.09  
     Discrete AIC: 12644.18  
 
     Mean discrete AIC per subject: 12.6442  
     Mean UACV per subject: 12.6439  
     Mean discrete LL per subject: -12.6182  
 
Maximum Likelihood Estimates: 
 
Fixed effects in the longitudinal model:

                              coef      Se    Wald p-value
intercept (not estimated)        0                        
age65                      0.42421 0.06279   6.756 0.00000
male                      -0.83140 0.19742  -4.211 0.00003
age65:male                 0.23371 0.10301   2.269 0.02327


Variance-covariance matrix of the random-effects:
          intercept  age65
intercept   1.89911       
age65      -0.39567 0.1711

Residual standard error (not estimated) = 1

Parameters of the link function:

               coef      Se    Wald p-value
I-splines1 -2.03816 0.13469 -15.132 0.00000
I-splines2  1.04627 0.02461  42.510 0.00000
I-splines3  0.74190 0.03773  19.665 0.00000
I-splines4  0.98399 0.03237  30.400 0.00000
I-splines5  1.55606 0.04480  34.735 0.00000
I-splines6  0.93273 0.16614   5.614 0.00000
I-splines7  1.38790 0.17687   7.847 0.00000
\end{Soutput}
\end{Schunk}

\begin{figure}[!ht]
\begin{center}
\includegraphics[width=0.9\textwidth]{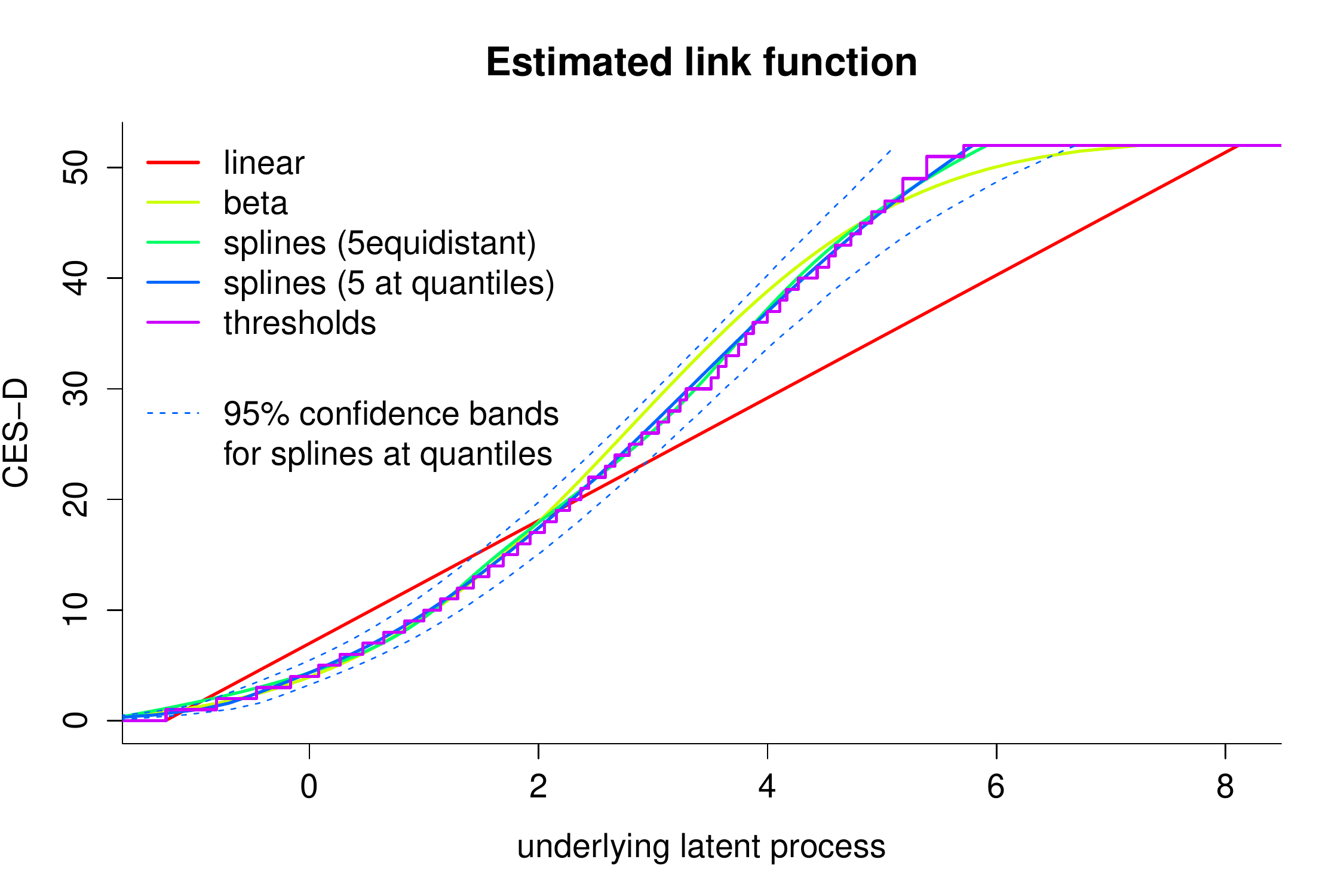}
\caption{\label{fig_lcmm_H} Estimated link functions for CES-D in a latent process mixed model using function \code{plot} with option \code{which="linkfunction"}, and function \code{predictlink}: linear, Beta CDF, I-splines with 5 knots at quantiles or equidistant, and thresholds. 95\% confidence bands are only given for the I-splines with 5 knots at quantiles.}
\end{center}
\end{figure}

As mentionned earlier, the intercept of the latent process and the standard error of the measurement errors are respectively constrained to 0 and 1, and the parameters involved in the link functions are given at the end. Models involving discrete and continuous link functions can be compared using the discrete AIC provided in the \code{summary}. In this case, the model with a link function approximated by I-splines with 5 knots placed at the quantiles provides the best fit (12644.18 \emph{versus} 12652.52 for the cumulative probit model, for example). To choose the optimal link function and further evaluate the nonlinearity of the relationship between the longitudinal marker and its underlying normal latent process, the estimated link functions are provided in output value \code{estimlink} and can be plotted as follows (plot in Figure \ref{fig_lcmm_H}). In this graph, confidence bands are obtained and plotted with function \code{predictlink} : 

\begin{Schunk}
\begin{Sinput}
R> col <- rainbow(5)
R> plot(mlin, which = "linkfunction", bty = 'l', ylab = "CES-D", lwd = 2,
+    col = col[1], xlab = "underlying latent process")
R> plot(mbeta, which = "linkfunction", add = T, col = col[2], lwd = 2)
R> plot(mspl, which = "linkfunction", add = T, col = col[3], lwd = 2)
R> plot(mspl5q, which = "linkfunction", add = T, col = col[4], lwd = 2)
R> plot(mord, which = "linkfunction", add = T, col = col[5], lwd = 2)
R> legend(x = "topleft", legend = c("linear", "beta",
+    "splines (5equidistant)", "splines (5 at quantiles)", "thresholds"), 
+    lty = 1, col = col, bty = "n", lwd = 2)
R> linkspl5q <- predictlink(mspl5q, ndraws = 2000)
R> plot(linkspl5q, add = TRUE, col = col[4], lty = 2)
R> legend(legend = c("95% confidence bands", "for splines at quantiles"),
+    x = "left", lty = c(2, NA), col = c(col[4], NA), bty = "n", lwd = 1)
\end{Sinput}
\end{Schunk}

As for \code{hlme}, plots for the residuals and the predictions \emph{versus} observations can be provided but they are now in the latent process scale (only the code is provided here):
\begin{Schunk}
\begin{Sinput}
R> plot(mspl5q)
R> plot(mspl5q, which = "fit", var.time = "age65", xlab = "(age - 65) / 10",
+    bty = "l", break.times = 8, ylab = "latent process", lwd = 2, marg = F,
+    ylim = c(-1, 2))
\end{Sinput}
\end{Schunk}

Predictions in the latent process scale and in the outcome scale are computed according to a profile of covariates using respectively functions \code{predictL} and \code{predictY} at the maximum likelihood estimates or using a MonteCarlo method with argument \code{draws=T} (by default 2000 draws). The following lines give the code for computing the predictions in the original scale (CES-D): 

\begin{Schunk}
\begin{Sinput}
R> datnew <- data.frame(age = seq(65, 95, length = 100))
R> datnew$age65 <- (datnew$age - 65) / 10
R> datnew$male <- 0
R> women <- predictY(mspl5q, newdata = datnew, var.time = "age", draws = T)
R> datnew$male <- 1
R> men <- predictY(mspl5q, newdata = datnew, var.time = "age", draws = T)
\end{Sinput}
\end{Schunk}

\begin{figure}[!ht]
\begin{center}
\includegraphics[width=0.9\textwidth]{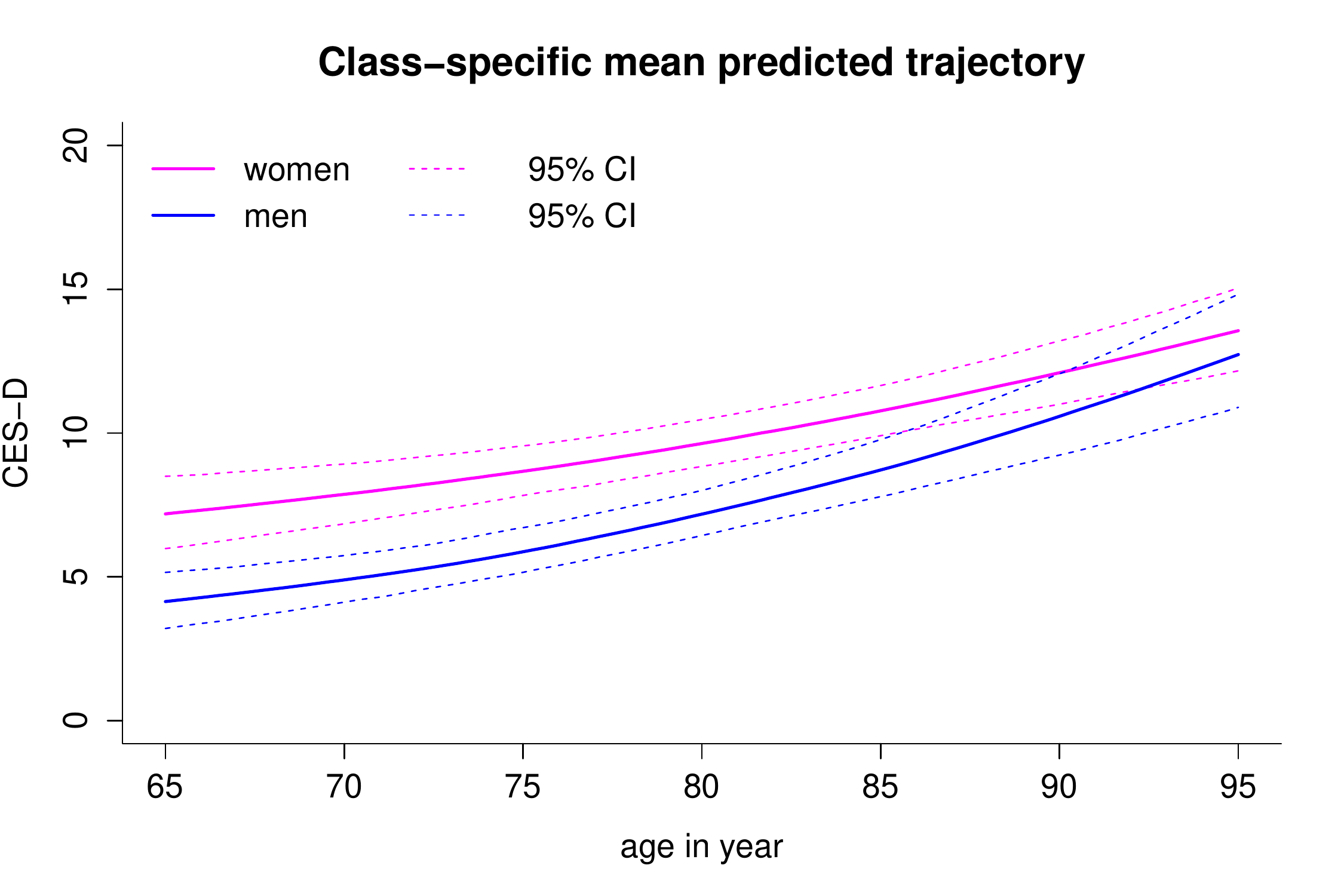}
\caption{\label{fig_lcmm_pred} Predicted CES-D trajectories versus age according to gender with 95\% confidence bands computed with function \code{predictY}.}
\end{center}
\end{figure}

The predicted trajectories can be plotted from these predictions as described below; the corresponding plot is displayed in Figure \ref{fig_lcmm_pred}.

\begin{Schunk}
\begin{Sinput}
R> plot(women, lwd = c(2, 1), type = "l", col = 6, ylim = c(0, 20),
+    xlab = "age in year", ylab = "CES-D", bty = "l", legend = NULL)
R> plot(men, add = TRUE, col = 4, lwd = c(2, 1))
R> legend(x = "topleft", bty = "n", ncol = 2, lty = c(1, 1, 2, 2), 
+    col = c(6, 4, 6, 4), lwd = c(2, 2, 1, 1),
+    legend = c("women", "men", "   95% CI", "   95% CI"))
\end{Sinput}
\end{Schunk}

In this example, only one latent class was assumed but a higher number of latent classes could be envisaged using the exact same syntax as shown in Section \ref{ex_hlme}. The corresponding post-fit functions also apply with the same syntax. 

\subsection[multlcmm examples]{\code{multlcmm}}\label{ex_multlcmm}

The latent process mixed models for multivariate longitudinal data implemented in \code{multlcmm} are illustrated by the study of the quadratic trajectory with time of the global cognitive level defined as the common factor underlying three psychometric tests: \code{MMSE}, \code{BVRT} and \code{IST}. Here the timescale is years since entry into the cohort and the model is adjusted for age at entry. To further investigate the effect of gender, both an effect on the common factor and differential effects (contrasts) on each marker are included. Correlated random effects on the time functions are considered, as well as a Brownian motion with \code{cor=BM(time)} and a marker-specific random intercept with \code{randomY=T}. The next lines estimate the corresponding latent process mixed model with Beta CDF link functions and provide the summary output:

\begin{Schunk}
\begin{Sinput}
R> paquid$time <- (paquid$age - paquid$age_init)
R> paquid$age0_centered <- paquid$age_init - 75
R> mult <- multlcmm(MMSE + IST + BVRT ~ age0_centered + male + contrast(male)
+    + time + I(time^2 / 10), random =~ time + I(time^2 / 10), subject = 'ID'
+    , data = paquid, randomY = T, cor = BM(time), link = c('beta', 'beta', 
+    'beta'))
R> summary(mult)
\end{Sinput}
\begin{Soutput}

General latent class mixed model 
     fitted by maximum likelihood method 
 
multlcmm(fixed = MMSE + IST + BVRT ~ age0_centered + male + contrast(male) + 
    time + I(time^2/10), random = ~time + I(time^2/10), subject = "ID", 
    randomY = T, link = c("beta", "beta", "beta"), cor = BM(time), 
    data = paquid)
 
Statistical Model: 
     Dataset: paquid 
     Number of subjects: 500 
     Number of observations: 6216 
     Number of latent classes: 1 
     Number of parameters: 30  
     Link functions: Standardised Beta CdF for MMSE  
                     Standardised Beta CdF for IST  
                     Standardised Beta CdF for BVRT  
 
Iteration process: 
     Convergence criteria satisfied 
     Number of iterations:  16 
     Convergence criteria: parameters= 1.2e-10 
                         : likelihood= 1.1e-08 
                         : second derivatives= 3.6e-13 
 
Goodness-of-fit statistics: 
     maximum log-likelihood: -14374.6  
     AIC: 28809.21  
     BIC: 28935.65  
 
Maximum Likelihood Estimates: 
 
Fixed effects in the longitudinal model:

                                  coef      Se   Wald p-value
intercept (not estimated)      0.00000                       
age0_centered                 -0.10941 0.01133 -9.657 0.00000
male                           0.14755 0.11882  1.242 0.21431
time                          -0.12226 0.01878 -6.511 0.00000
I(time^2/10)                  -0.02396 0.00975 -2.459 0.01394
Contrasts on male (p=0.00027)                                
MMSE                          -0.07482 0.06917 -1.082 0.27940
IST                           -0.26814 0.07853 -3.415 0.00064
BVRT**                         0.34296 0.08807  3.894 0.00010

Variance-covariance matrix of the random-effects:
(the variance of the first random effect is not estimated)
             intercept     time I(time^2/10)
intercept      1.00000                      
time           0.00385  0.01399             
I(time^2/10)  -0.00729 -0.00467      0.00298

                       coef      Se
BM standard error:  0.30856 0.04186

                                         MMSE      IST     BVRT
Residual standard error:              1.01369  0.92830  1.42295
Standard error of the random effect:  0.57499  0.85364  0.88877

Parameters of the link functions:

               coef      Se   Wald p-value
MMSE-Beta1  1.42040 0.07114 19.966 0.00000
MMSE-Beta2 -0.24691 0.08368 -2.951 0.00317
MMSE-Beta3  0.45613 0.02603 17.521 0.00000
MMSE-Beta4  0.06458 0.00580 11.125 0.00000
IST-Beta1  -0.03694 0.05647 -0.654 0.51295
IST-Beta2  -0.42183 0.07554 -5.584 0.00000
IST-Beta3   0.65192 0.01561 41.776 0.00000
IST-Beta4   0.08294 0.00676 12.262 0.00000
BVRT-Beta1  0.40165 0.08063  4.982 0.00000
BVRT-Beta2 -0.26390 0.12251 -2.154 0.03124
BVRT-Beta3  0.55731 0.01977 28.185 0.00000
BVRT-Beta4  0.06290 0.00632  9.952 0.00000

 ** coefficient not estimated but obtained from the others as minus the sum 
of them 
\end{Soutput}
\end{Schunk}

In this example, squared time was divided by 10 in order to avoid very small numbers for the corresponding fixed effect and variance of the random effect. 
The summary has the same appearance as the summaries of \code{hlme} or \code{lcmm} objects. In addition to the fixed effects on the latent process (common factor), the marker-specific contrasts are given and their global significance is tested with a multivariate Wald test. 
Then the estimated variance-covariance of the random effects and the standard error of the marker-specific intercept ("\texttt{Standard error of the random effect}") are given along with the standard error of the independent Gaussian error ("\texttt{Residual standard error}"). Finally, the marker-specific link function parameters are provided. \rev{Note that in \code{multlcmm}, the intercept of the latent process and the variance of the random intercept are respectively constrained to 0 and 1}.

\begin{figure}[!ht]
\begin{center}
\includegraphics{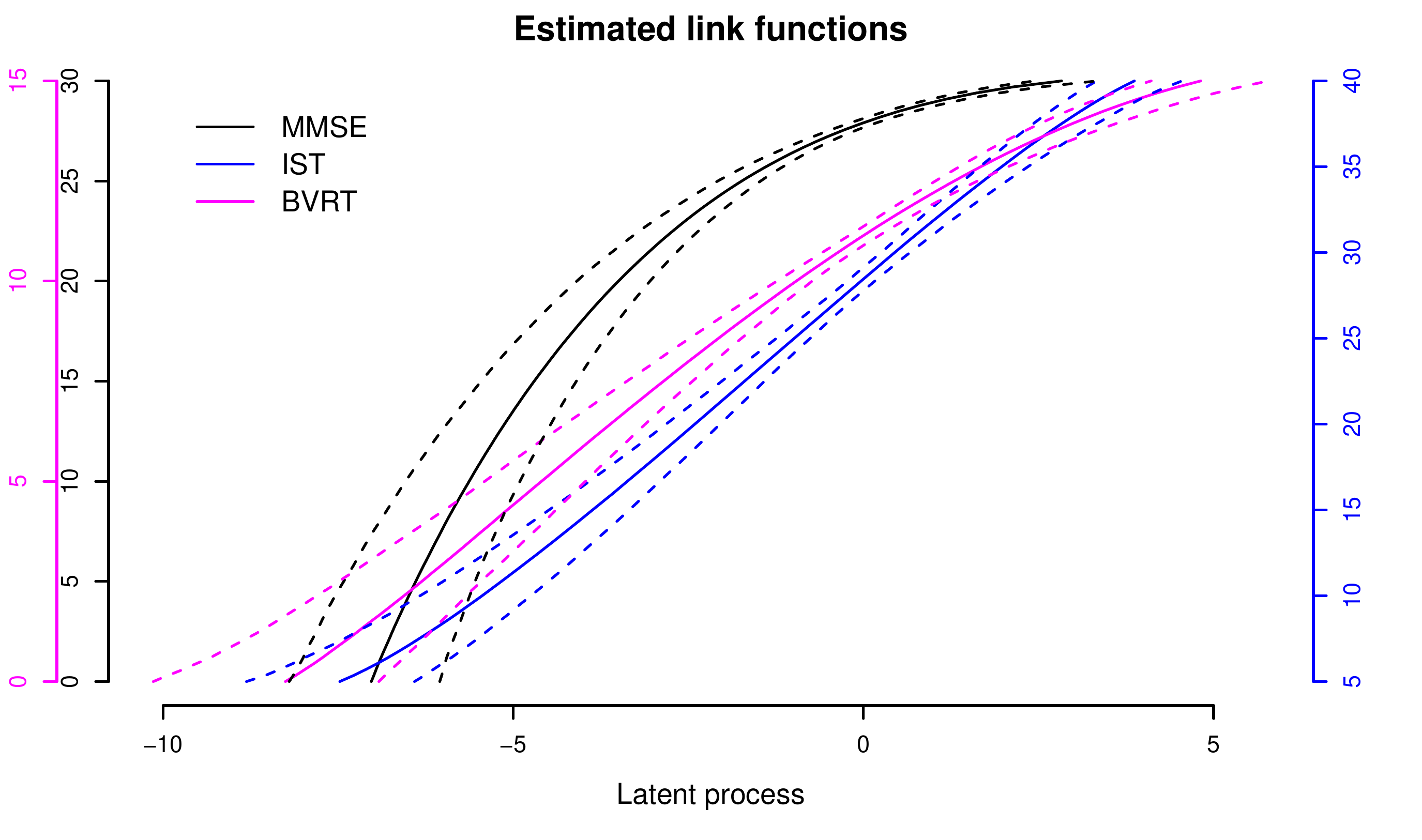}
\caption{\label{fig_multlcmm_H} Estimated link functions for \code{MMSE}, \code{BVRT} and \code{IST} with 95\% confidence bands using \code{predictlink}.}
\end{center}
\end{figure}

As for \code{lcmm} object, the estimated link function can be plotted with function \code{plot} and option \code{"linkfunction"} or with the 95\% confidence bands using \code{predictlink} as shown below and in Figure \ref{fig_multlcmm_H}:

\begin{Schunk}
\begin{Sinput}
R> plot(mult, which = "linkfunction", col = c(1, 4, 6), lwd = 2)
R> CI <- predictlink(mult)
R> plot(CI, col = c(1, 4, 6), lwd = 2)
R> head(CI$pred)
\end{Sinput}
\begin{Soutput}
  Yname   Yvalues transfY_50 transfY_2.5 transfY_97.5
1  MMSE 0.0000000  -7.039481   -8.272866    -6.088553
2  MMSE 0.3030303  -7.005178   -8.214649    -6.062041
3  MMSE 0.6060606  -6.969232   -8.158112    -6.040532
4  MMSE 0.9090909  -6.931033   -8.103282    -6.006305
5  MMSE 1.2121212  -6.895605   -8.055205    -5.975374
6  MMSE 1.5151515  -6.856988   -8.010372    -5.949715 
\end{Soutput}
\end{Schunk}

Note here that the predicted values of the link functions shown above are the median (50\%), and 2.5\%, 97.5\% of 2000 draws generated from the asymptotic distribution. As such, they may vary depending on the seed.

The percentage of variance explained by the common latent process (that is $\forall t$, $\frac{\text{Var}(\Lambda(t))}{\text{Var}(Y_k(t))}$) can be computed at a given time with function \code{VarExpl}. Note that \code{VarExpl} is also available with \code{lcmm}, \code{hlme} and \code{Jointlcmm} objects. In these cases, it computes the percentage of variance not explained by the measurement error. 

Here is the call for the explained variance at time 0 and time 5 in model \code{mult}:

\begin{Schunk}
\begin{Sinput}
R> VarExpl(mult, values = data.frame(time = 0))
\end{Sinput}
\begin{Soutput}
            class1
%Var-MMSE 42.40560
%Var-IST  38.60366
%Var-BVRT 26.21439
\end{Soutput}
\begin{Sinput}
> VarExpl(mult, values = data.frame(time = 5))
\end{Sinput}
\begin{Soutput}
            class1
%Var-MMSE 56.01423
%Var-IST  52.09580
%Var-BVRT 38.06074
\end{Soutput}
\end{Schunk}

In this example, the latent process explains between 26\% and 42\% of the total variance of the markers at time 0, and between 38\% and 56\% after 5 years. \\

The model was estimated with a unique latent class but the exact same function also applies with a higher number of latent classes specified in input of \code{multlcmm} according to the same syntax as explained in \ref{ex_hlme}. All the functions described in \code{hlme} and \code{lcmm} sections (i.e., Section \ref{ex_hlme} and Section \ref{ex_lcmm}) apply with a \code{multlcmm} object.

\subsection[Jointlcmm examples]{\code{Jointlcmm}}\label{ex_Jointlcmm}

The joint latent class mixed models implemented in \code{Jointlcmm} are illustrated by the study of the trajectories of \code{normMMSE} with age and the associated risk of dementia.
Indeed, cognitive change over time and the risk of dementia are two processes that are closely linked. Their joint study is useful both to better understand the natural history of cognitive aging and dementia, and to provide dynamic tools to evaluate the individual risk of dementia based on observed repeated cognitive measures. For the sake of simplicity and because this illustration only aims at explaining the function implementation, we do not take into account the competing risk of death although this could be done with the package, and neglecting death in the dementia context may lead to biased estimates of dementia incidence. 

In this example, we assumed class-specific quadratic trajectories of \code{normMMSE} with \code{age65} adjusted for \code{CEP}, and we jointly modelled the risk of dementia according to \code{CEP} and \code{male} assuming class-specific Weibull baseline risk functions with age. No covariates are included in the class-membership model. In this illustration, the delayed entry into the cohort at age \code{age\_init} was taken into account in the estimation process. The observed time of event was \code{agedem} and the indicator of event was \code{dem}. We selected the 499 subjects for whom \code{agedem}>\code{age\_init}:

\begin{Schunk}
\begin{Sinput}
paquidS <- paquid[paquid$agedem > paquid$age_init, ]
\end{Sinput}
\end{Schunk}

\rev{The next lines give the code for estimating the joint latent class mixed model with one latent class, i.e., the model assuming independence between the cognitive measures and time to dementia. }
\begin{Schunk}
\begin{Sinput}
R> mj1 <- Jointlcmm(normMMSE ~ poly(age65, degree = 2, raw = TRUE) + CEP,
+    random =~ poly(age65, degree = 2, raw = TRUE), 
+    survival = Surv(age_init, agedem, dem) ~ CEP + male, 
+    hazard = "Weibull", subject = 'ID', data = paquidS, ng = 1)

\end{Sinput}
\end{Schunk}

\rev{From this model, joint latent class mixed models with two to four classes are estimated using the automatic specification of the initial values:}
\begin{Schunk}
\begin{Sinput}
R> mj2 <- Jointlcmm(normMMSE ~ poly(age65, degree = 2, raw = TRUE) + CEP,
+    mixture =~ poly(age65, degree = 2, raw = TRUE), 
+    random =~ poly(age65, degree = 2, raw = TRUE), 
+    survival = Surv(age_init, agedem, dem) ~ CEP + male,
+    hazard = "Weibull", subject = 'ID', data = paquidS, ng = 2, B = mj1)
R> mj3 <- Jointlcmm(normMMSE ~ poly(age65, degree = 2, raw = TRUE) + CEP,
+    mixture =~ poly(age65, degree = 2, raw = TRUE), 
+    random =~ poly(age65, degree = 2, raw = TRUE), 
+    survival = Surv(age_init, agedem, dem) ~ CEP + male,
+    hazard = "Weibull", subject = 'ID', data = paquidS, ng = 3, B = mj1)
R> mj4 <- Jointlcmm(normMMSE ~ poly(age65, degree = 2, raw = TRUE) + CEP,
+    mixture =~ poly(age65, degree = 2, raw = TRUE), 
+    random =~ poly(age65, degree = 2, raw = TRUE), 
+    survival = Surv(age_init, agedem, dem) ~ CEP + male,
+    hazard = "Weibull", subject = 'ID', data = paquidS, ng = 4, B = mj1)

\end{Sinput}
\end{Schunk}

Function \code{summarytable} provides a table summarizing the results of these four models:

\begin{Schunk}
\begin{Sinput}
R> summarytable(mj1, mj2, mj3, mj4)
\end{Sinput}
\begin{Soutput}
    G    loglik npm      BIC   %class1   %class2  %class3 %class4
mj1 1 -9449.079  15 18991.35 100.00000                           
mj2 2 -9337.187  21 18804.84  78.95792  21.04208                 
mj3 3 -9449.079  27 19065.90   0.00000 100.00000  0.00000        
mj4 4 -9337.187  33 18879.39   0.00000  78.95792 21.04208       0
\end{Soutput}
\end{Schunk}

This table provides the maximum log-likelihood, the number of parameters \code{npm}, the BIC and the posterior proportion of each latent class. It is useful to compare the different models estimated and select the most appropriate one. Here, we first see that for each additional latent class, there is a 6-parameter increase. This corresponds to the additional class-specific parameters: the proportion of the class, the two Weibull parameters, and the three fixed effects for the quadratic trajectory (intercept, time and time squared). 

The two-latent-class model provides a better BIC than the one-class model and posterior classes with proportions 21\% and 79\%. With the automatic choice of initial values, models with three and four latent classes reached local maxima  (but without correct convergence thanks to the derivative criterion): \code{mj3} reached the one-class solution of \code{mj1}, and \code{mj4} reached the two-class solution of \code{mj2}. This illustrates once again that default initial values do not necessarily lead to a global maximum (and a convergence), and that multiple sets of initial values should be systematically tried. The models were thus reestimated with various sets of initial values specified in \code{B}. For example, the following code illustrates a reestimation of the three-class model using estimates of the two-class model as initial values along with arbitrary initial values for an additional class:

\begin{Schunk}
\begin{Sinput}
R> Binit <- rep(0, length(mj2$best) + 6)
R> Binit[c(2, 5:10, 12, 13, 15, 16, 18, 19:(length(Binit)))] <- mj2$best
R> Binit[c(1, 3, 4, 11, 14, 17)] <- c(0, 0.11, 4, 70, 0, 0)
R> mj3b <- Jointlcmm(normMMSE ~ poly(age65, degree = 2, raw = TRUE) + CEP,
+    mixture =~ poly(age65, degree = 2, raw = TRUE), 
+    random =~ poly(age65, degree = 2, raw = TRUE), 
+    survival = Surv(age_init, agedem, dem) ~ CEP + male,
+    hazard = "Weibull", subject = 'ID', data = paquidS, ng = 3, B = Binit)
\end{Sinput}
\end{Schunk}
Similarly for the four-class model:
\begin{Schunk}
\begin{Sinput}
R> Binit <- rep(0, length(mj3b$best) + 2 + 3 + 1)
R> Binit[c(1, 2, 4:7, 10:15, 17:19, 21:23, 25:length(Binit))] <- mj3b$best
R> Binit[c(3, 8, 9, 16, 20, 24)] <- c(0, 0.1, 10, 60, 5, -10)
R> mj4b <- Jointlcmm(normMMSE ~ poly(age65, degree = 2, raw = TRUE) + CEP,
+    mixture =~ poly(age65, degree = 2, raw = TRUE), 
+    random =~ poly(age65, degree = 2, raw = TRUE), 
+    survival = Surv(age_init, agedem, dem) ~ CEP + male,
+    hazard = "Weibull", subject = 'ID', data = paquidS, ng = 4, B = Binit)
\end{Sinput}
\end{Schunk}

\rev{A grid search can also be used for the three and four class models. The next lines provide the code with 30 random vectors of initial values and a maximum of 15 iterations for selecting the best log-likelihood from which the estimation procedure can be finalized.}

\begin{Schunk}
\begin{Sinput}
R> mj3c <- gridsearch(rep = 30, maxiter = 15, minit = mj1,
+    Jointlcmm(normMMSE ~ poly(age65, degree = 2, raw = TRUE) + CEP,
+    mixture =~ poly(age65, degree = 2, raw = TRUE), 
+    random =~ poly(age65, degree = 2, raw = TRUE), 
+    survival = Surv(age_init, agedem, dem) ~ CEP + male,
+    hazard = "Weibull", subject = 'ID', data = paquidS, ng = 3)) 
R> mj4c <- gridsearch(rep = 30, maxiter = 15, minit = mj1,
+    Jointlcmm(normMMSE ~ poly(age65, degree = 2, raw = TRUE) + CEP,
+    mixture =~ poly(age65, degree = 2, raw = TRUE), 
+    random =~ poly(age65, degree = 2, raw = TRUE), 
+    survival = Surv(age_init, agedem, dem) ~ CEP + male,
+    hazard = "Weibull", subject = 'ID', data = paquidS, ng = 4))
\end{Sinput}
\end{Schunk}

The trials can be summarized using:

\begin{Schunk}
\begin{Sinput}
R> summarytable(mj1, mj2, mj3, mj3b, mj3c, mj4, mj4b, mj4c)
\end{Sinput}
\begin{Soutput}
     G    loglik npm      BIC   %class1   %class2   %class3  %class4
mj1  1 -9449.079  15 18991.35 100.00000                             
mj2  2 -9337.187  21 18804.84  78.95792  21.04208                   
mj3  3 -9449.079  27 19065.90   0.00000 100.00000  0.000000         
mj3b 3 -9294.952  27 18757.64  68.33667  18.63727 13.026052         
mj3c 3 -9294.952  27 18757.64  18.63727  13.02605 68.336673         
mj4  4 -9337.187  33 18879.39   0.00000  78.95792 21.042084  0.00000
mj4b 4 -9267.481  33 18739.98  69.13828  13.82766  3.006012 14.02806
mj4c 4 -9285.206  33 18775.43  48.49699  27.65531  9.819639 14.02806
\end{Soutput}
\end{Schunk}

The model with four latent classes \code{mj4b} is selected as providing the lowest BIC. Note however that the model with five latent classes should also be estimated to ensure that the model with four latent classes provides the best BIC -we did not do it here to avoid additional computations. \rev{Note also that the grid search, as defined here with 30 replicates and 15 iterations, did not provide the global maximum for four classes. This might happen as the method only reduces the odds of converging towards a local maximum. Other grid searches could be tested by changing the number of replicates and/or iterations.}

The summary of the selected four-class joint model is given below:  

\begin{Schunk}
\begin{Sinput}
R> summary(mj4b)
\end{Sinput}
\begin{Soutput}
Joint latent class model for quantitative outcome and competing risks 
     fitted by maximum likelihood method 
 
Jointlcmm(fixed = normMMSE ~ poly(age65, degree = 2, raw = TRUE) + 
    CEP, mixture = ~poly(age65, degree = 2, raw = TRUE), random = ~poly(age65, 
    degree = 2, raw = TRUE), subject = "ID", ng = 4, survival = Surv(age_init, 
    agedem, dem) ~ CEP + male, hazard = "Weibull", data = paquidS)
 
Statistical Model: 
     Dataset: paquidS 
     Number of subjects: 499 
     Number of observations: 2213 
     Number of latent classes: 4 
     Number of parameters: 33  
     Event 1: 
        Number of events: 128
        Class-specific hazards and 
        Weibull baseline risk function 
 
Iteration process: 
     Convergence criteria satisfied 
     Number of iterations:  37 
     Convergence criteria: parameters= 9.1e-05 
                         : likelihood= 1.5e-07 
                         : second derivatives= 7.3e-11 
 
Goodness-of-fit statistics: 
     maximum log-likelihood: -9267.48  
     AIC: 18600.96  
     BIC: 18739.98  
     Score test statistic for CI assumption: 30.667 (p-value=0) 
 
Maximum Likelihood Estimates: 
 
Fixed effects in the class-membership model:
(the class of reference is the last class) 

                       coef       Se     Wald p-value
intercept class1    0.89729  0.20360    4.407 0.00001
intercept class2    0.24518  0.24919    0.984 0.32518
intercept class3   -1.19317  0.35708   -3.341 0.00083

Parameters in the proportional hazard model:

                                       coef       Se     Wald p-value
event1 +/-sqrt(Weibull1) class 1    0.10121  0.00039  259.466 0.00000
event1 +/-sqrt(Weibull2) class 1    7.04817  0.82050    8.590 0.00000
event1 +/-sqrt(Weibull1) class 2    0.10567  0.00037  284.423 0.00000
event1 +/-sqrt(Weibull2) class 2    6.91442  0.50276   13.753 0.00000
event1 +/-sqrt(Weibull1) class 3    0.11421  0.00080  142.606 0.00000
event1 +/-sqrt(Weibull2) class 3    5.57068  0.73739    7.555 0.00000
event1 +/-sqrt(Weibull1) class 4    0.10955  0.00039  278.049 0.00000
event1 +/-sqrt(Weibull2) class 4    6.39438  0.42368   15.092 0.00000
CEP                                -0.66581  0.26350   -2.527 0.01151
male                                0.43642  0.29473    1.481 0.13867

Fixed effects in the longitudinal model:

                                                  coef       Se     Wald
intercept class1                              65.30225  3.38062   19.317
intercept class2                              57.39464  5.26865   10.894
intercept class3                              83.22500  9.26143    8.986
intercept class4                              65.50576  4.93682   13.269
poly(age65, degree = 2, raw = TRUE)1 class1    4.62181  3.73950    1.236
poly(age65, degree = 2, raw = TRUE)1 class2   19.31410  6.05887    3.188
poly(age65, degree = 2, raw = TRUE)1 class3  -64.10563 16.27356   -3.939
poly(age65, degree = 2, raw = TRUE)1 class4   15.15505  6.72560    2.253
poly(age65, degree = 2, raw = TRUE)2 class1   -3.25830  1.08352   -3.007
poly(age65, degree = 2, raw = TRUE)2 class2  -11.44706  1.80143   -6.354
poly(age65, degree = 2, raw = TRUE)2 class3   16.81947  6.30089    2.669
poly(age65, degree = 2, raw = TRUE)2 class4  -16.26249  2.15066   -7.562
CEP                                           12.80547  1.20942   10.588
                                            p-value
intercept class1                            0.00000
intercept class2                            0.00000
intercept class3                            0.00000
intercept class4                            0.00000
poly(age65, degree = 2, raw = TRUE)1 class1 0.21648
poly(age65, degree = 2, raw = TRUE)1 class2 0.00143
poly(age65, degree = 2, raw = TRUE)1 class3 0.00008
poly(age65, degree = 2, raw = TRUE)1 class4 0.02424
poly(age65, degree = 2, raw = TRUE)2 class1 0.00264
poly(age65, degree = 2, raw = TRUE)2 class2 0.00000
poly(age65, degree = 2, raw = TRUE)2 class3 0.00760
poly(age65, degree = 2, raw = TRUE)2 class4 0.00000
CEP                                         0.00000


Variance-covariance matrix of the random-effects:
                                      intercept
intercept                             235.93031
poly(age65, degree = 2, raw = TRUE)1 -251.67613
poly(age65, degree = 2, raw = TRUE)2   74.24158
                                     poly(age65, degree = 2, raw = TRUE)1
intercept                                                                
poly(age65, degree = 2, raw = TRUE)1                             439.6308
poly(age65, degree = 2, raw = TRUE)2                            -139.0736
                                     poly(age65, degree = 2, raw = TRUE)2
intercept                                                                
poly(age65, degree = 2, raw = TRUE)1                                     
poly(age65, degree = 2, raw = TRUE)2                             45.49113

                              coef       Se
Residual standard error    9.94241  0.19331
\end{Soutput}
\end{Schunk}

The summary of a \code{Jointlcmm} object is very similar to the summaries of \code{hlme} or \code{lcmm} objects (depending on whether a link function was assumed in \code{Jointlcmm}). The main difference is that in addition to estimates of the multinomial model and of the mixed model, estimates from the survival model are also given. The summary also provides the statistic of a score test for the conditional independence assumption (see \cite{jacqmin-gadda2010,proust-lima2014} for more details). Note here that the conditional independence assumption between the longitudinal and survival processes given the latent classes is rejected although the statistic of the test was much lower with four classes than with three or two.  

\begin{figure}[!ht]
\begin{center}
\includegraphics{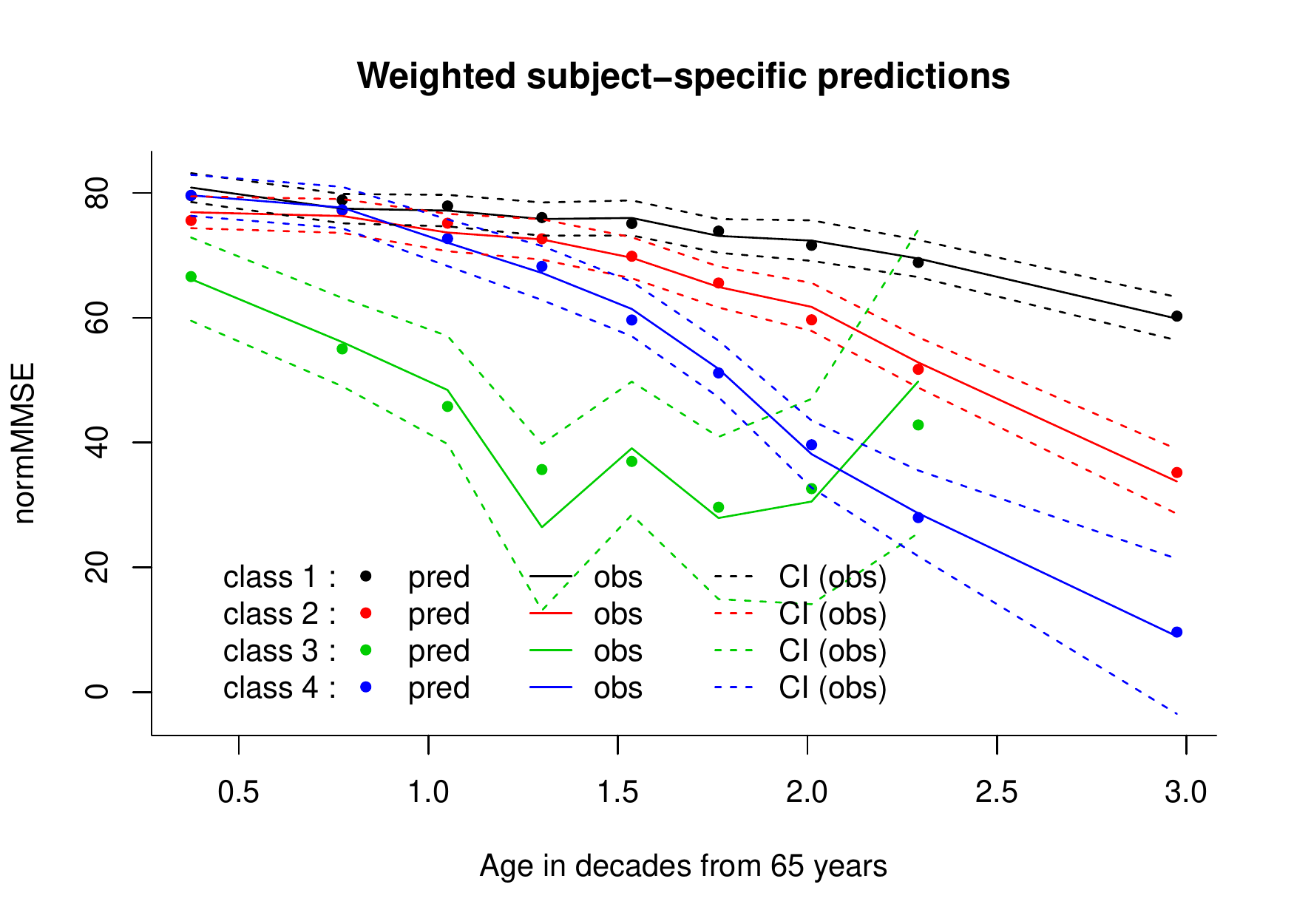}
\caption{\label{fig_Jointlcmm_fit} Weighted observations and weighted mean of subject-specific predictions from the joint model with four latent classes.}
\end{center}
\end{figure}

Postfit functions \code{plot} and \code{predictY} (along with its \code{plot} functionality) described in Section \ref{ex_hlme} are also available for \code{Jointlcmm} objects. \code{plot} provides longitudinal residuals (with option \code{which="residuals"}) and the comparison between observed and predicted longitudinal data (with option \code{which="fit"}) as shown below with subject-specific predictions (and in Figure \ref{fig_Jointlcmm_fit}):

\begin{Schunk}
\begin{Sinput}
R> plot(mj4b, which = "fit", var.time = "age65", marg = F, break.times = 10,
+    bty = "l", ylab = "normMMSE", xlab = "Age in decades from 65 years")
\end{Sinput}
\end{Schunk}

Class-specific predicted longitudinal trajectories can be computed using \code{predictY} for a given covariate profile, and can be plotted with \code{plot}. The baseline risks and baseline survival functions can also be plotted with function \code{plot} and options \code{which="hazard"} and \code{which="survival"}. Here is the code for the predicted mean longitudinal trajectories and baseline survival functions in each class of the four-class joint model:

\begin{Schunk}
\begin{Sinput}
R> datnew <- data.frame(age65 = seq(0, 3, length=100))
R> datnew$male <- 0
R> datnew$CEP <- 0
R> par(mfrow = c(1, 2))
R> mj4b.pred <- predictY(mj4b, newdata = datnew, var.time = "age65")
R> plot(mj4b.pred, bty = "l", ylim = c(0, 80), legend.loc = "bottomleft",
+    ylab = "normMMSE", xlab = "age in decades from 65 years", lwd = 2)
R> plot(mj4b, which = "survival", lwd = 2, legend.loc = F, bty = "l", 
+    xlab = "age in years", ylab = "dementia-free probability")
\end{Sinput}
\end{Schunk}
\begin{figure}[!ht]
\begin{center}
\includegraphics{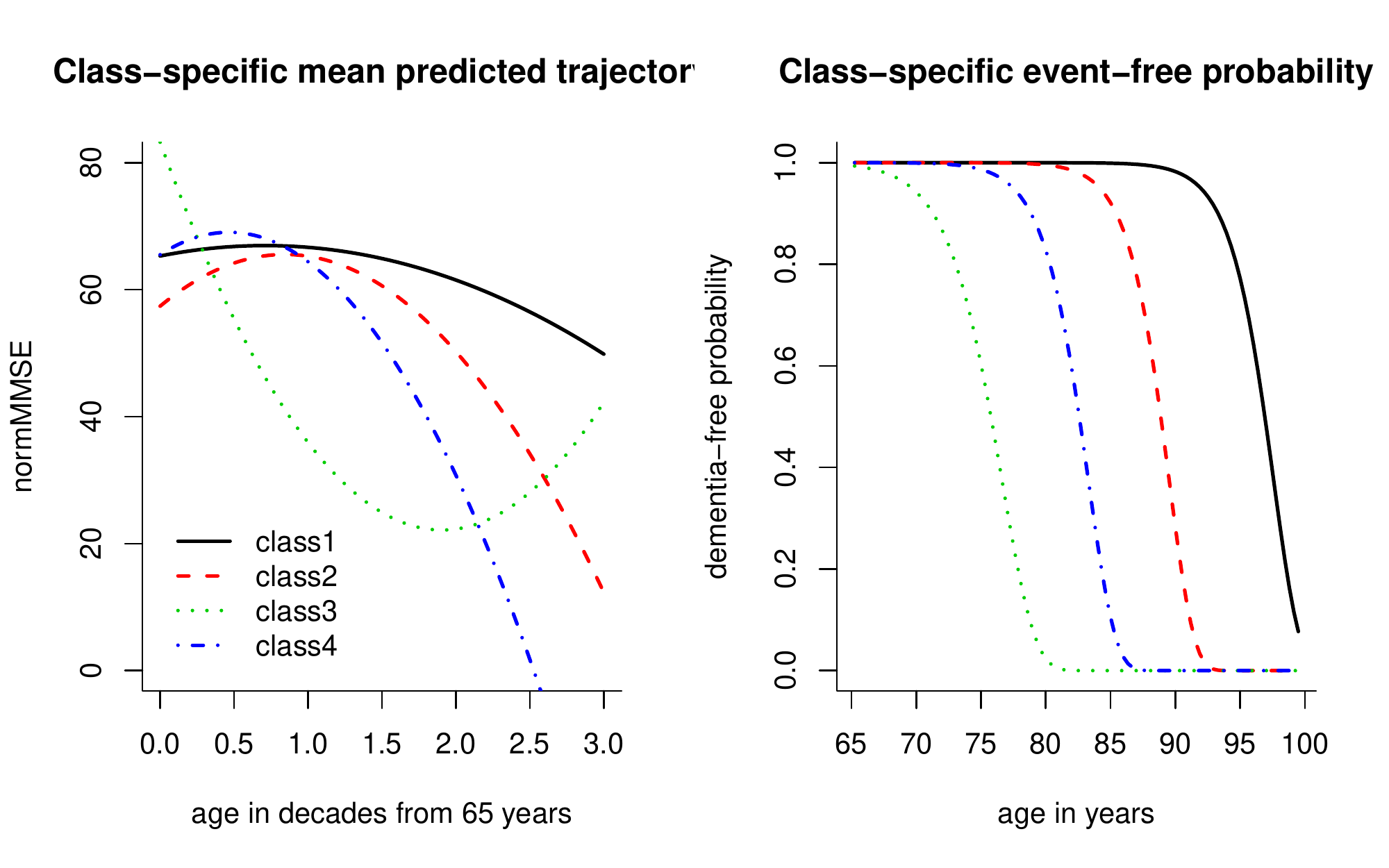}
\caption{\label{fig_pred_surv} Marginal predicted trajectories of normMMSE and associated survival curves in each of the four latent classes for a woman with low educational level.}
\end{center}
\end{figure}

Class-specific risk functions and cumulative risk functions are also provided in output value \code{predSurv}. 

As for other objects in package \pkg{lcmm}, the classification can be summarized with function \code{postprob}:

\begin{Schunk}
\begin{Sinput}
R> postprob(mj4b)
\end{Sinput}
\begin{Soutput}
Posterior classification based on longitudinal and time-to-event data: 
  class1 class2 class3 class4
N 345.00  69.00  15.00  70.00
%  69.14  13.83   3.01  14.03
 
Posterior classification table: 
     --> mean of posterior probabilities in each class 
        prob1  prob2  prob3  prob4
class1 0.7188 0.1902 0.0093 0.0817
class2 0.1155 0.8224 0.0000 0.0621
class3 0.0355 0.0254 0.8873 0.0518
class4 0.0565 0.1137 0.0319 0.7978
 
Posterior probabilities above a threshold (%): 
         class1 class2 class3 class4
prob>0.7  49.28  71.01  86.67  67.14
prob>0.8  40.87  65.22  86.67  54.29
prob>0.9  31.30  59.42  73.33  52.86
 
 
Posterior classification based only on longitudinal data: 
  class1 class2 class3 class4
N 344.00  71.00  16.00  68.00
%  68.94  14.23   3.21  13.63
\end{Soutput}
\end{Schunk}

The classification provided in the classification table is satisfactory with mean posterior probability in each class above 71\% and up to 88.7\%. Note that here, two posterior classifications are provided, the main one based on all the information, and the classification based only on the longitudinal information. The corresponding classifications along with the posterior class-membership probabilities are provided in output values \code{pprob} and \code{pprobY}. \\

One objective of joint models may be the dynamic prediction of the event. Function \code{epoce} computes the predictive ability of the models using the expected prognostic observed cross-entropy (EPOCE) at different landmark times. When comparing different models, \code{epoce} can be plotted to visualize the predictive abilities at different landmark times. The difference in the predictive abilities of the two models can also be computed with \code{Diffepoce} function and plotted with the associated \code{plot} function.
Here is an example of the code:

\begin{Schunk}
\begin{Sinput}
R> landmark <- c(70, 72, 75, 77, 80, 82, 85, 87, 90)
R> epoce1 <- epoce(mj1, pred.times = landmark, var.time = "age65",
+    fun.time = function(x) { 10 * x + 65 })
R> epoce2 <- epoce(mj2, pred.times = landmark, var.time = "age65",
+    fun.time = function(x) { 10 * x + 65 })
R> epoce3 <- epoce(mj3b, pred.times = landmark, var.time = "age65",
+    fun.time = function(x) { 10 * x + 65 })
R> epoce4 <- epoce(mj4b, pred.times = landmark, var.time = "age65",
+    fun.time = function(x) { 10 * x + 65})
R> diff23 <- Diffepoce(epoce2, epoce3)
R> diff34 <- Diffepoce(epoce3, epoce4)
R> par(mfrow = c(1, 2))
R> plot(epoce1, ylim = c(0.5, 1.5), main = "cross-validated EPOCE estimates",
+    bty = "l")
R> plot(epoce2, add = TRUE, col = 2, lty = 2)
R> plot(epoce3, add = TRUE, col = 3, lty = 3)
R> plot(epoce4, add = TRUE, col = 4, lty = 4)
R> legend("topright", legend = c("G=1", "G=2", "G=3", "G=4"), col = 1:4, 
+    lty = 1:4, bty = "n")
R> plot(diff23, main = "Difference in EPOCE estimates", lty = c(1, 2, 2),
+    pch = 20, ylim = c(-0.05, 0.30), bty = "l")
R> plot(diff34, add = T, main = "Difference in EPOCE estimates", col = 4, 
+    lty = c(1, 2, 2), pch = 18)
R> legend("topleft", legend = c("G=2/G=3", "G=3/G=4", "95%TI", "95%TI"), 
+    ncol = 2, col = c(1, 4, 1, 4), lty = c(1, 1, 2, 2), 
+    pch = c(20, 18, 20, 18), bty = "n")
\end{Sinput}
\end{Schunk}
\begin{figure}[!ht]
\begin{center}
\includegraphics{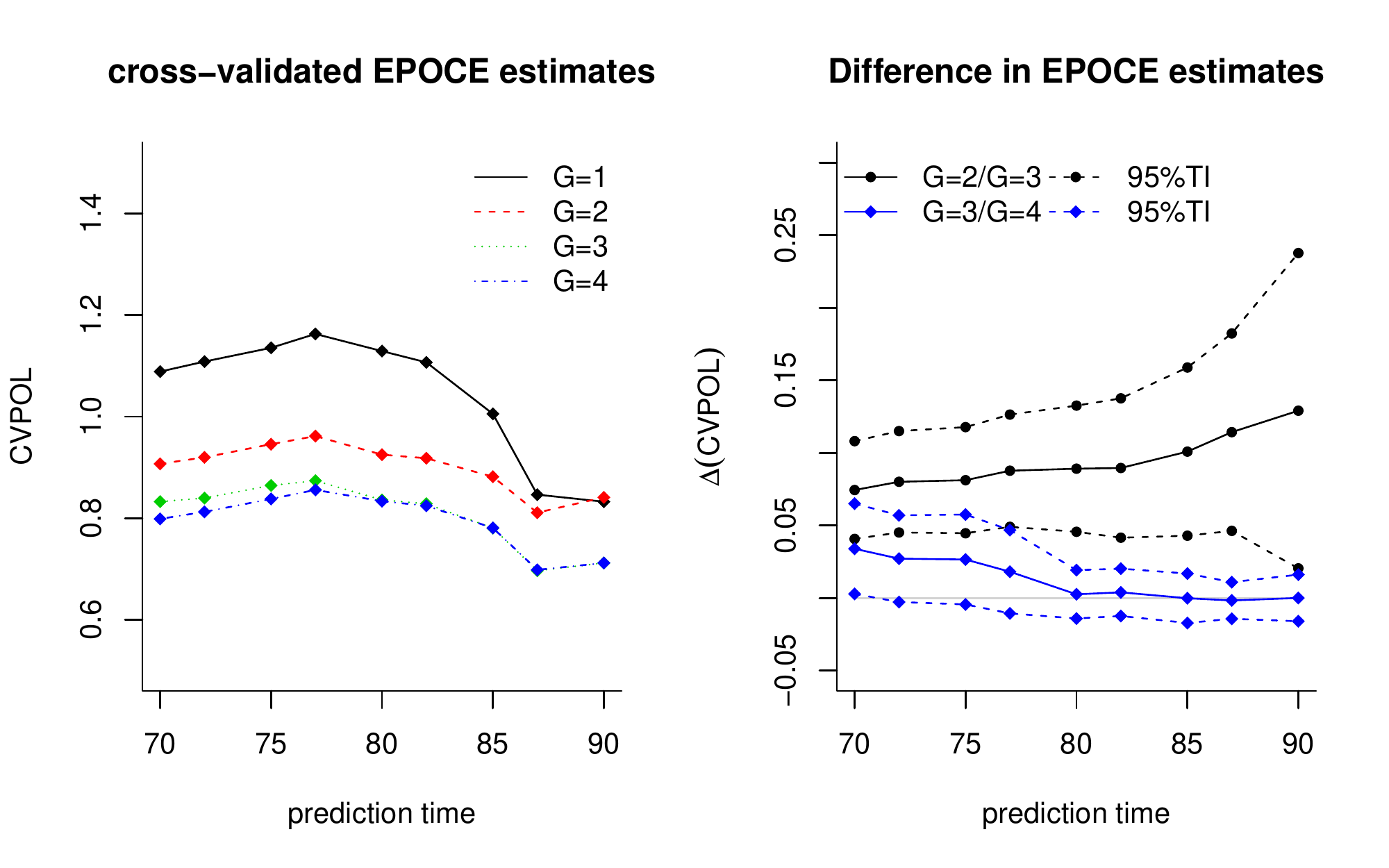}
\caption{\label{fig_epoce} Cross-validated EPOCE (expected prognostic observed cross-entropy) estimates for joint models with one to four classes (on the left) and difference in EPOCE estimates between joint models with two and three latent classes or three and four classes (on the right).}
\end{center}
\end{figure}

\rev{Joint models for \texttt{normMMSE} change and incidence of dementia give a better predictive ability, that is lower EPOCE, than the simple survival model for dementia (with one class). Although the four-class model provides the best goodness-of-fit in terms of BIC, the three-class model gives roughly the same predictive accuracy as the four-class model, especially after 80 years old (difference in EPOCE close to 0).} \\

\begin{figure}[!ht]
\begin{center}
\includegraphics{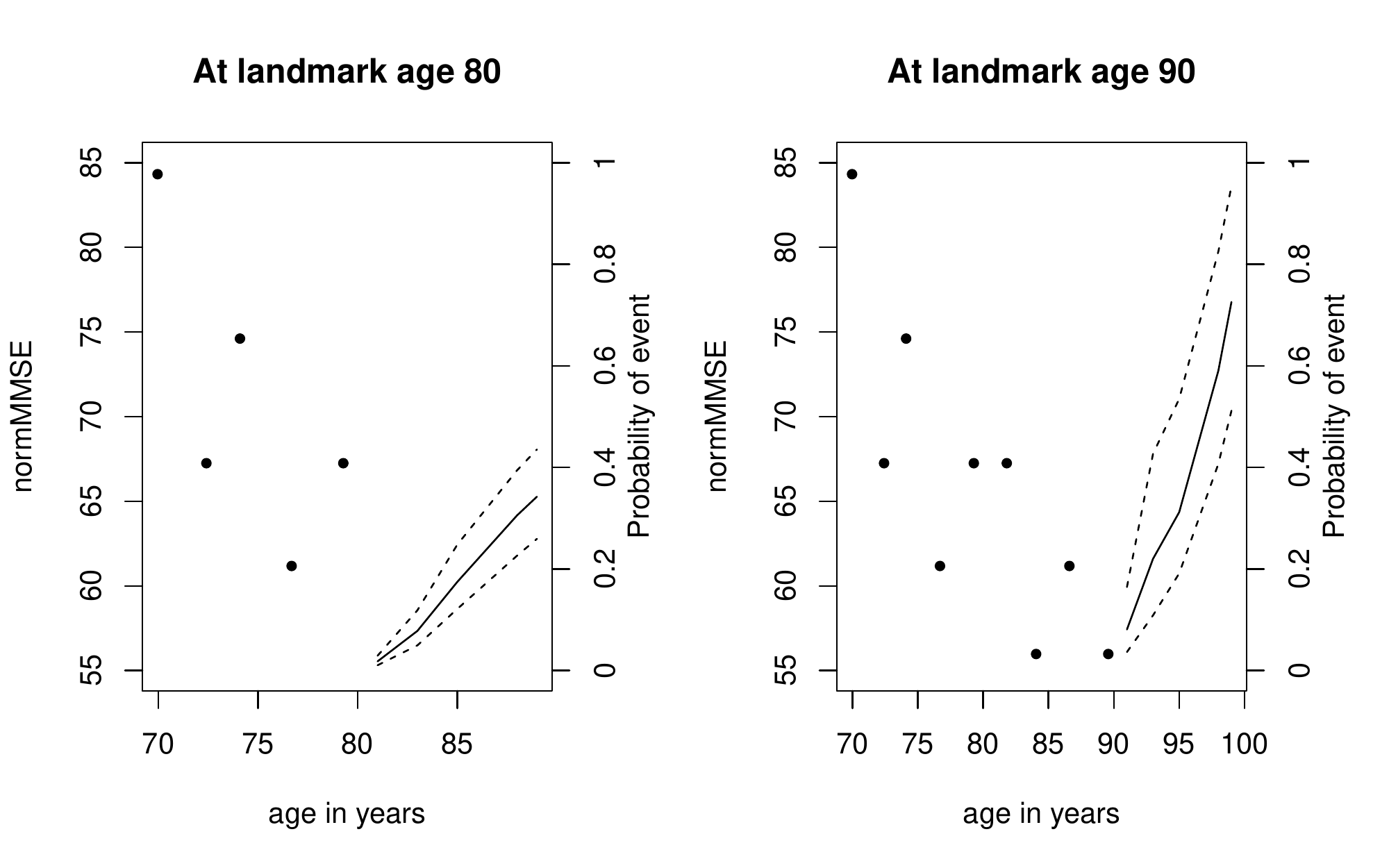}
\caption{\label{fig_dynpred} Individual dynamic prediction of dementia at landmark ages 80 and 90 years old and for horizon times 1, 3, 5, 8 and 9 years for subject 72 from the Paquid sample.}
\end{center}
\end{figure}

Finally, individual dynamic prediction of the event can be computed using \code{dynpred} function. For a specific subject whose data are provided in input, the probabilities of occurrence of the event from landmark times indicated in \code{landmark} and at horizons indicated in \code{horizon} are computed from the estimated model by using the longitudinal information up to the landmark times (for G>1). We give here an example (including the graphs in Figure \ref{fig_dynpred}) for a subject from the estimation data but the exact same computation could have be done for any individual (not included in the estimation data). We considered two landmark ages (80 and 90) and computed the probility of dementia derived from model \code{mj4b} for horizons of 1, 3, 5, 8 and 9 years:

\begin{Schunk}
\begin{Sinput}
R> paq72 <- paquid[which(paquid$ID == 72), ]
R> dynp <- dynpred(mj4b, paq72, landmark = c(80, 90), var.time = "age65", 
+    horizon = c(1, 3, 5, 8, 9), fun.time = function(x) { 10 * x + 65 }, 
+    draws = TRUE)
R> plot(dynp, landmark = 80, ylim = c(55, 85, 0, 1), col = 1, pch = 20, 
+    ylab = "normMMSE", main = "At landmark age 80", xlab = "age in years")
R> plot(dynp, landmark = 90, ylim = c(55, 85, 0, 1), col = 1, pch = 20, 
+    ylab = "normMMSE", main = "At landmark age 90", xlab = "age in years")
\end{Sinput}
\end{Schunk}

\section[Concluding remarks]{Concluding remarks}

The \pkg{lcmm} package provides a series of functions that extend the linear mixed model to various settings including specific types of nonlinear mixed models and multivariate mixed models, but also latent class mixed models and joint models. Although initially designed for the analysis of cognitive data in aging cohort studies such as those available in the dataset \code{paquid}, the functions also apply to many other settings. In particular, the latent process mixed model is designed for the longitudinal analysis of scales that usually have asymmetric distributions with possibly a ceiling effect, floor effects and unequal interval scaling, which was the case for the \code{CES-D} for depressive symptoms.

To our knowledge, no other programs estimate general (multivariate) latent process mixed models or joint latent class mixed models. However, 
some programs exist for fitting types of latent class mixed models. The most well-known software is \proglang{Mplus} \citep{muthen2001mplus} which more generally fits latent variable models from the structural equation modelling approach. Other free programs include macros \code{HETMIXED} \citep{HETMIXED} and \code{HETNLMIXED} \citep{HETNLMIXED} in  \proglang{SAS} \citep{SAS} that are numerically limited \citep{proust2005}, and the free \proglang{Fortran 90} program \code{HETMIXSURV} \citep{HETMIXSURV, proust-lima2015} that might be faster but is not user-friendly. To our knowledge, three packages can be used to fit latent class mixed models in \proglang{R} \citep{R}: function \code{GLMM\_MCMC} \citep{komarek2009} of package \pkg{mixAK} \citep{mixAK} fits latent class generalized linear mixed models with possibly multivariate longitudinal data based on MCMC estimation; package \pkg{FlexMix} \citep{flexmix,flexmix2} also proposes the estimation of classes of mixture models including the latent class linear mixed model with function \code{FLXMRlmer}; and \pkg{mixtools} \citep{mixtools} includes a functionality to fit latent class linear models with random effects. In \proglang{stata}, the program \code{GLLAMM} fits some latent class and latent variable models \citep{rabe2004}. Finally, when one is interested in classification of trajectories, exploratory methods such as the latent class growth analysis \citep{nagin1999} are usually preferred. They have especially become very popular in psychology \citep{bongers2004} and more recently in public health \citep{gill2010}. It should be stressed that the latent class growth analysis as implemented in \proglang{SAS} with \code{proc TRAJ} \citep{jones2001} is a specific case of latent class mixed models in which no random effect is included. As such, it assumes that given a specific latent class, the repeated measures of the same subject are independent. Although of possible interest in an exploratory analysis, the assumption that repeated measures are independent given a restricted number of latent groups is very strict and unlikely so inference based on this approach is usually impossible. 

Further developments of \pkg{lcmm}  will include the development of parallel computations. Indeed, although computationally efficient code in \proglang{Fortran 90} was used for the estimation procedure, the computation may still be long and could benefit from parallel computation, as was the case in the original \proglang{Fortran 90} executable \pkg{HETMIXSURV} available online \citep{HETMIXSURV}.

\section[Aknowledgements]{Acknowledgements}

The authors would like to thank Amadou Diakité and Lionelle Nkam for their contribution to the implementation, Hélène Jacqmin-Gadda and Daniel Commenges for their contribution to the implemented statistical methods, and Jean-Fran\c cois Dartigues and the Paquid Program for sharing a subsample of the Paquid cohort. The development of \pkg{lcmm} was funded by two French agencies: Agence Nationale de la Recherche [grant 2010 PRSP 006 01] and Institut de Recherche en Sant\'e Publique [grant AAP12CanBio16].

\bibliography{biblio}

\begin{thebibliography}{48}
\newcommand{\enquote}[1]{``#1''}
\providecommand{\natexlab}[1]{#1}
\providecommand{\url}[1]{\texttt{#1}}
\providecommand{\urlprefix}{URL }
\expandafter\ifx\csname urlstyle\endcsname\relax
  \providecommand{\doi}[1]{doi:\discretionary{}{}{}#1}\else
  \providecommand{\doi}{doi:\discretionary{}{}{}\begingroup
  \urlstyle{rm}\Url}\fi
\providecommand{\eprint}[2][]{\url{#2}}

\bibitem[{Benaglia \emph{et~al.}(2009)Benaglia, Chauveau, Hunter, and
  Young}]{mixtools}
Benaglia T, Chauveau D, Hunter DR, Young D (2009).
\newblock \enquote{\pkg{mixtools}: An \proglang{R} Package for Analyzing Finite
  Mixture Models.}
\newblock \emph{Journal of Statistical Software}, \textbf{32}(6), 1--29.

\bibitem[{Biernacki \emph{et~al.}(2003)Biernacki, Celeux, and
  Govaert}]{biernacki2003}
Biernacki C, Celeux G, Govaert G (2003).
\newblock \enquote{Choosing Starting Values for the {EM} Algorithm for Getting
  the Highest Likelihood in Multivariate {G}aussian Mixture models.}
\newblock \emph{Computational Statistics and Data Analysis}, \textbf{41}(3-4),
  561--575.

\bibitem[{Blanche \emph{et~al.}(2015)Blanche, Proust-Lima, Loubère, Berr,
  Dartigues, and Jacqmin-Gadda}]{blanche2014}
Blanche P, Proust-Lima C, Loubère L, Berr C, Dartigues JF, Jacqmin-Gadda H
  (2015).
\newblock \enquote{Quantifying and Comparing Dynamic Predictive Accuracy of
  Joint Models for Longitudinal Marker and Time-to-Event in Presence of
  Censoring and Competing Risks.}
\newblock \emph{Biometrics}, \textbf{71}(1), 102--113.

\bibitem[{Bongers \emph{et~al.}(2004)Bongers, Koot, van~der Ende, and
  Verhulst}]{bongers2004}
Bongers IL, Koot HM, van~der Ende J, Verhulst FC (2004).
\newblock \enquote{Developmental Trajectories of Externalizing Behaviors in
  Childhood and Adolescence.}
\newblock \emph{Child development}, \textbf{75}, 1523--37.

\bibitem[{Clauser and Mazor(1998)}]{clauser1998}
Clauser B, Mazor K (1998).
\newblock \enquote{Using Statistical Procedures to Identify Differentially
  Functioning Test Items.}
\newblock \emph{Educational Measurement: Issues and Practice}, \textbf{17},
  31--44.

\bibitem[{Commenges \emph{et~al.}(2012)Commenges, Liquet, and
  Proust-Lima}]{commenges_choice_2012}
Commenges D, Liquet B, Proust-Lima C (2012).
\newblock \enquote{Choice of Prognostic Estimators in Joint Models by
  Estimating Differences of Expected Conditional Kullback-Leibler Risks.}
\newblock \emph{Biometrics}, \textbf{68}(2), 380--387.

\bibitem[{Commenges \emph{et~al.}(2015)Commenges, Proust-Lima, Samieri, and
  Liquet}]{commenges2012}
Commenges D, Proust-Lima C, Samieri C, Liquet B (2015).
\newblock \enquote{A Universal Approximate Cross-Validation Criterion and Its
  Asymptotic Distribution.}
\newblock \emph{International Journal of Biostatistics}, (11(1)), 51--67.

\bibitem[{Fitzmaurice \emph{et~al.}(2009)Fitzmaurice, Davidian, Verbeke, and
  Molenberghs}]{fitzmaurice2009}
Fitzmaurice G, Davidian M, Verbeke G, Molenberghs G (2009).
\newblock \emph{Longitudinal Data Analysis}.
\newblock Handbooks of Modern Statistical Methods. {CRC} Press, Boca Raton.

\bibitem[{Fletcher(1987)}]{fletcher1987}
Fletcher R (1987).
\newblock \emph{Practical Methods of Optimization; (2nd Ed.)}.
\newblock John Wiley \& sons, New York, {NY}, {USA}.

\bibitem[{Genz and Keister(1996)}]{genz1996}
Genz A, Keister BD (1996).
\newblock \enquote{Fully Symmetric Interpolatory Rules for Multiple Integrals
  over Infinite Regions with Gaussian Weight.}
\newblock \emph{Journal of Computational and Applied Mathematics},
  \textbf{71}(2), 299--309.

\bibitem[{Gill \emph{et~al.}(2010)Gill, Gahbauer, Han, and Allore}]{gill2010}
Gill TM, Gahbauer EA, Han L, Allore HG (2010).
\newblock \enquote{Trajectories of Disability in the Last Year of Life.}
\newblock \emph{The New England Journal of Medicine}, \textbf{362}, 1173--80.

\bibitem[{Grün and Leisch(2008)}]{flexmix2}
Grün B, Leisch F (2008).
\newblock \enquote{\pkg{FlexMix} Version 2: Finite Mixtures with Concomitant
  Variables and Varying and Constant Parameters.}
\newblock \emph{Journal of Statistical Software}, \textbf{28}(4), 1--35.
\newblock ISSN 1548-7660.

\bibitem[{Hedeker and Gibbons(2006)}]{hedeker2006}
Hedeker D, Gibbons RD (2006).
\newblock \emph{Longitudinal Data Analysis}.
\newblock Wiley Series in Probability and Statistics. John Wiley \& Sons,
  Hoboken.

\bibitem[{Jacqmin-Gadda \emph{et~al.}(2010)Jacqmin-Gadda, Proust-Lima, Taylor,
  and Commenges}]{jacqmin-gadda2010}
Jacqmin-Gadda H, Proust-Lima C, Taylor JMG, Commenges D (2010).
\newblock \enquote{Score Test for Conditional Independence between Longitudinal
  Outcome and Time to Event Given the Classes in the Joint Latent Class Model.}
\newblock \emph{Biometrics}, \textbf{66}(1), 11--19.

\bibitem[{Jones \emph{et~al.}(2001)Jones, Nagin, and Roeder}]{jones2001}
Jones BL, Nagin DS, Roeder K (2001).
\newblock \enquote{A \proglang{SAS} Procedure Based on Mixture Models for
  Estimating Developmental Trajectories.}
\newblock \emph{Sociological Methods and Research}, \textbf{29}, 374--93.

\bibitem[{Komarek(2009)}]{komarek2009}
Komarek A (2009).
\newblock \enquote{A New \proglang{R} Package for Bayesian Estimation of
  Multivariate Normal Mixtures Allowing for Selection of the Number of
  Components and Interval-Censored Data.}
\newblock \emph{Computational Statistics and Data Analysis}, \textbf{53}(12),
  3932--3947.

\bibitem[{Komárek and Komárková(2014)}]{mixAK}
Komárek A, Komárková L (2014).
\newblock \enquote{Capabilities of \proglang{R} Package \pkg{mixAK} for
  Clustering Based on Multivariate Continuous and Discrete Longitudinal Data.}
\newblock \emph{Journal of Statistical Software}, \textbf{59}(12).
\newblock ISSN 1548-7660.

\bibitem[{Komárek and Verbeke(2002)}]{HETMIXED}
Komárek A, Verbeke G (2002).
\newblock \enquote{A \proglang{SAS} Macro for Linear Mixed Models with Finite
  Normal Mixtures as Random-Effect Distribution.}
\newblock
  \urlprefix\url{https://ibiostat.be/online-resources/online-resources/longitudinal#Mixturelin}.

\bibitem[{Laird and Ware(1982)}]{laird1982}
Laird N, Ware J (1982).
\newblock \enquote{Random-Effects Models for Longitudinal Data.}
\newblock \emph{Biometrics}, \textbf{38}, 963--74.

\bibitem[{Leisch(2004)}]{flexmix}
Leisch F (2004).
\newblock \enquote{\pkg{FlexMix}: A General Framework for Finite Mixture Models
  and Latent Class Regression in \proglang{R}.}
\newblock \emph{Journal of Statistical Software}, \textbf{11}(8), 1--18.

\bibitem[{Letenneur \emph{et~al.}(1994)Letenneur, Commenges, Dartigues, and
  Barberger-Gateau}]{letenneur1994}
Letenneur L, Commenges D, Dartigues JF, Barberger-Gateau P (1994).
\newblock \enquote{Incidence of Dementia and {A}lzheimer's Disease in Elderly
  Community Residents of South-Western {F}rance.}
\newblock \emph{International Journal of Epidemiology}, \textbf{23}(6),
  1256--61.

\bibitem[{Lin \emph{et~al.}(2002)Lin, Turnbull, McCulloch, and Slate}]{lin2002}
Lin H, Turnbull BW, McCulloch CE, Slate EH (2002).
\newblock \enquote{Latent Class Models for Joint Analysis of Longitudinal
  Biomarker and Event Process Data: Application to Longitudinal
  Prostate-Specific Antigen Readings and Prostate Cancer.}
\newblock \emph{Journal of the American Statistical Association}, \textbf{97},
  53--65.

\bibitem[{Little(1995)}]{little1995}
Little RJA (1995).
\newblock \enquote{Modeling the Drop-Out Mechanism in Repeated-Measures
  Studies.}
\newblock \emph{Journal of the American Statistical Association},
  \textbf{90}(431), 1112--1121.

\bibitem[{Marquardt(1963)}]{marquardt1963}
Marquardt D (1963).
\newblock \enquote{An Algorithm for Least-Squares Estimation of Nonlinear
  Parameters.}
\newblock \emph{Journal of the Society for Industrial and Applied Mathematics},
  \textbf{11}(2), 431--441.

\bibitem[{Muth\'{e}n and Muth\'{e}n(2001)}]{muthen2001mplus}
Muth\'{e}n B, Muth\'{e}n L (2001).
\newblock \emph{Mplus User's Guide}.
\newblock Muth\'{e}n and Muth\'{e}n, Los Angeles, CA.

\bibitem[{Muth\'{e}n and Shedden(1999)}]{muthen1999}
Muth\'{e}n B, Shedden K (1999).
\newblock \enquote{Finite Mixture Modeling with Mixture Outcomes Using the {EM}
  Algorithm.}
\newblock \emph{Biometrics}, \textbf{55}(2), 463--9.

\bibitem[{Nagin(1999)}]{nagin1999}
Nagin DS (1999).
\newblock \enquote{Analyzing Developmental Trajectories: A Semiparametric,
  Group-Based Approach.}
\newblock \emph{Psychological Methods}, \textbf{4}(2), 139--57.

\bibitem[{Philipps \emph{et~al.}(2014)Philipps, Amieva, Andrieu, Dufouil
  C.and~Berr, Dartigues, Jacqmin-Gadda, and C.}]{philipps2014}
Philipps V, Amieva H, Andrieu S, Dufouil Cand~Berr C, Dartigues JF,
  Jacqmin-Gadda H, C PL (2014).
\newblock \enquote{{Normalized MMSE for Assessing Cognitive Change in
  Population-Based Aging Studies}.}
\newblock \emph{NeuroEpidemiology}, \textbf{43}(1), 15--25.

\bibitem[{Proust and Jacqmin-Gadda(2005)}]{proust2005}
Proust C, Jacqmin-Gadda H (2005).
\newblock \enquote{Estimation of Linear Mixed Models with a Mixture of
  Distribution for the Random Effects.}
\newblock \emph{Computer Methods and Programs in Biomedicine}, \textbf{78}(2),
  165--173.

\bibitem[{Proust \emph{et~al.}(2006)Proust, Jacqmin-Gadda, Taylor, Ganiayre,
  and Commenges}]{proust2006}
Proust C, Jacqmin-Gadda H, Taylor JMG, Ganiayre J, Commenges D (2006).
\newblock \enquote{A Nonlinear Model with Latent Process for Cognitive
  Evolution Using Multivariate Longitudinal Data.}
\newblock \emph{Biometrics}, \textbf{62}(4), 1014--1024.

\bibitem[{Proust-Lima(2015)}]{HETMIXSURV}
Proust-Lima C (2015).
\newblock \enquote{\code{HETMIXSURV:} Program for the Analysis of Multivariate
  Curvilinear Possibly Heterogeneous Longitudinal Outcomes and a Time-to-Event
  Using a Latent Process Approach.}
\newblock \proglang{Fortran90} program version 2.0,
  \urlprefix\url{http://www.isped.u-bordeaux.fr/BIOSTAT}.

\bibitem[{Proust-Lima \emph{et~al.}(2013)Proust-Lima, Amieva, and
  Jacqmin-Gadda}]{proust-lima2012}
Proust-Lima C, Amieva H, Jacqmin-Gadda H (2013).
\newblock \enquote{Analysis of Multivariate Mixed Longitudinal Data: A Flexible
  Latent Process Approach.}
\newblock \emph{The British Journal of Mathematical and Statistical
  Psychology}, \textbf{66}(3), 470--487.

\bibitem[{Proust-Lima \emph{et~al.}(2011)Proust-Lima, Dartigues, and
  Jacqmin-Gadda}]{proust-lima_misuse_2011}
Proust-Lima C, Dartigues JF, Jacqmin-Gadda H (2011).
\newblock \enquote{Misuse of the Linear Mixed Model When Evaluating Risk
  Factors of Cognitive Decline.}
\newblock \emph{American Journal of Epidemiology}, \textbf{174}(9), 1077--1088.

\bibitem[{Proust-Lima \emph{et~al.}(2015)Proust-Lima, Dartigues, and
  Jacqmin-Gadda}]{proust-lima2015}
Proust-Lima C, Dartigues JF, Jacqmin-Gadda H (2015).
\newblock \enquote{Joint Modelling of Repeated Multivariate Cognitive Measures
  and Competing Risks of Dementia and Death: a latent process and latent class
  approach.}
\newblock \emph{Statistics in Medicine}, \textbf{early view}.

\bibitem[{Proust-Lima \emph{et~al.}(2009)Proust-Lima, Joly, and
  Jacqmin-Gadda}]{proust-lima2009}
Proust-Lima C, Joly P, Jacqmin-Gadda H (2009).
\newblock \enquote{Joint Modelling of Multivariate Longitudinal Outcomes and a
  Time-to-Event: A Nonlinear Latent Class Approach.}
\newblock \emph{Computational Statistics and Data Analysis}, \textbf{53},
  1142--54.

\bibitem[{Proust-Lima \emph{et~al.}(2014)Proust-Lima, S\`ene, Taylor, and
  Jacqmin-Gadda}]{proust-lima2014}
Proust-Lima C, S\`ene M, Taylor J, Jacqmin-Gadda H (2014).
\newblock \enquote{Joint Latent Class Models for Longitudinal and Time-to-Event
  Data: A Review.}
\newblock \emph{Statistical Methods in Medical Research}, \textbf{23}(1),
  74--90.

\bibitem[{Proust-Lima and Taylor(2009)}]{proust-lima2009biostat}
Proust-Lima C, Taylor JMG (2009).
\newblock \enquote{{Development and Validation of a Dynamic Prognostic Tool for
  Prostate Cancer Recurrence Using Repeated Measures of Posttreatment PSA: A
  Joint Modeling Approach.}}
\newblock \emph{Biostatistics (Oxford, England)}, \textbf{10}(3), 535--49.

\bibitem[{Rabe-Hesketh \emph{et~al.}(2004)Rabe-Hesketh, Skrondal, and
  Pickles}]{rabe2004}
Rabe-Hesketh S, Skrondal A, Pickles A (2004).
\newblock \enquote{Generalized Multilevel Structural Equation Modelling.}
\newblock \emph{Psychometrika}, \textbf{69}, 167--90.

\bibitem[{Ramsay(1988)}]{ramsay1988}
Ramsay J (1988).
\newblock \enquote{Monotone Regression Splines in Action.}
\newblock \emph{Statistical Science}, \textbf{3}(4), 425--461.

\bibitem[{{\proglang{R} Development Core Team}(2014)}]{R}
{\proglang{R} Development Core Team} (2014).
\newblock \emph{\proglang{R}: A Language and Environment for Statistical
  Computing}.
\newblock \proglang{R} Foundation for Statistical Computing, Vienna, Austria.
\newblock {ISBN} 3-900051-07-0, \urlprefix\url{http://www.R-project.org}.

\bibitem[{Redner and Walker(1984)}]{redner1984}
Redner R, Walker H (1984).
\newblock \enquote{Mixture Densities, Maximum Likelihood and the {EM}
  Algorithm.}
\newblock \emph{{SIAM} Review}, \textbf{26}(2), 195--239.

\bibitem[{Rizopoulos(2012)}]{Rizopoulos2012}
Rizopoulos D (2012).
\newblock \enquote{Joint Models for Longitudinal and Time-to-Event Data: With
  Applications in \proglang{R}.}
\newblock \emph{Chapman \& {Hall/CRC} Biostatistics Series}.

\bibitem[{{\proglang{SAS} Institute Inc.}(2003)}]{SAS}
{\proglang{SAS} Institute Inc} (2003).
\newblock \emph{\proglang{SAS/STAT} Software, Version~9.3}.
\newblock Cary, NC.
\newblock \urlprefix\url{http://www.sas.com/}.

\bibitem[{Spiessens \emph{et~al.}(2004)Spiessens, Verbeke, and
  Fieuws}]{HETNLMIXED}
Spiessens B, Verbeke Geert aKA, Fieuws S (2004).
\newblock \enquote{A \proglang{SAS} Macro for Nonlinear and Generalised Linear
  Mixed Models with Finite Normal Mixtures as Random-Effect Distribution.}
\newblock
  \urlprefix\url{https://ibiostat.be/online-resources/online-resources/longitudinal#MixedNonLin}.

\bibitem[{Therneau(2013)}]{survival-package}
Therneau TM (2013).
\newblock \emph{{A Package for Survival Analysis in \proglang{S}}}.
\newblock \proglang{R} package version 2.37-4,
  \urlprefix\url{http://CRAN.R-project.org/package=survival}.

\bibitem[{Verbeke and Lesaffre(1996)}]{verbeke1996}
Verbeke G, Lesaffre E (1996).
\newblock \enquote{A Linear Mixed-Effects Model with Heterogeneity in the
  Random-Effects Population.}
\newblock \emph{Journal of the American Statistical Association},
  \textbf{91}(433), 217--221.

\bibitem[{Verbeke and Molenberghs(2000)}]{verbeke2000}
Verbeke G, Molenberghs G (2000).
\newblock \emph{Linear Mixed Models for Longitudinal Data}.
\newblock Springer series in statistics. Springer-Verlag, New-York.

\bibitem[{Xu and Hedeker(2001)}]{xu2001}
Xu W, Hedeker D (2001).
\newblock \enquote{A Random-Effects Mixture Model for Classifying Treatment
  Response in Longitudinal Clinical Trials.}
\newblock \emph{Journal of Biopharmaceutical Statistics}, \textbf{11}(4),
  253--273.

\end{thebibliography}


\end{document}